\documentclass[twoside,11pt]{article}

\usepackage[preprint]{jmlr2e} 

\usepackage{amsfonts,amssymb,amsmath} 
\usepackage{sgame}
\usepackage{upgreek}
\usepackage{bm} 
\usepackage[linesnumbered, ruled,noend]{algorithm2e}
\usepackage{graphicx}
\usepackage{mathrsfs}  %
\usepackage{dutchcal} %
\usepackage{tikz}
\usepackage{cleveref}  %
\allowdisplaybreaks %

\renewcommand{\aa}{\mathbb{A}}
\newcommand{\ee}{\mathbb{E}}

\newcommand{\nn}{\mathbb{N}}

\newcommand{\rr}{\mathbb{R}}

\newcommand{\xx}{\mathbb{X}}
\newcommand{\yy}{\mathbb{Y}}
\newcommand{\zz}{\mathbb{Z}}

\renewcommand{\AA}{\mathcal{A}}
\newcommand{\BB}{\mathcal{B}}

\newcommand{\FF}{\mathcal{F}}
\newcommand{\GG}{\mathcal{G}}
\newcommand{\HH}{\mathcal{H}}
\newcommand{\JJ}{\mathcal{J}}
\newcommand{\KK}{\mathcal{K}}
\newcommand{\LL}{\mathcal{L}}
\newcommand{\MM}{\mathcal{M}}
\newcommand{\NN}{\mathcal{N}}
\newcommand{\PP}{\mathcal{P}}
\newcommand{\QQ}{\mathcal{Q}}
\newcommand{\TT}{\mathcal{T}}
\newcommand{\VV}{\mathcal{V}}
\newcommand{\WW}{\mathcal{W}}
\newcommand{\XX}{\mathcal{X}}
\newcommand{\YY}{\mathcal{Y}}

\newcommand{\newJ}{\mathscr{J}}
\newcommand{\newG}{\mathscr{G}}

\renewcommand{\P}{{\rm Pr}}
\newcommand{\prob}{{\rm Prob}}

\newcommand{\bpi}{\bm{\pi}}
\newcommand{\bPi}{\bm{\Pi}}

\newcommand{\bnu}{\bm{\nu}}

\newcommand{\bA}{\textnormal{\textbf{A}}}

\newcommand{\bG}{\textnormal{\textbf{G}}}
\newcommand{\bH}{\textnormal{\textbf{H}}}
\newcommand{\bM}{\textnormal{\textbf{M}}}

\newcommand{\bX}{\textnormal{\textbf{X}}}

\newcommand{\ba}{\textnormal{\textbf{a}}}
\newcommand{\bd}{\textnormal{\textbf{d}}}
\newcommand{\bh}{\textnormal{\textbf{h}}}
\newcommand{\bs}{\textnormal{\textbf{s}}}

\newcommand{\bx}{\textnormal{\textbf{x}}}

\newcommand{\uniform}{\text{Unif}}

\newcommand{\eq}[1][0]{\bm{\Pi}^{#1{\rm \text{-}eq}}   }

\newcommand{\BR}{{\rm BR}}

\newcommand{\loc}{{\rm loc}}
\newcommand{\mf}{{\rm MF}}

\newcommand{\Emp}{{\rm Emp}}
\newcommand{\sym}{{\rm sym}}
\newcommand{\Subj}{{\rm Subj}}
\newcommand{\SubjBR}{{\rm Subj\text{-}BR}}
\newcommand{\swap}{{\rm swap}}
\newcommand{\comp}{{\rm comp}}
\newcommand{\opt}{{\rm opt}}

\usepackage{xcolor}

\definecolor{ForestGreen}{RGB}{34,139,34}

\newtheorem{assumption}{Assumption}

\usepackage{lastpage}
\jmlrheading{25}{2024}{1-\pageref{LastPage}}{10/22; Revised 11/23}{10/24}{22-1207}{Bora Yongacoglu, G\"{u}rdal Arslan, and Serdar Y\"{u}ksel}

\ShortHeadings{Independent Learning in Mean-Field Games}{Yongacoglu, Arslan, and Y\"{u}ksel}
\firstpageno{1}

\begin{document}

\title{Mean-Field Games With Finitely Many Players: Independent Learning and Subjectivity}

\author{\name Bora Yongacoglu \email bora.yongacoglu@utoronto.ca \\
       \addr Department of Electrical and Computer Engineering\\
        University of Toronto \\
       \AND
       \name G\"{u}rdal Arslan \email gurdal@hawaii.edu \\
       \addr Department of Electrical Engineering \\
       University of Hawaii at Manoa \\
       \AND
       \name Serdar Y\"{u}ksel \email yuksel@queensu.ca \\
       \addr Department of Mathematics and Statistics\\
        Queen's University\\
       }

\editor{Csaba Szepesvari}

\maketitle

\begin{abstract} %
Independent learners are agents that employ single-agent algorithms in multi-agent systems, intentionally ignoring the effect of other strategic agents. This paper studies mean-field games from a decentralized learning perspective with two aims: (i) to identify structure that can guide algorithm design, and (ii) to understand emergent behaviour in systems of independent learners. We study a new model of partially observed mean-field games with finitely many players, local action observability, and partial observations of the global state. Specific observation channels considered include (a) global observability, (b) mean-field observability, (c) compressed mean-field observability, and (d) only local observability. We establish conditions under which the control problem of a given agent is equivalent to a fully observed MDP, as well as conditions under which the control problem is equivalent only to a POMDP. Using the connection to MDPs, we prove the existence of perfect equilibrium among memoryless stationary policies under mean-field observability.  Using the connection to POMDPs, we prove convergence of learning iterates obtained by independent learners under any of our observation channels. We interpret the limiting values as subjective value functions, which an agent believes to be relevant to its control problem. These subjective value functions are used to propose subjective Q-equilibrium, a new solution concept whose existence is proved under mean-field or global observability. Furthermore, we provide a decentralized independent learning algorithm, and by adapting the recently developed theory of satisficing paths to allow for subjectivity, we prove that it drives play to subjective Q-equilibrium. Our algorithm is decentralized, in that it uses only local information for learning and allows players to use different, heterogeneous policies during play. As such, it departs from the conventional representative agent approach common to other algorithms for mean-field games.
\end{abstract}

\begin{keywords}
Multi-agent reinforcement learning, independent learners, learning in games, mean-field games, decentralized systems
\end{keywords}

\section{Introduction}

Mean-field games are a theoretical framework for studying strategic environments with a large number of weakly coupled agents \cite[]{huang2006large, huang2007large, lasry2007mean}. In a mean-field game, the cost and state dynamics of any particular agent are influenced by the collective behaviour of others only through a distributional \textit{mean-field} term. Due to the ubiquity of large-scale decentralized systems in modern engineering, mean-field games have been used to model a diverse range of applications, such as resource management \cite[]{bauso2012robust}, social conventions \cite[]{gomes2014socio}, traffic control \cite[]{chevalier2015micro}, and opinion dynamics \cite[]{stella2013opinion}, among many others. 

Multi-agent reinforcement learning (MARL) is the study of the emergent behaviour in systems of interacting learning agents, with stochastic games serving as the most popular framework for modelling such systems \cite[]{zhang2021multi}. In recent years, there has been a considerable amount of research in MARL that has aimed to produce algorithms with desirable system-wide performance and convergence properties. While these efforts have led to a number of empirically successful algorithms, there are fewer works that offer formal convergence analyses of their algorithms, and the bulk of existing work is suitable only for systems with a small number of agents.

There are several competing paradigms for studying multi-agent learning. The rational learning paradigm, popularized by \cite[]{kalai1993rational}, studies Bayesian agents who maintain beliefs on the strategies (also called policies) used by other agents. Learners select policies to best-respond to their beliefs, while beliefs are updated as new information arrives during the course of play. In this framework, under conditions aligning beliefs with the actual policies used, several results on the convergence of play to equilibrium were proved. However, this framework has two major shortcomings. First, the paradigm does not lend itself to tractable algorithms, as players must maintain beliefs over the strategies used by every other agent, and these strategies can be arbitrarily complex and history-dependent. The second shortcoming of rational learning has to do with the restrictive conditions under which its results hold \cite[]{nachbar2005beliefs}.

One desirable feature of the Bayesian framework is its explicit  modelling of beliefs. Subjectivity and beliefs are relevant in MARL as they allow for players to be uncertain of the model of the game being played. In \cite[]{kalai1995subjective}, the authors proposed a definition for subjective equilibrium in which players behave optimally with respect to their subjective beliefs, while each player's subjective beliefs satisfy a consistency condition with respect to the observed trajectory of play. Competing notions of subjective equilibrium have been proposed, some of which are structural and non-Bayesian: agents believe their system has a particular structure, and agents act in accordance with their assumptions. We refer the reader to \cite[]{arslan2023subjective} for a review.  In this paper, we propose \emph{subjective Q-equilibrium,} a new non-Bayesian notion of subjective equilibrium that is well-suited to decentralized learning in large-scale systems, where agents suffer from a high degree of uncertainty and may need to make structural assumptions for tractability. This equilibrium definition contributes to an emerging line of work that studies simple agents in complex, possibly non-Markovian environments with a perceived local state \cite[]{dong2022simple,chandak2024reinforcement,kara2022convergence, kara2024qlearning}.

Outside of the Bayesian learning paradigm, the majority of theoretical contributions in MARL have focused on highly structured classes of stochastic games, such as two-player zero-sum games \cite[]{sayin2021decentralized,sayin2022fictitious} and $n$-player stochastic teams and their generalizations \cite[]{ding2022independent, fox2022independent, leonardos2022global, mguni2021learning, unlu2023episodic,zhang2022global}. In much of the existing literature on MARL, a great deal of information is assumed to be available to agents while they learn. These assumptions, such as full state observability (as assumed by \cite[]{daskalakis2020independent} among others) or action-sharing among all agents (as assumed by \cite[]{littman1996generalized,Hu2003} and \cite[]{Littman2001ffq}), are appropriate in some settings but are unrealistic in many large-scale, decentralized systems. One issue with designing MARL algorithms that use global information about the local states and actions of all players is that such algorithms do not scale with the number of players. The so-called \textit{curse of many agents} is a widely cited challenge to MARL \cite[]{wang2020breaking}. 

\textit{Independent learners} \cite[]{matignon2009coordination, matignon2012survey} are a class of MARL algorithms that are characterized by intentional obliviousness to the strategic environment: independent learners ignore the presence of other players, effectively treating other players as part of the environment rather than as non-stationary learning agents. By naively running a single-agent reinforcement learning algorithm using only local information, independent learners are relieved of the burden of excessive information, which may lead to scalable algorithms for large-scale decentralized systems. However, additional care must be taken when designing independent learners, as direct application of single-agent reinforcement learning has had mixed success in small empirical studies \cite[]{condon1990,tan1993multi,Sen94,Claus1998}. Here, we offer a new perspective on independent learners by framing their obliviousness as a subjective structural belief on the system within which they exist. Accordingly, agents may act rationally with respect to their (possibly incorrect) subjective beliefs. Building on this interpretation, we propose a notion of subjective equilibrium that is inspired by, and well-suited to, the analysis of independent learners in decentralized MARL. Structural properties of such subjective equilibria, such as their (non-) existence and their comparison to traditional solution concepts, may explain why independent learners lead to stable outcomes in certain MARL applications and not others.

\subsubsection*{The Learning Setting}

We study independent learners in partially observed $n$-player mean-field games. Our model is closely related to that of \cite[]{saldi2018markov}, but allows for a variety of observation channels through which agents can observe non-local information. Decentralization and learning are our primary focuses, and we are interested in online algorithms that involve minimal coordination between agents.  Some salient characteristics of our model are the following:

\vspace{2.5pt}
	
\textit{Limited View of the Global State:} Using a single, unified model, we consider four alternative observation channels by which players observe non-local state information.

\vspace{2.5pt}

\textit{Local Action Information:} A given player does not know the policies used by the remaining players, and the player does not observe the actions generated by these policies.

\

\textit{Decentralized, Online Feedback:} In this study, players to not have access to the model governing the system, nor do they have access to a simulator/oracle/generative model for sampling data in an arbitrary order, nor do they have access to offline data sets that can be used for training. Instead, we assume that players view only the stream of data encountered during play, which is not shared across agents.

\vspace{2.5pt}
	
\textit{Decentralized Training:} This work does not belong to the paradigm of \textit{centralized training with decentralized execution}  \cite[]{lowe2017multi, foerster2018counterfactual}, in which a global value function is learned during training and players select policies that can be implemented in a decentralized manner informed by this global quantity. 

\vspace{2.5pt}

\textit{Heterogeneous Policies:} Since agents learn using different local information encountered during play, we avoid the common requirement that all agents employ the same policy at any given time.

\ \\
\textbf{Contributions:}

\vspace{2.5pt}

1) In Section~\ref{sec:model}, we propose a new model of partially observed $n$-player mean-field games. This model is suitable for the study of decentralized learning and is flexible enough to model various alternative observation channels. In particular, each agent observes its own local actions and partially observes the global state. Observation channels considered include (a)~global state observability, (b) (local state and) mean-field observability (c) compressed mean-field observability, and (d) only local state observability. 

\vspace{2.5pt}

2) In Section~\ref{sec:objective-stationary-equilibrium}, we show that if the remaining players follow (memoryless) stationary policies, then an individual faces a control problem equivalent to a POMDP (Lemma~\ref{lemma:POMDP}). Moreover, under mean-field observability and a symmetry condition on the policies of other players, Theorem~\ref{theorem:mean-field-MDP} states that this control problem is equivalent to a fully observed MDP. We also show that the MDP equivalence does not hold in general when one relaxes either mean-field observability or symmetry of policies. This non-equivalence to an MDP is an important distinction between the model studied here and the limit model with infinitely many agents. In Theorem~\ref{theorem:mean-field-equilibrium}, for the case of mean-field observability, we prove the existence of perfect equilibrium in the set of (memoryless) stationary policies. This result builds on Theorem~\ref{theorem:mean-field-MDP}, and is of independent interest.

\vspace{2.5pt}

3) We study learning iterates obtained by independent learners in a partially observed $n$-player mean-field game. In Theorem~\ref{theorem:naive-learning}, we show that when each agent uses a (memoryless) stationary policy and naively runs single-agent value estimation algorithms (Algorithm~\ref{algo:Q-learning}), its learning iterates converge almost surely under mild assumptions. In Appendix~\ref{appendix:implied-MDPs}, we characterize these almost sure limits in terms of underlying, implied MDPs. We offer a new perspective on independent learning, interpreting it as a subjective model adopted by the learning agent. The limiting values of the agent's learning iterates are then interpreted as subjective value functions, in analogy to objective value functions in single-agent MDPs. Then, we define subjective $\epsilon$-best-responding in Definition~\ref{def:subjective-BR} and we use this to define subjective $\epsilon$-equilibrium (also called subjective Q-equilibrium) in Definition~\ref{def:subjective-equilibrium}. In Lemma~\ref{lemma:mean-field-subjective-equilibrium}, we show that subjective $\epsilon$-equilibrium exists under mean-field observability. 

\vspace{2.5pt}

4) We present Algorithm~\ref{algo:main}, a decentralized independent learning algorithm for partially observed $n$-player mean-field games. We use a modified theory of satisficing paths to show that Algorithm~\ref{algo:main} drives play to subjective $\epsilon$-equilibrium under mean-field observability (Theorem~\ref{theorem:mean-field-state-observability}). Analogous results for other observation channels can be found in Appendix~\ref{appendix:learning-results-under-other-observation-channels}.

\subsection*{Organization}

In \S\ref{ss:related-work} and \S\ref{ss:applications}, we survey related literature. The formal model is presented in Section~\ref{sec:model}. In Section~\ref{sec:objective-stationary-equilibrium}, we examine the connection between partially observed $n$-player mean-field games and partially/fully observed Markov decision problems (POMDP/MDPs), and we prove several structural results. Independent learners and subjectivity are discussed in Section~\ref{sec:naive-learning}, where we study the convergence and interpretation of learning iterates under partial observability. We subsequently define our notion of subjective equilibrium and state an existence result under mean-field observability.  Algorithmic results, including a decentralized learning algorithm (Algorithm~\ref{algo:main}) and its analysis, are given in Section~\ref{sec:satisficing-paths}.  Results of a simulation study are presented in Section~\ref{sec:simulation}.  Proofs, background material, and further results can be found in the expansive appendix sections, including an additional result on objective $\epsilon$-equilibrium under local observability that is presented in Appendix~\ref{appendix:large-N}. 

\vspace{7.5pt}

\noindent\textbf{Notation:} We use $\P$ to denote a probability measure on some underlying probability space, with additional superscripts and subscripts as needed. For a finite set $S$, we let $\rr^{S}$ denote the real vector space of dimension $|S|$ where vector components are indexed by elements of $S$. We let $0 \in \rr^{S}$ denote the zero vector of $\rr^{S}$ and $1 \in \rr^{S}$ denote the vector in $\rr^{S}$ for which each component is 1. For standard Borel sets $S, S'$, we let $\Delta ( S)$ denote the set of probability measures on $S$ with the Borel $\sigma$-algebra on $S$, and we let $\PP ( S' | S)$ denote the set of transition kernels on $S'$ given $S$. We use $Y \sim \mu$ to denote that the random variable $Y$ has distribution $\mu$. For an event $\{ \cdot \}$, we let $\textbf{1} \{ \cdot \} $ denote the indicator function of the event's occurrence.  If a probability distribution $\mu$ is a mixture of distributions $\mu_1, \dots, \mu_n$ with mixture weights $p_1, \dots, p_n$, we write $Y \sim \sum_{i =1}^{n} p_i \mu_i$. For $s \in S$, we use $\delta_{s} \in \Delta ( S)$ to denote the Dirac measure centered at $s$.

\subsection{Related Work} \label{ss:related-work}

A number of research items attempt to approximate solution concepts for mean-field games using learning algorithms. By and large, these works approach learning in mean-field games by analyzing the single-agent problem for a representative agent in the limiting case as $n \to \infty$, and equilibrium is defined using a best-responding condition as well as a consistency condition; see, for instance, \cite[Definition 2.1]{subramanian2019reinforcement}. The solution concepts sought in these works are inherently symmetric: at equilibrium, all agents use the same policy as the representative agent. As a result, such equilibria may be difficult to learn in decentralized learning settings, where agents will use heterogeneous policies during the learning process. We now describe a representative sample of works in this centralized learning tradition, and we refer the reader to \cite[]{lauriere2022learning} for a survey.

In  \cite[]{guo2019learning}, existence and uniqueness of mean-field equilibrium is shown under stringent assumptions, and an algorithm is proposed and analyzed under an assumption that a generative model for sampling transition and cost data is available.

Stationary mean-field games are studied in \cite[]{subramanian2019reinforcement}. Optimality and equilibrium are defined in terms of the limiting invariant distribution of the mean-field state. The authors present multiple notions of equilibrium and study two-timescale policy gradient methods, where a representative agent updates its policy to best respond to its (estimated) environment on the fast timescale and updates its estimate for the mean-field flow on the slow timescale. With access to a simulator for obtaining data, convergence to a local notion of equilibrium is proved. 

A variant of fictitious play for mean-field games is proposed in  \cite[]{elie2020convergence}, though the question of learning to best-respond is black-boxed. Other algorithms based on fictitious play have also been proposed in \cite[]{mguni2018decentralised} and  \cite[]{xie2021learning}.

A value iteration algorithm for computing stationary mean-field equilibrium in both discounted and average-cost mean-field games is presented in \cite[]{anahtarci2020value} and extended in \cite[]{anahtarci2023learning}. Regularized mean-field games are studied and a learning algorithm based on fitted Q-learning is proposed in \cite[]{anahtarci2023q}. Algorithms and operators based on regularization are also studied in \cite[]{cui2021approximately}. The preceding analytical works are closely related to the two-timescale algorithm of \cite[]{zaman2023oracle}. A different two-timescale algorithm is considered in \cite[]{angiuli2022unified}.

In the preceding works, the authors largely restrict their attention to Markov policies that depend only on local state observations, without dependence on the mean-field distribution. At least two research items have considered learning with a focus on \emph{population-dependent policies}, which condition one's action choice on both the local state and the mean-field distribution. In this vein, the existence of so-called master policies is studied and a learning algorithm for their computation is proposed in  \cite[]{perrin2022generalization}. An algorithm for learning mean-field equilibrium in finite horizon games is considered in \cite[]{mishra2023model}, using results from \cite[]{vasal2023sequential}.

Most of the works cited above are \textit{centralized} methods for a decentralized system, as they consider a population of agents using symmetric policies at all times. As a result, the problems studied are single-agent rather than multi-agent in flavour, due to the lack of strategic interaction in the mean-field limit. The principal aim of these papers is to use learning techniques to compute a (near) equilibrium for the mean-field games they study. By contrast, this paper aims to understand patterns of behaviour that emerge when agents use reasonable (if limited) learning algorithms in a shared environment. Our focus, then, is less computational and more descriptive in nature.  

In many realistic settings, agents may employ different learning algorithms for a variety of reasons, such as differing prior beliefs. Moreover, since distinct agents observe distinct local observation histories and feed these local observation histories to their learning algorithms, distinct agents may use radically different policies over the course of learning. Work in the computational tradition largely avoids such learning dynamics, and therefore does not encounter plausible equilibrium policies consisting of various heterogeneous policies used by a population of homogeneous players. In this paper, we depart from the traditional approach of mandating all agents to follow the same policy during learning. 

The learning environment studied in \cite[]{yardim2023policy} is perhaps the closest to the one considered in this paper. \cite[]{yardim2023policy} study a decentralized learning problem in an $n$-player symmetric anonymous game with local observability and propose a mirror descent algorithm. Unlike the papers described earlier, the $n$ agents are allowed to use heterogeneous policies during the course of learning. Under Lipschitz continuity assumptions on various operators, the authors show convergence to equilibrium in the regularized game. 

The primary objective of the literature surveyed above is to provide algorithms for the approximation of conventional variants of mean-field equilibrium. These works consider a variety of models for system interaction and a variety of objective criteria and time horizons. To facilitate the design of such algorithms, many of these works conduct structural analysis on mean-field games, typically in the limit with infinitely many players, and almost always under symmetry conditions mandating that the entire population uses the policy of the representative agent. 

This paper, too, makes contributions on the structure and analysis of mean-field games, focusing on the case with finitely many players and obtaining clear insights both with and without symmetry in policies. This is illustrated by the results and examples in Section~\ref{sec:objective-stationary-equilibrium}. On the other hand, unlike the papers surveyed above, the goal of this paper is not the approximation of mean-field equilibria \textit{per se}. Instead, the goal here is to better understand the outcomes of decentralized learning processes in large-scale systems. Since players use heterogeneous, asymmetric policies during the learning process and iteratively adjust their behaviour as new information arrives, we do not restrict ourselves to the symmetric solution concepts of mean-field equilibrium. In Section~\ref{sec:naive-learning}, we propose a novel solution concept that is well-suited for the study of decentralized learners. In considering alternative solution concepts for mean-field games, this paper is also related to the works \cite[]{lee2021reinforcement,subramanian2022decentralized,muller2022learning,campi2022correlated}. 

In \cite[]{lee2021reinforcement}, mean-field games with strategic complementarities are studied and a trembling hand equilibrium concept is proposed. The authors of  \cite[]{subramanian2022decentralized} propose a definition of decentralized mean-field equilibrium that attempts to capture heterogeneous policies while preserving the overall consistency-optimality form of conventional mean-field equilibrium. Mean-field correlated (and coarse correlated) equilibrium are proposed and studied in  \cite[]{muller2022learning}. A related solution concept called the \emph{correlated solution} of a mean-field game is proposed by \cite[]{campi2022correlated}.

\subsection{Applications of Mean-Field Games} \label{ss:applications}

We now describe some relevant recent applications. For a survey of some classical applications of mean-field games, see \cite[]{gomes2014mean}.

A peer-to-peer energy market is modelled as a mean-field game in \cite[]{xia2019small}. The mean-field quality of the market owes to the large population size and the random matching of consumers and producers, according to which a player's best response depends on the distribution of other agents' budgets (states). 

In \cite[]{li2019efficient}, vehicle dispatch in a ride-sharing application is modelled as independent learning in a mean-field game. The approach used there is well-aligned with the present paper: players observe an aggregate environmental state (describing weather conditions, traffic congestion, and so on) as well as local information about rides in their vicinity, and use this information to guide their action choice. The authors propose an independent MARL algorithm based on Q-learning, which manipulates an object resembling a Q-function: for agent $i$, they consider a quantity denoted $Q_i ( o_i, a_i )$, which attempts to assign a value to playing action $a_i$ after observing the symbol $o_i$. Strictly speaking, this object is not a Q-function, since the agent does not face an MDP. Nevertheless, the authors observe that this approach leads to tractable, decentralized algorithms that outperform their centralized predecessors. 

Traffic routing is often modelled as a mean-field game, where overall travel time depends on the distribution of agents through the traffic network. In \cite[]{salhab2018mean}, traffic routing is modelled as a mean-field game with mean-field observability, where availability of the mean-field is justified through the use of a mobile application. In a related study, \cite[]{cabannes2022solving}, traffic routing is modelled as a mean-field game with only local observability.

Another application of mean-field games involves resource allocation when a large number of agents compete for limited resources, often through auctions. Examples include service in a cellular network \cite[]{manjrekar2019mean}, auctions for advertisements \cite[]{iyer2014mean}, resource distribution in smart grids \cite[]{li2018mean}, and device-to-device wireless transmission \cite[]{li2016incentivizing}.  

\section{Model}   \label{sec:model}
In this section, we present our model of partially observed $n$-player mean-field games, where several strategic agents interact in a shared environment. In subsequent sections, we will investigate the deep connection between our model of mean-field games and the single-agent model of partially observed Markov decision problems (POMDPs). Background material on POMDPs and fully observed MDPs can be found in Appendix~\ref{ss:POMDPs}.

\subsection{Partially Observed Mean-Field Games} \label{ss:MFGs}

For $n \in \nn$, a partially observed $n$-player mean-field game is described by a list $\bG$:
\begin{equation} \label{eq:mean-field-game} 
\bG = ( \NN, \xx , \yy,  \aa,   \{ \varphi^i \}_{i \in \NN}, c , \gamma, P_{\loc}, \nu_0 ). 
\end{equation}

The game $\bG$ consists of $n$ players/agents, indexed by $\NN = \{1, \dots, n \}$. At time $t \in \zz_{\geq 0}$, the local state for player $i \in \NN$ is denoted by $x^i_t \in \xx$, where $\xx$ is a finite set of \emph{local states}. We let $\bX := \times_{i \in \NN} \xx$ denote the $n$-fold Cartesian product of $\xx$, indexed by $\NN$. An element of $\bX$ is called a \emph{global state}, and the global state at time $t$ is denoted by $\bx_t = ( x^i_t )_{i \in \NN}$. The \emph{mean-field} associated with global state $\bx_t$ is denoted by $\mu_t \in \Delta ( \xx )$ and defined as the empirical measure of $\bx_t$: 
\[
	\mu_t ( B ) := \frac{1}{n} \sum_{i \in \NN} \delta_{ x^i_t } ( B ), \quad \forall B \subseteq \xx. 
\]
Each player $i \in \NN$ observes the global state $\bx_t$ indirectly, making the local observation $y^i_t = \varphi^i ( \bx_t )$, where $\yy$ is a finite set and $\varphi^i : \bX \to \yy$ is player $i$'s observation function. Each player $i$ uses its locally observable history variable $h^i_t \in H_t$, to be defined shortly, to select an action $a^i_t \in \aa$, where $\aa$ is a finite set of actions. We denote the \emph{joint action} of all players by $\ba_t = (a^i_t )_{i \in \NN}$, and note that player $i$ only observes its own action $a^i_t$. After choosing action $a^i_t$, player $i$ incurs a stage cost of $c^i_t = c ( x^i_t, \mu_t, a^i_t )$, where $c : \xx \times \Delta( \xx ) \times \aa \to \rr$. Player $i$'s local state then evolves according to $x^i_{t+1} \sim P_{\loc} ( \cdot | x^i_t, \mu_t, a^i_t )$, where $P_{\loc} \in \PP ( \xx | \xx \times \Delta ( \xx ) \times \aa )$. Each player discounts its stream of costs using discount factor $\gamma \in ( 0,1)$, and the initial global state $\bx_0$ has distribution $\nu_0$. The process repeats at time $t+1$ and so on.

\subsubsection*{Observation Channels and Local Information}

In what follows, we refer to the pair $( \yy, \{ \varphi^i \}_{i \in \NN} )$ as the \emph{observation channel} for the game $\bG$. To simplify the statement of assumptions on the observation channel, we define a function $\upmu : \bX \to \Delta( \xx )$ that maps a global state to its empirical distribution:
\[
\left( \upmu ( \bs ) \right) \left( B \right) := \frac{1}{n} \sum_{i \in \NN} \delta_{s^i} ( B) , \quad B \subseteq \xx . 
\]

With this notation, we have the mean-field state $\mu_t = \upmu ( \bx_t )$ in the description of play above. We also define a finite subset $\Emp_n \subset \Delta ( \xx )$ as $\Emp_n := \{ \upmu ( \bs ) : \bs \in \bX \}$. 

\

In order to cover a diverse range of large, decentralized learning environments with a unified model, we now state some alternative observation channels that may be assumed.

{
\begin{assumption}[Global Observability]  \label{ass:global-state-observability} 
\emph{$\yy  = \bX$ and $\varphi^i ( \bs ) = \bs $} for each global state \emph{$\bs \in \bX$} and player \emph{$i \in \NN$.}
\end{assumption}
}

\begin{assumption}[Mean-Field Observability]   \label{ass:mean-field-state-observability}
$\yy  = \xx \times \Emp_n$ and \emph{$\varphi^i ( \bs ) = (s^i , \upmu (  \bs )) $} for each global state \emph{$\bs \in \bX$} and player $i \in \NN$. 
\end{assumption}

\begin{assumption}[Compressed Observability]   \label{ass:compressed-observability}
For $k \in \nn$, let $[k] := \{ 1, 2, \dots, k \}$ and let $f : \Delta( \xx  ) \to [ k ] $. Then, $\yy = \xx \times [k ]$ and for each $i \in \NN$, \emph{$\bs \in \bX$}, we have \emph{$ \varphi^i ( \bs ) = ( s^i , f ( \upmu ( \bs )) .$}
\end{assumption}

\begin{assumption}[Local Observability]   \label{ass:local-state-observability}
$\yy = \xx $ and $ \varphi^i ( \bs ) = s^i $ for each $i \in \NN, \bs \in \bX$. 
\end{assumption}

Assumptions~\ref{ass:global-state-observability}--\ref{ass:local-state-observability} are presented in order of increasing decentralization. In practice, the particular choice of informational assumption will depend on one's application area: in some instances, there will be a natural restriction of information leading to a particular observation channel. In other instances, agents may voluntarily compress a more informative observation variable for the purposes of function approximation. We now briefly describe these alternative observation channel assumptions. Additional discussion on this topic is provided in Section~\ref{sec:discussion}.

In decentralized systems, global observability is often unrealistic, but this case is included in our discussion as it is helpful for developing insights. This observation channel has previously been used in some works on linear-quadratic mean-field games  \cite[]{zaman2020reinforcement,zaman2020approximate}. Mean-field observability is perhaps the most commonly considered observation channel in works on mean-field games: see \cite[]{saldi2018markov, zaman2023oracle} and the references therein. Local observability is also a commonly adopted assumption (see, for instance, \cite[]{muller2022learning,yardim2023policy} and the references therein). 

Compressed observability is an intermediate setting between mean-field and local observability. To our knowledge, this has not been explicitly considered in prior theoretical work. Compressed observability can be motivated using the discussion above: it serves to lessen the computational burden at a given learning agent in a partially observed $n$-player mean-field game and, as we discuss in Section~\ref{sec:discussion}, may be a more appropriate modelling assumption in some applications. By taking $k = 1$, we see that local state observability is in fact a special case of compressed state observability, where the compressed information about the mean-field state is uninformative. We include Assumption~\ref{ass:local-state-observability} separately to highlight the importance of this set-up. 

Where possible, we conduct our analysis in a unified manner, without specifying the observation channel. In some instances, stronger results can be proved under richer observation channels. In such cases, our exposition and analysis centres on mean-field observability, with discussion on other information structures either omitted or postponed. 

\subsubsection*{Local Histories and Policies}  We now formalize the action selection process by defining policies. We make distinctions between the overall system history and each player's locally observable histories. For any $t \in \zz_{\geq 0}$, we define the sets
\begin{align*}
	\bH_t 	&:= \left( \bX \times \bA \right)^t \times \bX ,  \\
 	H_t		&:= \Delta ( \bX ) \times \left( \yy \times \aa \times \rr  \right)^t \times \yy.
\end{align*}

For any $t \geq 0$, the set $ \bH_t $ represents the set of overall system histories of length $t$, while the set $ H_t $ is the set of histories of length $t$ that an individual player in the game $\bG$ may observe. Elements of $\bH_t$ are called \textit{system histories of length $t$}, and elements of $H_t$ are called \emph{observable histories of length $t$}. 

For each $t \geq 0$ and player $i \in \NN$, we let 
\begin{align*}
\bh_t 	&:= ( \bx_0 , \ba_0, \cdots, \bx_{t-1}, \ba_{t-1}, \bx_t) ,  \\
 h^i_t 	&:= ( \nu_0, y^i_0, a^i_0, c^i_0, \cdots, y^i_{t-1}, a^i_{t-1}, c^i_{t-1}, y^i_t )
\end{align*}
denote the $t^{\rm th}$ \textit{system history variable} and player $i$'s $t^{\rm th}$ \emph{locally observable history variable}, respectively. Note that $\bh_t$ is a random quantity taking values in $\bH_t$, while $h^i_t$ is a random quantity taking values in $H_t$. 

\begin{definition}
A sequence $\pi^i = ( \pi^i_t )_{t \geq 0}$ with $\pi^i_t \in \PP ( \aa | H_t )$ for every $t \geq 0$ is called a \emph{policy} for player $i$. We let $\Pi^i$ denote the set of all policies for player $i$. 
\end{definition}

\begin{definition}
A policy $\pi^i \in \Pi^i$ is called \emph{(memoryless) stationary (or simply stationary)} if, for some $f^i \in \PP ( \aa | \yy )$, the following holds: for any $t \geq 0$ and any $\tilde{h}_t = ( \nu, \tilde{y}_0, \dots, \tilde{y}_t ) \in H_t$, we have $\pi^i_t ( \cdot | \tilde{h}_t ) = f^i ( \cdot | \tilde{y}_t ). $ We let $\Pi^i_{S}$ denote the set of stationary policies for player $i$.
\end{definition}

\noindent \textbf{Remark:} The set of policies -- and thus learning algorithms -- available to an agent depends on the set of locally observable histories, which itself depends on the observation channel $( \yy , \{ \varphi^i \}_{i \in \NN })$. In this paper, our focus is on \emph{independent learners}, which are learners that do not use the joint action information in their learning algorithms. Here, we have chosen to incorporate this constraint into the information structure. Moreover, to underscore the importance of learning in our study, we also do not assume that the players know the cost function $c$. Instead, we assume only that they receive feedback costs in response to particular system interactions. These assumptions on the information structure resemble those of other work on independent learners \cite[]{daskalakis2020independent,sayin2021decentralized, YAY-TAC, matignon2009coordination, matignon2012survey, wei2016lenient,yongacoglu2023satisficing}. A rather different information structure is that of \emph{joint action learners}, studied by  \cite[]{Claus1998} among others, where the locally observable history variables also include the joint action history. 

\ \\
\textbf{Notation:} We let $\bPi := \times_{i \in \NN} \Pi^i$ denote the set of \textit{joint policies}. To isolate player $i$'s component in a particular joint policy $\bpi \in \bPi$, we write $\bpi = ( \pi^i, \bpi^{-i} )$, where $-i$ is used in the agent index to represent all agents other than $i$. Similarly, we write the joint policy set as $\bPi = \Pi^i \times \bPi^{-i}$, a joint action may be written as $\ba = (a^i , \ba^{-i}) \in \bA := \aa^{\NN}$, and so on.

\subsubsection*{Value Functions, Best-Responding, and Objective Equilibrium} 

For each player $i \in \NN$, we identify the set of stationary policies $\Pi^i_{S}$ with the set $\PP ( \aa | \yy )$ of transition kernels on $\aa$ given $\yy$. When convenient, a stationary policy $\pi^i \in \Pi^i_{S}$ is treated as if it were an element of $\PP ( \aa | \yy )$, and reference to the locally observable history variable is omitted. For each $i \in \NN$, we introduce the metric $d^i$ on $\Pi^i_{S}$, defined by 
\[
d^i ( \pi^i, \tilde{\pi}^i ) := \max \{ | \pi^i ( a^i | y ) - \tilde{\pi}^i ( a^i | y ) | : y \in \yy, a^i \in \aa \}, \quad \forall \pi^i, \tilde{\pi}^i \in \Pi^i_{S}.
\]
We metrize the set of stationary joint policies $\bPi_{S}$ with a metric $\bd$, defined as
\[
\bd ( \bpi , \tilde{\bpi} ) := \max_{i \in \NN} d^i ( \pi^i , \tilde{\pi}^i ), \quad \forall \bpi, \tilde{\bpi} \in \bPi_{S}. 
\] 

A metric $\bd^{-i}$ for the set $\bPi^{-i}_{S}$ is defined analogously to $\bd$. We have that the sets $\{ \Pi^i_{S} \}_{ i \in \NN}$, $\{ \bPi^{-i}_{S} \}_{i \in \NN}$ and $\bPi_{S}$ are all compact in the topologies induced by the corresponding metrics.

\ \\
For any joint policy $\bpi = ( \pi^i )_{i \in \NN}$ and initial distribution $\nu \in \Delta ( \bX )$, there exists a unique probability measure $\P^{\bpi}_{\nu}$ on trajectories in $( \bX \times \bA )^{\infty}$ satisfying the following, for all $t \geq 0$:
(i) $\P^{\bpi}_{\nu} \left( \bx_0 \in \cdot \right) = \nu ( \cdot ) $. (ii) For any $i \in \NN$, $\P^{\bpi}_{\nu} \left( a^i_t \in \cdot \middle| h^i_t  \right) =  \pi^i_t \left( \cdot | h^i_t  \right)$. (iii) The collection $\{ a^j_t \}_{j \in \NN}$ is jointly independent given $\bh_t$. (iv) For any $i \in \NN$ and $t \geq 0$, local states evolve according to $\P^{\bpi}_{\nu} \left( x^i_{t+1} \in \cdot \middle| \bh_t, \ba_t     \right) = P_{\loc} \left( \cdot \middle| x^i_t, \mu_t , a^i_t \right)$. (v) The collection $\{ x^j_{t+1} \}_{j \in \NN}$ is jointly independent given $(\bh_t, \ba_t) $.

\vspace{2.5pt}

We let $E^{\bpi}_{\nu}$ denote the expectation associated with $\P^{\bpi}_{\nu}$ and we use it to define player $i$'s objective function, also called the (state) value function:
\[
J^i ( \bpi,  \nu ) := E^{\bpi}_{\nu}  \left[ \sum_{t = 0}^\infty \gamma^t c^i_t \right] = E^{\bpi}_{\nu} \left[ \sum_{t = 0}^{\infty} \gamma^t c ( x^i_t, \mu_t, a^i_t ) \right]  .
\]

\begin{lemma} \label{lemma:continuous-cost}
For any initial measure $\nu \in \Delta (\bX)$ and any player $i \in \NN$, the mapping $\bpi \mapsto J^i ( \bpi , \nu )$ is continuous on $\bPi_{S}$. 
\end{lemma}

The proof is omitted, as it resembles that of \cite[Lemma 2.10]{yongacoglu2023satisficing}. 

\vspace{2.5pt}

From the final expression in the definition of $J^i ( \bpi , \nu)$, one can see that player $i$'s objective is only weakly coupled with the rest of the system: player $i$'s costs depend on the global state and joint action sequences $\{ \bx_t, \ba_t \}_{t \geq 0}$ only through player $i$'s components $\{ x^i_t, a^i_t \}_{t \geq 0}$, the mean-field state sequence $\{ \mu_t \}_{t \geq 0}$, and the subsequent influence that $\{ \mu_t \}_{t \geq 0}$ has on the evolution of $\{ x^i_t \}_{t \geq 0}$. Nevertheless, player $i$'s objective function does depend on the policies of the remaining players. This motivates the following definitions. 

\begin{definition} \label{def:best-response}
Let $\epsilon \geq 0$, $\nu \in \Delta ( \bX )$, $i \in \NN$, and $\bpi^{-i} \in \bPi^{-i}$. A policy $\pi^{*i} \in \Pi^i$ is called an $\epsilon$-\emph{best-response to $\bpi^{-i}$ with respect to $\nu$} if 
\[
J^i ( \pi^{*i}, \bpi^{-i} ,  \nu ) \leq \inf_{ \pi^i \in \Pi^i } J^i ( {\pi}^i  , \bpi^{-i} ,  \nu ) + \epsilon. 
\] 
\end{definition}

For $i \in \NN$, $\bpi^{-i} \in \bPi^{-i}$, $\epsilon \geq 0$, and $\nu \in \Delta ( \bX )$, we let $\BR^i_{\epsilon} ( \bpi^{-i}, \nu ) \subseteq \Pi^i$ denote player $i$'s set of $\epsilon$-best-responses to $\bpi^{-i}$ with respect to $\nu$. If, additionally, $\pi^i \in \BR^i_{\epsilon} ( \bpi^{-i} , \nu )$ for all $\nu \in \Delta (\bX)$, then $\pi^i$ is called a \emph{uniform $\epsilon$-best-response to $\bpi^{-i}$}. The set of uniform $\epsilon$-best-responses to a policy $\bpi^{-i}$ is denoted $\BR^i_{\epsilon} ( \bpi^{-i} )$.

\begin{definition}
Let $\epsilon \geq 0$, $\nu \in \Delta ( \bX)$, and $\bpi^{*} \in \bPi$. The joint policy $\bpi^{*}$ is called an \emph{$\epsilon$-equilibrium with respect to $\nu$}  if $\pi^{*i}$ is an $\epsilon$-best-response to $\bpi^{*-i}$ with respect to $\nu$ for every player $i \in \NN$. Additionally, if $\bpi^{*} \in \bPi $ is an $\epsilon$-equilibrium with respect to every $\nu \in \Delta ( \bX)$, then $\bpi^{*}$ is called a \emph{perfect $\epsilon$-equilibrium.}
\end{definition}

For $\epsilon \geq 0$ and $\nu \in \Delta  ( \bX )$, we let $\eq[\epsilon] ( \nu ) \subset \bPi $ denote the set of $\epsilon$-equilibrium policies with respect to $\nu$, and we let $\eq[\epsilon]$ denote the set of perfect $\epsilon$-equilibrium policies. Furthermore, we let $\eq[\epsilon]_{S} (\nu)  := \eq[\epsilon] (\nu ) \cap \bPi_{S}$ for each $\nu \in \Delta ( \bX )$ and we let $\eq[\epsilon]_{S} := \eq[\epsilon] \cap \bPi_{S}$.

\subsubsection*{Further Comments on Models of Mean-Field Games}

The model above differs from the classical model of mean-field games, which assumes a continuum of agents. Here, we consider models with a possibly large but finite number of symmetric, weakly coupled agents. Our model closely resembles that of \cite[]{saldi2018markov}, which studies existence of equilibrium and allows for general state and actions spaces. Unlike \cite[]{saldi2018markov}, we consider games with finite state and action spaces, and we consider a variety of observation channels.  Mean-field games can be viewed as limit models of $n$-player symmetric stochastic games, where players are exchangeable and symmetric. A number of papers have formally examined the connection between games with finitely many players and the corresponding limit model, including the works of \cite[]{fischer2017connection, saldi2018markov,sanjari2022optimality}.


\section{From Games to MDPs and Back: Existence of Equilibrium}    \label{sec:objective-stationary-equilibrium}

In this section, we explore the deep connection between POMDPs and $n$-player mean-field games. We begin by observing that, in any partially observed $n$-player mean-field game, player $i$ faces a stochastic control problem equivalent to a POMDP whenever its counterparts use a stationary policy. Next, under mean-field observability and a symmetry condition on the joint policy $\bpi^{-i}$, we prove that player $i$ faces a problem equivalent to a \emph{fully observed} MDP. This result is of independent interest, but it additionally enables a proof that, under mean-field observability, the set of stationary joint policies admits a perfect equilibrium. 
 
\begin{lemma} \label{lemma:POMDP}
Let $\bG$ be a partially observed $n$-player mean-field game. Fix $i \in \NN$ and let $\bpi^{-i} \in \bPi^{-i}_{S}$ be a stationary policy for the remaining players. Then, player $i$ faces a partially observed Markov decision problem $\MM_{\bpi^{-i}}$, with partially observed state process \emph{$\{ \bx_t \}_{t \geq 0}$ and observation process $\{ y^i_t \}_{t \geq 0}$.} 
\end{lemma}

Lemma~\ref{lemma:POMDP}, whose proof is straightforward and omitted, gives conditions under which a player faces a POMDP. Under certain additional conditions, such as those presented in Theorem~\ref{theorem:mean-field-MDP}, one can show that player $i \in \NN$ faces a fully observed MDP. When player $i$ faces an MDP in its observation variable, the classical theory of MDPs and reinforcement learning can be brought to bear on player $i$'s optimization problem, leading to results on the existence of certain equilibrium policies and characterization of one's best-response set.

\subsection{An MDP Reduction under Mean-Field Observability}

We now state and prove Theorem~\ref{theorem:mean-field-MDP}, which provides sufficient conditions for a given player to face an MDP in the mean-field game. Theorem~\ref{theorem:mean-field-MDP} will later play a critical role in the proof of Theorem~\ref{theorem:mean-field-equilibrium}, the second main result of this section, which asserts that the game admits a perfect equilibrium in the set of stationary policies.

\begin{definition}
For players $i, j \in \NN$, let $\pi^i \in \Pi^i_{S}$, $\pi^j \in \Pi^j_{S}$ be stationary policies. We say that the policies $\pi^i$ and $\pi^j$ are \emph{symmetric} if both are identified with the same transition kernel in $\PP ( \aa | \yy )$. For any subset of players $I \subset \NN$, a collection of policies $\{ \pi^i \}_{i \in I}$ is called symmetric if, for every $i, j \in I$, we have that $\pi^i$ and $\pi^j$ are symmetric. 
\end{definition}

\begin{theorem}  \label{theorem:mean-field-MDP}
Let $\bG$ be a partially observed $n$-player mean-field game with mean-field observability (that is, Assumption~\ref{ass:mean-field-state-observability} holds). Let $i \in \NN$. If $\bpi^{-i} \in \bPi^{-i}_{S}$ is symmetric, then $i$ faces a fully observed MDP $\MM_{\bpi^{-i}}$ with controlled state process $\{ y^i_t \}_{t \geq 0}$, where \emph{$y^i_t = \varphi^i ( \bx_t ) = ( x^i_t , \upmu( \bx_t ))$} for all $t \geq 0$. 
\end{theorem}

For the proof of Theorem~\ref{theorem:mean-field-MDP}, we introduce the following notation: $\Sigma_{\NN}$ is the set of permutations on $\NN$. For a given permutation $\sigma \in \Sigma_{\NN}$ and a global state $\bs = (s^i)_{i\in\NN}$, let $\sigma(\bs)$ be the global state in which $\sigma(\bs)^i := s^{\sigma(i)}$ for each player $i \in \NN$. In other words, the local state of player $i$ in global state $\sigma(\bs)$ is given by the local state of player $\sigma(i)$ in global state $\bs$. Similarly, for a joint action $\ba = (a^i)_{i \in \NN}$, we let $\sigma(\ba)^i := a^{\sigma(i)}$ for each player $i \in \NN$.

\vspace{10pt} 

\begin{proof} Fix $i \in \NN$. Recall that $\yy = \xx \times \Emp_n$ and player $i$'s observations are given by $\varphi^i ( \bs) = ( s^i , \upmu ( \bs ))$ for any $\bs \in \bX$. For any $k \geq 0$, we let $y^i_{k} := \varphi^i ( \bx_{k}) = ( x^i_{k}, \upmu ( \bx_{k} ))$. 


Let $\nu \in \Delta (\bX)$ be any initial distribution of the global state variable and let $\bpi^{-i} \in \bPi^{-i}_{S}$ be a symmetric policy for the remaining players. Let $\pi^i \in \Pi^i$ be any policy for player $i$, and let $\bpi = ( \pi^i, \bpi^{-i})$. Note that player $i$'s cost at time $t$, $c^i_t$, is a measurable function of $(y^i_t, a^i_t)$:
\[
c^i_t =  c ( x^i_t, \upmu ( \bx_t ) , a^i_t ) = c ( y^i_t, a^i_t ), \, \forall t \geq 0.
\]

That is, player $i$'s observation variable $y^i_t$ summarizes the cost-relevant part of the system history for player $i$. Therefore, we must show that the following holds for all $t \geq 0$, any $\Upsilon \subseteq \yy$, and in a time invariant manner:
\begin{align*}
\P^{\bpi}_{\nu} \left[	y^i_{t+1} \in \Upsilon \middle| \{ y^i_{k}, a^i_{k}  : 0 \leq k \leq t  \}	\right] = \P^{\bpi}_{\nu} \left[  y^i_{t+1} \in \Upsilon \middle| y^i_t, a^i_t  \right],
\end{align*}
$\P^{\bpi}_{\nu}$-almost surely.\footnote{For brevity, we omit the qualifier ``$\P^{\bpi}_{\nu}$-almost surely'' for all subsequent equalities involving conditional expectations. The time invariance described here is for time homogeneity of the MDP, and requires that the right side of this expression does not depend on $t$.}

Fix $t \geq 0$. For any $\Upsilon \subseteq \yy$, let $\textbf{1}_{\Upsilon}$ denote the indicator function of the event $\{ y^i_{t+1} \in \Upsilon \}$. Using the law of iterated expectations and conditioning on the (finer) $\sigma$-algebra generated by the random variables $\{ \bx_k, a^i_k : 0 \leq k \leq t \}$, we have
\begin{align*}
E^{\bpi}_{\nu} \left(   \textbf{1}_{\Upsilon}   \middle| \{ y^i_k , a^i_k : 0 \leq k \leq t	 \} \right) = &E^{\bpi}_{\nu} \left[ E^{\bpi}_{\nu} \left(  \textbf{1}_{\Upsilon}   \middle| \{ \bx_k , a^i_k : 0 \leq k \leq t \}	\right) \middle|  \{ y^i_k , a^i_k : 0 \leq k \leq t \} \right] \\
= &E^{\bpi}_{\nu} \left[ E^{\bpi}_{\nu} \left(	  \textbf{1}_{\Upsilon}   \middle| \bx_t , a^i_t  \right) \middle|  \{ y^i_k , a^i_k : 0 \leq k \leq t \} \right].
\end{align*}
Thus, it suffices to show that 
\begin{equation} \label{eq:sufficient-condition-for-MDP-reduction}
E^{\bpi}_{\nu} \left( \textbf{1}_{\Upsilon} \middle| \bx_t , a^i_t \right) = E^{\bpi}_{\nu} \left( \textbf{1}_{\Upsilon} \middle| y^i_t, a^i_t \right)
\end{equation} 
holds for all $\Upsilon \subseteq \yy$: since the left side of \eqref{eq:sufficient-condition-for-MDP-reduction} does not vary with $t$, this will establish that $\{y^i_t\}_{t \geq 0}$ is a (time homogeneous) controlled Markov process controlled by $\{ a^i_t \}_{t \geq 0}$. Moreover, since $\yy$ is a finite set, it suffices to show that this holds for all singletons $\Upsilon = \{ w \}$, $w \in \yy$.

Fix $w \in \yy$ and note that the events $\{ y^i_{t+1} = w \}$ and $\{ \bx_{t+1} \in \left(\varphi^i\right)^{-1} ( w ) \}$ are equivalent. We claim that 
\begin{align} \label{eq:mean-field-independence} 
	&\P^{\bpi}_{\nu} \left[ \bx_{t+1} \in \left(\varphi^i\right)^{-1} ( w ) \middle| \bx_t = \bs , a^i_t = a^i \right]   \nonumber \\   
	=&\P^{\bpi}_{\nu} \left[ \bx_{t+1} \in \left(\varphi^i\right)^{-1}  ( w ) \middle| \bx_t \in  \left(\varphi^i\right)^{-1} ( \varphi^i ( \bs )) , a^i_t = a^i \right]
\end{align}
holds for any $\bs \in \bX$ and any $a^i \in \aa$, where $ \left(\varphi^i\right)^{-1} ( \varphi^i ( \bs )) $ denotes the pre-image of $\varphi^i(\bs)$. By mean-field observability, we can explicitly characterize the set  $ \left(\varphi^i\right)^{-1} ( \varphi^i ( \bs )) $ as
\begin{align} \label{eq:characterize-observation-set}
\left(\varphi^i\right)^{-1} ( \varphi^i ( \bs ))  &= \left\{ \tilde{\bs} \in \bX : \varphi^i ( \tilde{\bs} ) = \varphi^i (\bs ) \right\} 		\nonumber \\
 								&= \left\{ \tilde{\bs} \in \bX : \tilde{s}^i = s^i \text{ and } \upmu ( \tilde{\bs} ) = \upmu (\bs) \right\}  \nonumber \\
								&= {\left\{ \sigma(\bs) \middle| \sigma \in \Sigma_{\NN} : \sigma(i) = i \right\} .}   
\end{align}

The final expression comes from the fact that $\varphi^i ( \bs ) = \varphi^i (\tilde{\bs} )$ $\iff$ $\tilde{\bs}$ can be obtained from $\bs$ by permuting the local states of other players while leaving $i$'s local state fixed.

To verify the claim in \eqref{eq:mean-field-independence}, let $\tilde{\bs} \in \left(\varphi^i\right)^{-1} ( \varphi^i ( \bs) )$ be an arbitrary global state with $\varphi^i ( \tilde{\bs} ) = \varphi^i ( \bs )$. Let $\sigma \in \Sigma_{\NN}$ be a permutation of the players such that $\sigma ( i ) = i$ and $\tilde{s}^j = s^{\sigma(j)}$ for all $j \in \NN \setminus \{ i \}$. Using the notational conventions described earlier, we have that $\tilde{\bs} = \sigma ( \bs )$.

Since the policies $\pi^{j}$ and $\pi^{\sigma(j)}$ are symmetric, for any $\bs \in \bX$ and $\ba \in \bA$, one has 
\begin{equation}  \label{eq:action-permutation}
   \P^{\bpi}_{\nu} \left( \ba_t = \ba \middle| \bx_t =  \bs , a^i_t = a^i \right) = \P^{\bpi}_{\nu} \left( \ba_t = \sigma( \ba ) \middle| \bx_t = \sigma( \bs ), a^i_t = a^i \right) 
\end{equation}

Since the local states $\{ x^j_{t+1} \}_{j \in \NN}$ are conditionally independent given $( \bx_t , \ba_t )$, the global state transition function has a product structure, which is invariant under permutations: 
\begin{align} \label{eq:global-state-transition-permutation} 
\P^{\bpi}_{\nu} \left( \bx_{t+1} = \bs' \middle| \bx_t = \bs, \ba_t = \ba \right) &= \prod_{j \in \NN} P_{\loc} \left( s'^{j} \middle| s^j, \upmu ( \bs ), a^j \right)  \nonumber  \\
														&{= \prod_{j \in \NN} P_{\loc} \left( s'^{\sigma(j) } \middle| s^{\sigma(j)}, \upmu ( \sigma( \bs) ), a^{\sigma(j)} \right) }  \nonumber \\
														&{= \P^{\bpi}_{\nu} \left( \bx_{t+1} = \sigma(\bs') \middle| \bx_t = \sigma(\bs) , \ba_t = \sigma(\ba) \right) }.
\end{align}

\noindent Using the law of total probability, conditioning on $\ba_t$, and recalling that $\tilde{\bs} = \sigma (\bs )$, it follows that for any $\bs' \in \bX$,
\begin{align} \label{eq:state-permutation}
\P^{\bpi}_{\nu} \left[   \bx_{t+1} = \bs' \middle| \bx_t = \bs , a^i_t = a^i \right]  = \P^{\bpi}_{\nu} \left[ \bx_{t+1} = \sigma( \bs' ) \middle| \bx_t = \tilde{\bs}, a^i_t = a^i \right].  
\end{align}

\noindent Noting that $\upmu ( \bs' ) = \upmu ( \sigma ( \bs' ) )$, we have that $\bs' \in \left(\varphi^i\right)^{-1} ( w ) \iff \sigma (\bs' ) \in \left(\varphi^i\right)^{-1} (w )$. Thus,
\begin{align}   \label{eq:permutation-representation-of-pre-image}
{\left(\varphi^i\right)^{-1} ( w ) =  \bigcup_{ \bs' \in \left(\varphi^i\right)^{-1} ( w ) } \{   \bs'   \} = \bigcup_{ \bs' \in \left(\varphi^i\right)^{-1} ( w ) } \{ \sigma ( \bs' )  \}.}  
\end{align}

\noindent  It follows that 
\begin{align*}
\P^{\bpi}_{\nu} \left( \bx_{t+1} \in \left(\varphi^i\right)^{-1} ( w ) \middle| \bx_t = \bs , a^i_t = a^i \right ) 
= &\sum_{ \bs' \in \left(\varphi^i\right)^{-1} ( w ) } \P^{\bpi}_{\nu}  \left( \bx_{t+1} = \bs' \middle| \bx_t = \bs , a^i_t = a^i 	\right)  \\
= &\sum_{ \bs' \in \left(\varphi^i\right)^{-1} ( w ) } \P^{\bpi}_{\nu} \left( \bx_{t+1} = \sigma(\bs ') \middle| \bx_t = \sigma( \bs ) , a^i_t = a^i 	\right) \\
= &\quad \P^{\bpi}_{\nu} \left( \bx_{t+1} \in \left(\varphi^i\right)^{-1} ( w ) \middle| \bx_t = \sigma ( \bs ) , a^i_t = a^i \right).
\end{align*}

\noindent Since $\tilde{\bs} \in \left(\varphi^i\right)^{-1} ( \varphi^i (\bs))$ was arbitrary, using \eqref{eq:characterize-observation-set}, we see that
\begin{align} \label{eq:equal-conditional-probabilities} 
\P^{\bpi}_{\nu} \left( \bx_{t+1} \in \left(\varphi^i\right)^{-1} ( w ) \middle| \bx_t = \bs , a^i_t = a^i \right)  = \P^{\bpi}_{\nu} \left( \bx_{t+1} \in \left(\varphi^i\right)^{-1} ( w ) \middle| \bx_t = \bar{\bs} , a^i_t = a^i \right)   ,
\end{align}
for any $\bar{\bs} \in \left(\varphi^i\right)^{-1} ( \varphi^i ( \bs ))$. We conclude that \eqref{eq:mean-field-independence} holds by applying iterated expectations to its right-hand side, conditioning  on $\{ \bx_t  , a^i_t \}$, and using \eqref{eq:equal-conditional-probabilities} to simplify the resulting summation.

Finally, the event $\{ \bx_t \in \left(\varphi^i\right)^{-1} ( \varphi^i ( \bs )) \}$ is equivalent to the event $\{ y^i_t = \varphi^i ( \bs ) \}$ and $\{ \bx_{t+1} \in \left(\varphi^i\right)^{-1} ( w ) \}$ is equivalent to $\{ y^i_{t+1} = w \}$. We have therefore shown that
\begin{align*}
\P^{\bpi} \left[ \bx_{t+1} \in \left(\varphi^i\right)^{-1} ( w ) \middle| \bx_t = \bs , a^i_t = a^i \right]      
= \P^{\bpi} \left[ y^i_{t+1} =  w  \middle| y^i_t = \varphi^i ( \bs ) , a^i_t = a^i \right],
\end{align*}
for any $ w \in \yy$, $\bs \in \bX$, and $a^i \in \aa$. The result follows.
\end{proof}


\subsubsection*{Comments on the Proof of Theorem~\ref{theorem:mean-field-MDP}}

The proof of Theorem~\ref{theorem:mean-field-MDP} depends critically on the connection between local observations and hidden global states described in \eqref{eq:characterize-observation-set} : under mean-field observability, two global states $\bs, \tilde{\bs} \in \bX$ give rise to the same local observation for player $i$, i.e. $\varphi^i ( \bs ) = \varphi^i ( \tilde{\bs})$, if and only if they agree on the local state of player $i$ (i.e. $s^i = \tilde{s}^i$) \emph{and} one can be obtained from the other by permuting the local states of the remaining agents. This specific characterization of pre-images of $\varphi^i $ holds only for mean-field observability and fails under compressed or local observability. As a result, the proof above cannot be used in those settings. Indeed, we will show by counterexample in Section~\ref{ss:counterexamples} that both mean-field observability and symmetry of the policy $\bpi^{-i}$ were necessary in Theorem~\ref{theorem:mean-field-MDP} and cannot be relaxed without loss of generality. 

A byproduct of the proof of Theorem~\ref{theorem:mean-field-MDP} concerns the conditional distribution of player $i$'s observation-action trajectories, $\{y^i_t, a^i_t \}_{t \geq 0}$, conditional on the initial observation $y^i_0$. In particular, with $\pi^i$ arbitrary and $\bpi^{-i} \in \bPi^{-i}_{S}$ symmetric, putting $t = 0$ in equation \eqref{eq:sufficient-condition-for-MDP-reduction} and then invoking Theorem~\ref{theorem:mean-field-MDP}, one obtains the following equality for any initial measures $\nu , \nu' \in \Delta( \bX )$: 
\begin{align} \label{eq:arbitrary-nu}
	&\P^{\bpi}_{\nu} \left(   \{ y^i_t, a^i_t \}_{t \geq 0 } \in \cdot   \: \middle| y^i_0 = y, a^i_0 = a^i \right)  \nonumber  \\
	= &\P^{\bpi}_{\nu'} \left(   \{ y^i_t, a^i_t \}_{t \geq 0 } \in \cdot   \:   \middle| y^i_0 = y, a^i_0 = a^i \right) , \quad \forall (y, a^i ) \in \yy \times \aa. 
\end{align}

\subsubsection*{Q-Functions Under Mean-Field Observability and Symmetry}

In light of Theorem~\ref{theorem:mean-field-MDP}, we define the Q-function for player $i$ when playing $\bG$ against a symmetric policy $\bpi^{-i} \in \bPi^{-i}_{S}$ as
\[
Q^{*i}_{\bpi^{-i}} ( y, a^i ) := E^{ ( \pi^{*i}, \bpi^{-i} ) }_{\nu} \left[ \sum_{t= 0}^{\infty} \gamma^t c ( x^i_t, \upmu( \bx_t ) ,  a^i_t ) \middle| y^i_0 = y , a^i_0 = a^i \right], \quad \forall a^i \in \aa,
\]
for every $y \in \varphi^i ( \bX ) = \{ \varphi^i ( \bs ) : \bs \in \bX \}$, where $\pi^{*i} \in \BR^i_0 ( \bpi^{-i} ) \cap \Pi^i_{S}$ is a best response to $\bpi^{-i}$ and $\nu \in \Delta (\bX)$ is arbitrary.\footnote{That $\nu$ can be arbitrarily chosen follows from the preceding discussion culminating in \eqref{eq:arbitrary-nu}.} For elements $y \in \yy \setminus \varphi^i ( \bX)$, which do not arise as the observation output of any global state, we may define $Q^{*i}_{\bpi^{-i}} ( y, a^i )$ arbitrarily, say $Q^{*i}_{\bpi^{-i}} ( y, \cdot ) \equiv 0$. 

For any player $i \in \NN$, we let $\bPi^{-i}_{S, \sym} \subset \bPi^{-i}_{S}$ denote the set of symmetric joint policies for the remaining players, and we let $\bPi_{S,\sym} \subset \bPi_{S}$ denote the set of stationary joint policies that are symmetric. We note that the sets $\Pi^i_{S}$ and $\bPi^{-i}_{S, \sym}$ are in bijection, and we define $\sym^i : \Pi^i_{S} \to \bPi^{-i}_{S, \sym}$ by $\sym^i ( \pi^i ) = \left( \pi^i \right)_{j \in \NN \setminus \{i \} }$, for all $\pi^i \in \Pi^i_{S}.$ We metrize $\bPi^{-i}_{S, \sym}$ using the metric $d^i$ on $\Pi^i_{S}$: 
\begin{equation} \label{eq:symmetric-metric}
\bd^{-i}_{\sym} \left( \sym^i ( \pi^i ) , \sym^i ( \tilde{\pi}^i ) \right)  := d^i ( \pi^i, \tilde{\pi}^i), \quad \forall \pi^i, \tilde{\pi}^i \in \Pi^i_{S}.
\end{equation}

We note that the metric $\bd^{-i}_{\sym}$ is equivalent to the metric $\bd^{-i}$ restricted to $\bPi^{-i}_{S,\sym}$. We also define an analogous metric $\bd_{\sym}$ on $\bPi_{S, \sym}$. 

\vspace{5pt}

We now state some lemmas on the continuity of various functions, where continuity is with respect to the metrics defined above. {The proofs of these lemmas resemble those of  \cite[Lemmas 2.10-2.13]{yongacoglu2023satisficing} and are omitted. 
}
In each of the lemmas below, we let $\bG$ be an $n$-player mean-field game with mean-field observability (Assumption~\ref{ass:mean-field-state-observability}), and let $i \in \NN$ be any player.

\begin{lemma}  \label{lemma:mean-field-continuous-Q-factors}
Fix $(y, a^i) \in \yy \times \aa$. The mapping $\bpi^{-i} \mapsto Q^{*i}_{\bpi^{-i}} ( y, a^i )$ is continuous on $\bPi^{-i}_{S,\sym}$. 

\end{lemma}

\begin{lemma}  \label{lemma:mean-field-continuous-min-Q-factors}
Fix $y \in \yy$. The mapping $\bpi^{-i} \mapsto \min_{a^i \in \aa} Q^{*i}_{\bpi^{-i}} ( y, a^i )$ is continuous on $\bPi^{-i}_{S,\sym}$. 
\end{lemma}

\begin{lemma}    \label{lemma:mean-field-continuous-max-cost-difference}
Fix $\bpi^{-i} \in \bPi^{-i}_{S, \sym}$. The following mapping is continuous on $\Pi^{i}_{S}$:
\[
\pi^i \mapsto \max_{s \in \bX } \left( J^i  (\pi^i, \bpi^{-i}  ,  \bs )  - \min_{a^i \in \aa } Q^{*i}_{\bpi^{-i}} ( \varphi^i ( \bs ) , a^i ) \right ) .
\]
\end{lemma}

\subsection{Existence of Stationary Equilibrium under Mean-Field Observability}

Theorem~\ref{theorem:mean-field-equilibrium}, stated below, asserts that a stationary perfect equilibrium exists when the game has mean-field observability. The proof technique parallels that of \cite[Theorem 2]{fink1964equilibrium}, making the required modifications to account for partial observability of the global state. 

\begin{theorem} \label{theorem:mean-field-equilibrium}
Let $\bG$ be a partially observed $n$-player mean-field game satisfying Assumption~\ref{ass:mean-field-state-observability}. For any $\epsilon \geq 0$, there exists a perfect $\epsilon$-equilibrium policy in $\bPi_{S}$. That is, $ \eq[\epsilon]_{S} \not= \varnothing.$
\end{theorem}

\begin{proof} Fix player $i \in \NN$. We define a point-to-set mapping $\BB^i : \Pi^i_{S} \to 2^{\Pi^i_{S}}$ with
\[
\BB^i ( \pi^i ) = \BR^i_{0} ( \sym^i ( \pi^i ) ) \cap \Pi^i_{S}, \quad \forall \pi^i \in \Pi^i_{S}.
\]
By Theorem~\ref{theorem:mean-field-MDP}, player $i$ is facing an MDP with finite state and action spaces when playing against $\sym^i(\pi^i)$, and the set of stationary optimal policies for this MDP is given by $\BB^i ( \pi^i ) \subseteq \Pi^i_{S} $. Thus,  $\BB^i ( \pi^i ) $ is  non-empty, convex, and compact for each $\pi^i \in \Pi^i_{S}$.

If $( \sym^i ( \tilde{\pi}^i_k ) )_{k \geq 0}$ is a sequence of symmetric joint policies in $\bPi^{-i}_{S,\sym}$ converging to some $\sym^i ( \tilde{\pi}^i_{\infty}) \in \bPi^{-i}_{S,\sym}$ and $(\pi^i_k)_{k \geq 0}$ is a sequence in $\Pi^i_{S}$ such that (1) $\lim_{k \to \infty} \pi^i_k = \pi^i_{\infty} \in \Pi^i_{S}$ and (2) for every $k \geq 0$, we have that $\pi^i_k \in \BR^i_{0} ( \sym^i (\tilde{\pi}^i_k) )$, then one can use Lemma~\ref{lemma:continuous-cost}, Lemmas~\ref{lemma:mean-field-continuous-Q-factors}--\ref{lemma:mean-field-continuous-max-cost-difference}, and Lemma~\ref{lemma:stopping-condition} to conclude that $\pi^i_{\infty} \in \BR^i_{0} ( \sym^i ( \tilde{\pi}^i_{\infty}) ) \cap \Pi^i_{S} = \BB^i (  \tilde{\pi}^i_{\infty} )$. This implies that the point-to-set mapping $\BB^i$ is upper hemicontinuous. 

We invoke Kakutani's fixed point theorem on $\BB^i$ to obtain a fixed point 
\[
\pi^{*i} \in \BB^i ( \pi^{*i} ) = \BR^i_{0} ( \sym^i ( \pi^{*i} ) ) \cap \Pi^i_{S}.
\]
By symmetry, $\left( \pi^{*i} , \sym^{i} ( \pi^{*i} ) \right) \in \Pi^i_{S} \times \bPi^{-i}_{S,\sym} \subset \bPi_{S}$ is a perfect $0$-equilibrium for $\bG$, and \textit{a fortiori} a perfect $\epsilon$-equilibrium for $\bG$. 
\end{proof}

To the best of our knowledge, this result is new. It differs from earlier results in that it considers a setting with finitely many players, discrete time, discounted costs, and partial observability. The (person-by-person) optimality of each player's policy holds without approximation ($\epsilon = 0$) and holds over all admissible policies and not merely over the player's stationary policies. Moreover, this optimality holds for any initial distribution, and for any (finite) number of players. To better situate this result in the literature, we now describe some related results.

In the continuous time literature, \cite[]{arapostathis2017solutions} prove the existence of equilibrium in games with finitely many players and an ergodic (average) cost criterion. \cite[]{lacker2015mean} studies existence of Markov equilibrium in the limiting regime with infinitely many players. For games with finitely many players, several results about equilibria among feedback strategies with global observability are given in \cite[]{lacker2020convergence}. Further discussion on the topic can be found in 	 \cite[]{fischer2017connection}. 

In discrete time, \cite[]{perrin2022generalization} consider the limiting regime and study what they call population-dependent policies, with the goal of obtaining solutions that perform well across initial distributions. This consideration of the initial distribution is the defining feature of uniform best-responding and perfect equilibrium as studied here. In our framework, their notion of population-dependent policies corresponds to our notion of stationary policies under mean-field observability. 

Also in the discrete-time setting, \cite[]{biswas2015mean} studies mean-field games with finitely many players and local observability. Using the ergodic (average) cost criterion,  the existence of a certain kind of equilibrium is proved. The equilibrium considered by \cite[]{biswas2015mean} is somewhat less demanding than the one considered here: in our definition of a perfect equilibrium, each player $i$'s policy must be optimal over its complete set of admissible policies $\Pi^i$, rather than optimal over the restricted set $\Pi^i_{S} \subsetneq \Pi^i$.

\subsection{Counterexamples to MDP Equivalence} \label{ss:counterexamples}

We now provide explicit counterexamples showing that, in general, player $i$ will not face an MDP in $\{ y^i_t \}_{t \geq 0}$ when either mean-field observability or symmetry of $\bpi^{-i}$ fail to hold.

\subsubsection*{Example 1: On Relaxing the Symmetry of $\bpi^{-i}$}

\begin{figure}
\centering
\begin{tikzpicture}[->,>=stealth,shorten >=1pt,auto,node distance=3cm,semithick]
    \node[circle, draw] (0) at (0,0) {$0$};
    \node[circle, draw] (1) at (-2,-2) {$1$};
    \node[circle, draw] (2) at ( 2, -2 ) {$2$};
    
    \path (0) edge [loop above] node {$2\epsilon$} (0) 
    	  (0) edge [above left] node {$0.5 - \epsilon$} (1)
          (0) edge node {$0.5 - \epsilon $} (2)
          (1) edge [loop left] node {$a=1$} (1)
          (1) edge [bend left] node {$a=2$} (2)
          (2) edge [loop right] node {$a=2$} (2)
          (2) edge [bend left] node {$a=1$} (1);
\end{tikzpicture}
\caption{Local State Transition Probabilities for Example~1} \label{figure:cannot-relax-symmetry}
\end{figure}
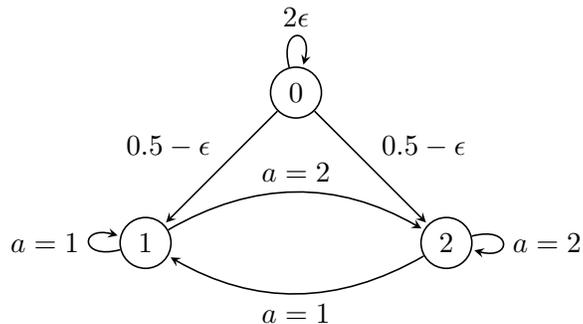

If one relaxes the symmetry assumption in Theorem~\ref{theorem:mean-field-MDP}, then the controlled Markov property of $\{ y^i_t \}_{t \geq 0}$ is lost in general, even under mean-field observability. We illustrate this using a simple example, with $\xx = \aa = \{ 0 , 1 , 2 \}$ and $n = 3$, though the example extends to any number of players. Local state transitions depend only on local state and action without dependence on the mean-field term. The relevant probabilities are summarized in Figure~\ref{figure:cannot-relax-symmetry}, with $\epsilon \in (0, 0.5)$ arbitrary and transitions from states 1 or 2 to state 0 omitted for  clarity. In words, the transition probabilities out of state 0 are independent of both action and the mean-field state, with transitions to states 1 and 2 being equally likely, with probability $0.5-\epsilon$. Transitions out of states 1 and 2 are deterministic, with $x^j_{t+1} = a^j_t$ for any agent $j$. That is, $P_{\loc} ( a | s , \mu, a ) = 1$, for $s \in \{1,2\}$, $\mu \in \Emp_n$, and $a \in \{0,1,2\}$.

\vspace{0pt}

We define a cost function $c ( x, \mu, a ) :=  \kappa \cdot \textbf{1} \left\{ \mu(x) \geq 0.5 \right\} $ for any $ (x, \mu, a ) \in \xx \times \Delta(\xx ) \times \aa$, where $\kappa > 0$ is a penalty for being in the same state as the majority of all agents. Although the mean-field term $\mu_t$ does not appear in its local state transition probabilities, the mean-field term is nevertheless relevant to a given player's cost. 

We define two stationary policies, denoted $\pi_{\rm stay} \in \PP ( \aa | \yy )$ and $\pi_{\rm go 2} \in \PP ( \aa | \yy )$. These policies are given by  $\pi_{\rm stay}  ( \cdot | s , \mu )  := \delta_{s}$ and $\pi_{\rm go 2} ( \cdot | s , \mu )  := \delta_{2}$, for all $(s,\mu) \in \yy$. In words, the policy $\pi_{\rm stay}$ always chooses $a^j_t = x^j_t$, while $\pi_{\rm go 2}$ always chooses $a^j_t = 2$. 

\vspace{5pt}

We fix $i = 3$ to be the index of the last player in the game, and we consider the \emph{non-symmetric} stationary joint policy $\bpi^{-i} \in \bPi^{-i}_S$ given by $\pi^1 = \pi_{\rm stay}$ and $\pi^2 = \pi_{\rm go2}$,   and we let $\bpi = ( \pi^i , \bpi^{-i} )$, where $\pi^i \in \Pi^i$ is arbitrary. 

Suppose that the initial distribution $\nu \in \Delta (\bX )$ is such that each player $j$'s initial state is $x^j_0 = 0$ with probability 1. Writing an element $( x^i, \mu ) \in \yy$ as $\left( x^i , [ \mu(0), \mu(1), \mu(2) ] \right)$, one can establish the following inequality
\begin{align*}
\P^{\bpi}_{\nu} \left(  y^i_{t+1} = \left( 2, [0,0,1] \right) \middle| y^i_t = \left( 0, \left[ \frac{1}{3}, \frac{1}{3}, \frac{1}{3} \right]  	\right) , a^i_t = a^i 		\right) > 0 .
\end{align*}

\vspace{2.5pt}

On the other hand, conditioning also on $y^i_{t-1}$, one has the following:
\begin{align*}
\P^{\bpi}_{\nu} \left(  y^i_{t+1} = \left( 2, [0,0,1] \right) \middle| y^i_t = y^i_{t-1} = \left( 0,  \left[ \frac{1}{3}, \frac{1}{3}, \frac{1}{3} \right]   	\right)		\right) = 0 .
\end{align*}

The condition $y^i_t = y^i_{t-1}$ here implies that (a) player 2 (who changes state from 1 to 2 given the chance) was already in state 2 at time $t-1$, and (b) player 1 (who remains in whatever state it is already in) was in state 1 at time $t-1$. Hence, neither will choose to change its state at time $t$.

In summary, as $a^i_t$ did not feature in the analysis above, we have observed that 
\[
\P^{\bpi}_{\nu} \left( y^i_{t+1} \in \cdot \middle| y^i_t, a^i_t \right) \not= \P^{\bpi}_{\nu} \left( y^i_{t+1} \in \cdot | \left\{ y^i_k, a^i_k : 0 \leq k \leq t \right\} \right),
\]
which establishes that $\{ y^i_t , a^i_t \}_{t \geq 0}$ is not a controlled Markov process.  \hfill $\diamond$

\subsubsection*{Example 2: On Relaxing Mean-Field Observability}

We now show that if one relaxes the mean-field observability assumption in Theorem~\ref{theorem:mean-field-MDP}, then player $i$ does not face an MDP with state $\{ y^i_t \}_{t \geq 0}$. In this example, one sees that the local observation can fail to capture the cost-relevant history and furthermore can fail to satisfy the controlled Markov property, even under symmetry of the joint policy $\bpi^{-i}$. Thus, both defining features of MDPs are violated in the counterexample below. 

Consider an $n$-player mean-field game $\bG$ with local observability, $\xx = \{ 1, \dots,  10 \}$, and $\aa = \Emp_n$. Define the cost function as $c(s,\mu,a) = \| \mu - a \|_2$ for each $(s,\mu,a) \in \xx \times \Delta(\xx) \times \aa$. Fixing $\delta > 0$, the transition kernel $P_{\loc}$ is given by
\[
P_{\loc} ( s' | s, \mu, a ) = 	\begin{cases}
					\delta , &\text{if } s' = 10 \\
					1 - \delta, &\text{if } s' = \left \lceil \sum_{\bar{s} \in \xx} \bar{s} \mu(\bar{s}) \right \rceil \\
					0, &\text{else.}
					\end{cases}	
\]

In words, player $i$'s local state transitions either to the mean of the mean-field (rounding up to the nearest integer) or to the maximum value, 10. Decreases to one's local state inform the player about the mean of the preceding mean-field term: if $x^i_{t+1} = \omega+1 < x^i_t$, then the mean of $\mu_t$ lies in $( \omega, \omega + 1]$. Note also that the means of $\{ \mu_t \}_{t \geq 0}$ are non-decreasing.

The discount factor $\gamma \in (0,1)$ is set arbitrarily, and the initial distribution $\nu_0 \in \Delta(\bX )$ is chosen such that $\{ x^j_0 \}_{j \in \NN}$ is independently and identically distributed according to the uniform distribution on $\xx$.

Each player $i$'s cost depends only on its chosen action and the empirical distribution associated with the global state. Thus, each player $i$ minimizes its costs by tracking the (unobserved) mean-field process $\{ \mu_t \}_{t \geq 0}$. It is clear that player $i$'s cost at time $t$ is not measurable with respect to $(y^i_t , a^i_t ) = (x^i_t, a^i_t )$. By itself, this shows that $\{ y^i_t \}_{t \geq 0}$ is not the state variable for an MDP, since it does not summarize the cost-relevant history.

Let $\bpi$ be \emph{any} joint policy. We have  $ \P^{\bpi}_{\nu_0} \left( x^i_{t+1} \leq 4 \middle| x^i_t = 10 \right) > 0 .$ However, decreases in one's local state are informative about the mean-field, so conditioning on $y^i_{t-1}$ and $y^i_{t-2}$ can change this strict inequality to an equality: 
\[
\P^{\bpi}_{\nu_0} \left( x^i_{t+1} = 4 \middle| x^i_t = 10, x^i_{t-1} = 5 , x^i_{t-2} = 10  \right) = 0.
\]
This shows that $\{ y^i_t = x^i_t \}_{t \geq 0}$ also fails to satisfy the controlled Markov property.

In summary, the discussion above shows that the control problem faced by player $i$ in a partially observed $n$-player mean-field game with compressed or local observability is not generally equivalent to a fully observed MDP. This may occur either because the observation variable fails to summarize the cost-relevant history of the system, because the observation variable fails to satisfy the controlled Markov property, or both. \hfill $\diamond$

\subsection{Further Comments on Equilibrium in $n$-Player Mean-Field Games}

\textit{On MDP Equivalence and Equilibrium Under Global Observability:} When an $n$-player mean-field game has global observability, the result is a fully observed stochastic game. In this case, results analogous to Theorems~\ref{theorem:mean-field-MDP} and \ref{theorem:mean-field-equilibrium} are well-known \cite[]{fink1964equilibrium}. The results above do not follow from this well-known theory due to the partial observability of the global state variable.

\vspace{5pt}

\noindent  \textit{Toward Equilibrium under Compressed Observability:} Under compressed observability, one cannot guarantee the existence of a perfect equilibrium among the stationary policies without loss of generality, as evidenced by the game in Example 2. Nevertheless, it may be desirable to identify (restrictive) sufficient conditions for existence. One approach involves taking the number of players to be large enough that the strategic coupling is negligible and the mean-field sequence $\{ \mu_t \}_{t \geq 0}$ is essentially constant over the effective planning horizon. For a result on existence of $\epsilon$-equilibrium under Assumption~\ref{ass:local-state-observability} as well as further discussion, see Appendix~\ref{appendix:large-N}.

\section{Independent Learning and Subjectivity} \label{sec:naive-learning} %

In this section, we study independent learning in $n$-player mean-field games by harnessing the connections between (PO)MDPs and the control problem of a player in the game.

\subsection{Independent Learning} 

In large, decentralized systems,  agents have limited information about the overall system. For instance, the policies, actions, and local states of other agents may be unobservable or difficult to estimate using local data. Moreover, maintaining estimates of various global quantities introduces a massive computational and communication burden at each agent. 

Independent learning is one approach to learning in decentralized settings. In this paradigm, agents use only local information and run single-agent learning algorithms, treating their environment as though it were a fully observed (single-agent) MDP. Such intentional obliviousness is characteristic of independent learners and is used to alleviate the computational burden at each agent. In effect, each independent learning agent makes an (erroneous) simplifying assumption about its environment to obtain tractable, scalable algorithms in complex multi-agent settings. In this paper, we interpret this simplifying assumption as implying \emph{subjective beliefs} held by the player about its system.

The independent learning paradigm is similar to various algorithms designed for single-agent, non-Markovian environments. In such settings, optimal policies are generally non-stationary and history-dependent. This leads to learning and planning problems that are generally intractable. Recently, some authors have proposed use of simpler surrogates for the full system history, in an effort to inform tractable algorithm design in complex environments. Examples of work in this line include the agent state concept introduced in  \cite[]{dong2022simple} and leveraged in \cite[]{sinha2024periodic}, the approximate information state concept of  \cite[]{subramanian2022approximate}, the approximate belief state of \cite[]{kara2022convergence}, and the quantized belief state of \cite[]{kara2024qlearning}. Viewed as an instance of this approach, independent learners in games are simple agents in complex, non-stationary environments and local observations can be seen as agent states or approximate belief states.

In the literature on multi-agent reinforcement learning with independent learning agents,  learning and policy adjustment are typically interleaved: agents follow a particular stationary policy to collect feedback data for a number of interactions, and then update their policy parameters using estimates of value functions or some other object. (For example, see the work on regret testers by \cite[]{foster2006regret}.) It is therefore important to understand the properties of estimates arising from the learning process. In this section, we study the convergence properties of value function estimates obtained by independent learners in $n$-player mean-field games. A full learning algorithm, with policy dynamics interleaved with value estimation, is presented in Section~\ref{sec:satisficing-paths}.

\subsection{Single-Agent Learning Estimates under Stationary Policies}

We now study the learning iterates of agents who attempt to naively estimate a Q-function and state value function, treating their local observations as though they were state variables for an MDP. We study the evolution of these learning iterates under stationary policies, leaving aside the challenge of non-stationarity that arises when policies are adjusted during play, which will be considered later.

In Algorithm~\ref{algo:Q-learning}, below, each player runs two stochastic approximation algorithms, one resembling Q-learning, the other resembling value function estimation. Since player $i$ does not generally face an MDP, the Q-function and state value function of single-agent theory need not be expressible as a function of player $i$'s local observations. For this reason, we interpret Algorithm~\ref{algo:Q-learning} as learning \emph{subjective} value functions that are compatible with player $i$'s beliefs but need not have a meaningful connection to the objective function.

\begin{algorithm}[h] 
	\SetAlgoLined
	\DontPrintSemicolon
	\SetKw{Receive}{Receive}
	\SetKw{parameters}{Set Parameters}
	\SetKw{initialize}{Initialize}
	\SetKwBlock{For}{for}{end} 
	
	\initialize Soft $\bpi \in \bPi_{S}$, $\bar{Q}^i_0 = 0 \in \rr^{\yy \times \aa}$ and $\bar{J}^i_0 = 0 \in \rr^{\yy}$  \;

	\BlankLine
	
	\For($t \geq 0$ ($t^{th}$ stage{)}){ 
		
		\BlankLine 
		Simultaneously, each player $j \in \NN$ selects $a^j_t \sim \pi^j ( \cdot | y^j_t )$ \; 
		
		
		Player $i$ observes $y^i_{t+1}$ and cost $c^i_t := c ( x^i_t, \upmu (  \bx_t ) , a^i_t )$\;
		
		$n^i_t := \sum_{ k = 0}^t \textbf{1} \{ ( y^i_k , a^i_k ) = ( y^i_t, a^i_t ) \}$ \;
		
		$m^i_t := \sum_{ k = 0}^t \textbf{1} \{ 	y^i_k = y^i_t	\}  $\;
		
		\BlankLine
		
		Q-factor update:		\begin{align*}
		\bar{Q}^i_{t+1} ( y^i_t , a^i_t ) = \left( 1 - \frac{1}{n^i_t } \right) \bar{Q}^i_t ( y^i_t, a^i_t )   + \frac{1}{n^i_t}  \left( c^i_t + \gamma \min_{ a^i \in \aa } \bar{Q}^i_t ( y^i_{t+1} , a^i ) \right) ,
		\end{align*}
		
		and $\bar{Q}^i_{t+1} ( y, a^i ) = \bar{Q}^i_t (y, a^i ) $ for all $(y,a^i ) \not= (y^i_t, a^i_t)$. \;
		\BlankLine 
		Value function update: \begin{align*}
		\bar{J}^i_{t+1} ( y^i_t ) = \left(1 - \frac{1}{m^i_t} \right) \bar{J}^i_t ( y^i_t ) + \frac{1}{m^i_t} \left( c^i_t + \gamma \bar{J}^i_t ( y^i_{t+1} ) \right),
		\end{align*}
		and $\bar{J}^i_{t+1} ( y) = \bar{J}^i_t (y)$ for all $y \not= y^i_t$.

	}
	
	\caption{Independent Learning of (Subjective) Value Functions} \label{algo:Q-learning}
\end{algorithm}

We now study the convergence properties of the iterate sequences $\{ \bar{Q}^i_t ,\bar{J}^i_t \}_{t \geq 0}$, which are produced by Algorithm~\ref{algo:Q-learning}, and we provide sufficient conditions for their almost sure convergence. As we will describe, these conditions are relatively mild and can be relaxed further if desired. Importantly, the convergence of iterates does not depend on whether or not player $i$ faces an MDP in its observation variable $\{ y^i_t \}_{t \geq 0}$.

We begin with an assumption on the transition kernel $P_{\loc}$. For intuitive simplicity, we state this assumption in terms of the underlying state process.

\begin{assumption} \label{ass:state-visitation} 
Under any stationary policy $\bpi \in \bPi_{S}$ and initial distribution $\nu \in \Delta ( \bX)$, the global state process \emph{$\{ \bx_t \}_{t \geq 0}$} is an irreducible, aperiodic Markov chain on $\bX$. 
\end{assumption}

Assumption~\ref{ass:state-visitation} is satisfied, for instance, when $P_{\loc} ( s' | s, \mu, a ) > 0$ for any arguments $(s', s, \mu, a ) \in \xx \times \xx \times \Delta(\xx) \times \aa$, but applies much more broadly. Assumption~\ref{ass:state-visitation} is sufficient but not necessary for the following results: it can be relaxed by assuming only that, for each soft policy $\bpi \in \bPi_{S}$, there exists a unique probability measure $\nu^{\infty}_{\bpi} \in \Delta ( \bX )$ such that the time averaged occupancy measure converges to $\nu^{\infty}_{\bpi}$.\footnote{A policy is soft if it places positive probability on each action in any context. See Definition~\ref{def:soft-policies}.} That is, one can relax Assumption~\ref{ass:state-visitation} by assuming only that, for any initial measure $\nu_0$, we have 
\[
\frac{1}{T} \sum_{t =0}^{T-1} \P^{\bpi}_{\nu_0} \left[ \bx_t \in \cdot \right] \to \nu^{\infty}_{\bpi} , \text{ as } T \to \infty .
\]

\begin{theorem}  \label{theorem:naive-learning}
Let $\bG$ be a partially observed $n$-player mean-field game for which Assumption~\ref{ass:state-visitation} holds, and let $\bpi \in \bPi_{S}$ be soft. Suppose player $i \in \NN$ uses Algorithm~\ref{algo:Q-learning}. For deterministic functions ${V}^{*i}_{\bpi} : \yy \to \rr $ and ${W}^{*i}_{\bpi} : \yy \times \aa \to \rr$, defined in Appendix~\ref{appendix:implied-MDPs}, we have the following:
	\begin{enumerate}
		\setlength \itemsep{0pt}
		\item For any $\nu \in \Delta(\bX)$, $\P^{\bpi}_{\nu}$ almost surely, we have 
			\[
			\lim_{ t \to \infty}  \bar{J}^i_t = V^{*i}_{\bpi} \quad \text{and} \quad \lim_{t \to \infty} \bar{Q}^i_t = {W}^{*i}_{\bpi} .
			\]
		
		\item If Assumption~\ref{ass:global-state-observability} holds, then {${V}^{*i}_{\bpi} ( \bs ) = J^i ( \bpi,  \bs )$} for all \emph{$\bs \in \bX$} and ${W}^{*i}_{\bpi} = Q^{*i}_{\bpi^{-i}}$.

		\item Under mean-field observability (Assumption~\ref{ass:mean-field-state-observability}), if $\bpi^{-i}$ is symmetric,  then {$ {V}^{*i}_{\bpi} ( \varphi^i ( \bs  ))  =  J^i ( \bpi ,  \bs )  $} for all {$\bs \in \bX$} and ${W}^{*i}_{\bpi} = Q^{*i}_{\bpi^{-i}}$.
	\end{enumerate}
\end{theorem}

\noindent The proof of Theorem~\ref{theorem:naive-learning} can be found in Appendix~\ref{appendix:subjective-iterates}. The first part of Theorem~\ref{theorem:naive-learning} holds for any observation channel and any soft stationary joint policy $\bpi \in \bPi_{S}$. In particular, it holds under mean-field, compressed, or local observability and even when $\bpi$ is not symmetric, and thus holds when player $i$ faces a POMDP and not an MDP. As a result, the convergence of the iterates $\{ \bar{Q}^i_t ,\bar{J}^i_t \}_{t \geq 0}$ is not a simple consequence of celebrated results in stochastic approximation, such as those of \cite[]{tsitsiklis1994asynchronous}.

\vspace{5pt}

\noindent \textbf{Remarks:} {In Appendix~\ref{appendix:implied-MDPs}, we explicitly define the functions $V^{*i}_{\bpi}$ and $W^{*i}_{\bpi}$.} In \cite[]{kara2022convergence}, it is shown that the function ${W}^{*i}_{\bpi}$, restricted to $ \{ \varphi^i ( \bs ) : \bs \in \bX \} \times \aa $, is in fact the Q-function for \emph{some} fully observed Markov decision problem with state space $ \varphi^i ( \bX) := \{ \varphi^i (\bs ) : \bs \in \bX \}$. The same argument used there can be used to establish that ${V}^{*i}_{\bpi}$, restricted to $\varphi^i (\bX)$, is the state value function for the \emph{same} MDP. A complete specification of this MDP is deferred to Appendix~\ref{appendix:implied-MDPs}. We note:
\begin{itemize}
	
	\setlength \itemsep{0pt}
	
	\item[(i)] The MDP with state space $\varphi^i ( \bX)$ described above is an instance of what \cite[]{kara2022convergence} calls \emph{an approximate belief MDP with memory length 0}. 
	
	\item[(ii)] Under Assumption~\ref{ass:state-visitation}, each soft policy $\tilde{\bpi} \in \bPi_{S}$ gives rise to a unique invariant measure $\nu_{\tilde{\bpi}} \in \Delta ( \bX )$, and we have that, for any $\nu \in \Delta ( \bX )$, $\P^{\tilde{\bpi}}_{\nu} \left( \bx_t \in \cdot \right) \to \nu_{\tilde{\bpi}} ( \cdot )$ in total variation as $t \to \infty$. The remark after Assumption~\ref{ass:state-visitation} says that it suffices to replace this convergence of laws by the convergence of their ergodic averages to $\nu_{\tilde{\bpi}}$.
	
	\item[(iii)] Under compressed observability, the limiting quantities of Algorithm~\ref{algo:Q-learning} depend, in general, on the policy $\pi^i$ used by player $i$. This is in contrast to MDP settings (such as the specific cases in Parts 2 and 3 of Theorem~\ref{theorem:naive-learning}), where the limiting values of Q-learning are the same for different (soft/sufficiently exploratory) policies. 
	
	\item[(iv)] In general, player $i$ does not actually face the particular MDP on $\varphi^i ( \bX ) \subseteq \yy$ described in item (iii). The limiting quantities ${V}^{*i}_{\bpi}$ and ${W}^{*i}_{\bpi}$ do not, in general, have any inherent relevance to player $i$'s objective function in the game $\bG$. These quantities should instead be interpreted as the subjective beliefs of player $i$, which were arrived at through a naive independent learning process.
\end{itemize}

\subsection{Independent Learners and Subjectivity} 

In an $n$-player mean-field game, each cost minimizing player's true objective is to select a best-response to the policies of its counterparts. The solution concepts of ($\epsilon$-) equilibrium, both perfect and with respect to an initial distribution, are then defined in terms of these objective functions. 
These definitions are inherently \emph{objective} notions: optimality and equilibrium are defined without any approximation, estimation, or modelling of beliefs. When studying the game $\bG$ as an environment for decentralized multi-agent learning, one remarks that a given agent $i$ may not be able to objectively verify whether a given policy $\pi^i$ is an $\epsilon$-best-response to a policy $\bpi^{-i}$ of its counterparts. Indeed, player $i$ does not know the system data defining $\bG$, does not know the joint policy $\bpi^{-i}$, and does not even observe the actions of other agents.\footnote{We reiterate that we wish to avoid the \emph{representative agent} approach. Players are free to employ different policies. As a result, a given agent $i$ will not know another agent $j$'s policy.} In this section, we formalize our notion of \emph{subjective Q-equilibrium} and motivate it as a learning-relevant solution concept that accounts for decentralized information and independent learning in mean-field games.

Independent learners model their environment as an MDP and run single-agent algorithms, treating their local observations as though they are controlled Markovian state variables. This simplifying assumption is made to reduce the computational burden, so that tractable and scalable -- if suboptimal -- algorithms can be employed. By making a simplifying assumption on its environment, player $i$ effectively adopts a \emph{subjective model} of the system. Due to model uncertainty and decentralized information inherent to its environment, player $i$ cannot verify whether a given policy is an objective $\epsilon$-best-response to the behaviour of others. On the other hand, player $i$ can check whether a given policy is \emph{subjectively} $\epsilon$-optimal for its assumed model, an MDP, using reinforcement learning techniques to estimate its (subjective) Q-function and value function.

\begin{definition}[Subjective Function Family]  \label{def:subjective-function-family}
For each player $i \in \NN$ and stationary joint policy $\bpi \in \bPi_{S}$, let $V^{*i}_{\bpi}$ and $W^{*i}_{\bpi}$ be the functions appearing in Theorem~\ref{theorem:naive-learning}, which are defined in Appendix~\ref{appendix:implied-MDPs}. We define two families of functions
\[
\VV^{*} := \left\{ V^{*i}_{\bpi} : \yy \to \rr \middle|  i \in \NN, \bpi \in \bPi_{S}	\right\} \quad \text{and} \quad \WW^{*} := \left\{ W^{*i}_{\bpi} : \yy \times \aa \to \rr \middle| i \in \NN , \bpi \in \bPi_{S} \right\}
\]
The pair $(\VV^{*}, \WW^{*})$ is called the {\emph{subjective function family for $\bG$.}}
\end{definition}

\begin{definition}[Subjective Best-Responding] \label{def:subjective-BR}
Let $i \in \NN$, $\epsilon \geq 0$, $\bpi^{-i} \in \bPi^{-i}_{S}$. A policy $\pi^{*i} \in \Pi^i_{S}$ is called a \emph{$(\VV^{*}, \WW^{*})$-subjective $\epsilon$-best-response to $\bpi^{-i}$} if we have
\[
V^{*i}_{( \pi^{*i}, \bpi^{-i} )} ( y ) \leq \min_{ a^i \in \aa } W^{*i}_{ ( \pi^{*i} , \bpi^{-i} ) } ( y, a^i ) + \epsilon , \quad \forall y \in \yy. 
\]
\end{definition}

\begin{definition}[Subjective (Q-) Equilibrium]  \label{def:subjective-equilibrium}
Let $\epsilon \geq 0$. A joint policy $\bpi^{*} \in \bPi_{S}$ is called a $(\VV^{*}, \WW^{*})$-subjective $\epsilon$-equilibrium for $\bG$ if $\pi^{*i}$ is a $(\VV^{*}, \WW^{*})$-subjective $\epsilon$-best-response to $\bpi^{*-i}$ for every $i \in \NN$.
\end{definition}

We will alternate between several terms for this solution concept: it will be called $(\VV^{*}, \WW^{*})$-subjective $\epsilon$-equilibrium when ambiguity is possible; it may be called subjective Q-equilibrium to distinguish it from earlier notions of subjective equilibrium; or it may simply be called subjective ($\epsilon$-) equilibrium. Similarly, we will often refer to $(\VV^{*}, \WW^{*})$-subjective ($\epsilon$-) best-responding simply as subjective ($\epsilon$-) best-responding.

The (possibly empty) set of $ ( \VV^{*} ,\WW^{*}) $-subjective $\epsilon$-best-responses for player $i \in \NN$ against stationary $\bpi^{-i} \in \bPi^{-i}_{S}$ will be denoted by $\SubjBR^i_{\epsilon} ( \bpi^{-i},   \VV^{*} ,\WW^{*})   \subseteq \Pi^i_{S} .$ Similarly, we let $\Subj_{\epsilon} ( \VV^{*} ,\WW^{*}) \subseteq \bPi_{S}$ denote the (possibly empty) set of $ ( \VV^{*} ,\WW^{*})$-subjective $\epsilon$-equilibrium policies for $\bG$.

\subsection{Interpretations of Subjective Best-Responding and Equilibrium}

Below, we describe some desirable qualities of our definition of subjective equilibrium. 

\vspace{5pt} 

\noindent  \textit{Centrality of Local Information and Subjectivity:} Independent learning agents make a simplifying assumption about their environment so as to obtain tractable, scalable algorithms. Since the independent learner in a mean-field game subjectively models its environment as an MDP, we have defined subjective $\epsilon$-best-responding in analogy to a characterization of $\epsilon$-optimality of MDPs (Lemma~\ref{lemma:stopping-condition}). That is, the definition of subjective best-responding is given with reference to the world model adopted by the learning agent, which has implications for the explanatory power of the definition.

\vspace{5pt}

\noindent \textit{Room for Asymmetry:} In online, decentralized learning, different players observe different feedback trajectories, which will then be used to update their policies in an uncoordinated manner. As a result, in the absence of centralization or significant communication, it is unreasonable to expect the whole population to follow the same policy. The subjective equilibrium of Definition~\ref{def:subjective-equilibrium} allows for asymmetry of policies, and applies equally well to a heterogeneous population of agents, each using a distinct policy. Thus, it is suitable for online, decentralized learning, unlike various prior mean-field solution concepts.

\vspace{5pt}

\noindent \textit{Explanatory and Predictive Power:}  The merits and advantages of considering subjective notions of best-responding and equilibrium do not lie in their ability to approximate their objective analogs. Rather, these notions have merit because of their explanatory and predictive power when studying independent learners in decentralized settings. Independent learners engage in an iterative learning process to (subjectively) evaluate their policies. If the joint policy constitutes a subjective equilibrium, then the learning process and modelling assumption of each agent jointly ensure that the agent will assess itself to be behaving optimally. As a result, each agent will continue to use the policy it is already using. In this sense, this definition of subjective equilibrium is self-reinforcing and individually rational given one's subjective beliefs. Indeed, in the coming sections, we will show that simple independent learners converge to subjective equilibrium. That is, one sees that subjective equilibrium policies are stable points for independent learning, and stability may be desirable from a system design point of view. With this perspective, the subjective framework can serve to explain the advantages and shortcomings of existing works that apply single-agent algorithms to multi-agent settings.

\subsection{Existence of Subjective Equilibrium}

\begin{lemma} \label{lemma:mean-field-subjective-equilibrium}
Let $\bG$ be a partially observed $n$-player mean-field game with mean-field observability (Assumption~\ref{ass:mean-field-state-observability}), and suppose Assumption~\ref{ass:state-visitation} holds. Let $\epsilon > 0$. Then, there exists a $(\VV^{*} , \WW^{*})$-subjective $\epsilon$-equilibrium. That is, $\Subj_{\epsilon} ( \VV^{*}, \WW^{*} ) \not= \varnothing$.  
\end{lemma}

\noindent A proof of  Lemma~\ref{lemma:mean-field-subjective-equilibrium}  can be found in Appendix~\ref{appendix:mean-field-subjective-equilibrium}. Lemma~\ref{lemma:mean-field-subjective-equilibrium} does not rule out the existence of \textit{subjective-but-not-objective} $\epsilon$-equilibrium policies, i.e. policies in $\Subj_{\epsilon} ( \VV^{*}, \WW^{*} ) \setminus \eq[\epsilon] $. 

A result analogous to Lemma~\ref{lemma:mean-field-subjective-equilibrium} holds when mean-field observability (Assumption~\ref{ass:mean-field-state-observability}) is strengthened to global observability (Assumption~\ref{ass:global-state-observability}). We refrain from including this result, since it mirrors the result above. Moreover, under global observability, the notion of subjective equilibrium essentially coincides with objective equilibrium.

\subsection{On Subjective Equilibrium under Compressed Observability}

The proof techniques used to guarantee the existence of both objective and subjective equilibria under Assumption 1 and 2 do not apply to partially observed $n$-player mean-field games with the observation channels of compressed observability (Assumption~\ref{ass:compressed-observability}) or local observability (Assumption~\ref{ass:local-state-observability}). In the convergence analysis of the coming sections, we will assume the existence of subjective equilibria when one of Assumptions~\ref{ass:compressed-observability} or \ref{ass:local-state-observability} holds.

Establishing existence of subjective equilibrium under compressed or local observability is an open problem, the resolution of which is closely related to other solution concepts in both $n$-player mean-field games and the $n \to \infty$ limit models. Under various names, several solution concepts have been proposed for mean-field games, and a number of approximation results relate these solutions concepts to one another. Some of these approximation results relate solution concepts for models with finitely many players to solution concepts for models with an infinite number of players, while others relate alternative solution concepts within a single model. For a sampling of works in this line, we cite \cite[]{weintraub2005oblivious, weintraub2008markov, adlakha2015equilibria, saldi2018markov} and \cite[]{arslan2023subjective} and the references therein. 

Due to similarities between the model considered here and the model considered in \cite[]{arslan2023subjective}, it may be possible to modify the analysis of \cite[Theorem 4.1]{arslan2023subjective} and use that line of argument to establish the existence of subjective equilibrium when $n$ is sufficiently large by relying on the existence of (objective) equilibrium in the associated limit model. In Appendix~\ref{appendix:large-N} of this paper, we present a result of the aforementioned type (Theorem~\ref{theorem:large-N-equilibrium}). We also describe how this approximation result may be used to establish the existence of subjective $\epsilon$-equilibrium in some partially observed $n$-player mean-field games.

\section{Subjective Satisficing: Algorithm and Convergence Result}  \label{sec:satisficing-paths}

In this section, we present an independent learning algorithm suitable for decentralized learning in partially observed $n$-player mean-field games. This algorithm belongs to the paradigm of \emph{win-stay, lose-shift algorithms}, in which agents assess the performance of a given policy by learning, and then update their policies in response to this assessment. This paradigm is popular in works on independent learners, as the structure is relatively simple and (subjectively) individually rational. Due to its win-stay, lose-shift structure, the algorithm we present in Section~\ref{ss:learning-algorithm} is representative of a variety of independent learning algorithms. Analysis of the convergence behaviour of Algorithm~\ref{algo:main} is, therefore, informative for interpreting or predicting the behaviour of independent learners and win-stay, lose-shift algorithms in large-scale, decentralized systems. We will show that Algorithm~\ref{algo:main} drives policies to subjective $\epsilon$-equilibrium, which illustrates the predictive and explanatory power of this new solution concept. 

\noindent For clarity of presentation, this section studies games with mean-field observability. Analogous structural results (without learning) under global, compressed, and local observability are given in Appendix~\ref{appendix:satisficing-under-other-observation-channels}. Analogous learning results under these observation channels are given in Appendix~\ref{appendix:learning-results-under-other-observation-channels}.

\subsection{Win-Stay, Lose-Shift Algorithms}

\emph{Win-Stay, Lose-Shift Algorithms} are a class of algorithms with the following form: players do not switch policies when they are satisfied with their performance (winning), but are free to experiment with other policies when they are unsatisfied (losing). This approach is based on the \textit{satisficing} principle \cite[]{simon1956rational}, which posits that a boundedly rational agent may tolerate some suboptimality in its performance. Thus, if an agent \emph{thinks} that its performance is $\epsilon$-optimal, it will not switch policies. Otherwise, if the agent thinks its performance is not $\epsilon$-optimal, it may switch policies in an exploratory manner. In this work, we use the term \emph{satisficing} synonymously with win-stay, lose-shift. 

The overarching structure described above allows for different satisfaction criteria, and unifies a number of different MARL algorithms \cite[]{chien2011convergence, Chasparis2013aspiration, candogan2013near, posch1999win, YAY-TAC}.  In what follows, we adopt the satisfaction criterion of subjective $\epsilon$-best-responding, using the subjective functions of $(\VV^{*}, \WW^{*})$ described in Section~\ref{sec:naive-learning} and defined in Appendix~\ref{appendix:implied-MDPs}. We will focus on win-stay, lose-shift algorithms in which players select their successor policies from a subset of their stationary policies. In symbols, player $i$ will select its policies from some set $\widehat{\Pi}^i \subset \Pi^i_{S}$. This restriction can be motivated either by bounded rationality or by player $i$'s ability to represent policies with only finite precision.  With these specifications in mind, the win-stay, lose-shift dynamics we consider take the following form: at time $k$, player $i \in \NN$ selects its policy $\pi^i_{k+1}$ with reference to $\bpi_k = ( \pi^i_k, \bpi^{-i}_k)$ as follows:
\[
\pi^i_{k+1} =
	\begin{cases}
		\pi^i_k, 								&\text{ if } \pi^i_k \in \SubjBR^i_{\epsilon} ( \VV^{*} ,\WW^{*} )  	\\
		\tilde{\pi}^i \sim \uplambda^i ( \cdot | \bpi_k )	&\text{ else, }
	\end{cases}
\]
where $\uplambda^i ( \cdot | \bpi_k )$ is a distribution over $\widehat{\Pi}^i$ that may depend on the joint policy $\bpi_k$.

\subsection{A Subjective Satisficing Algorithm with Oracle Access}

This section studies win-stay, lose-shift algorithms in a simplified setting, where an oracle provides each player with the relevant subjective functions for its policy update. A full learning algorithm is presented in Section~\ref{ss:learning-algorithm}. The complete learning algorithm can be interpreted as a noisy analog of the oracle algorithm below.


\begin{algorithm}[h] \label{algo:oracle}
	\SetAlgoLined
	\DontPrintSemicolon
	\SetKw{Receive}{Receive}
	\SetKw{parameters}{Set Parameters}
	\SetKw{initialize}{Initialize}
	\SetKwBlock{For}{for}{end} 
	
	\parameters    \;
	\Indp 
	$e^i \in (0,1)$: experimentation probability when not subjectively $\epsilon$-best-responding \;
	$\widehat{\Pi}^i \subset \widehat{\Pi}^i_{S}$: a subset of stationary policies. \;
	\Indm 
	
	\initialize $\pi_0^i \in \widehat{\Pi}^i$: initial policy \;

	\BlankLine
	
	\For($k \geq 0$ ($k^{th}$ policy update{)}){ 
		
		\BlankLine 
		\Receive $V^{*i}_{\bpi_k}$ and $W^{*i}_{\bpi_k}$ from oracle    \; 
		
		\vspace{5pt}
		
		\If	{    $V^{*i}_{\bpi_k } ( y ) \leq \min_{a^i} W^{*i}_{\bpi_k} ( y, a^i ) + \epsilon$ for all $y \in \yy$	   } { $\pi^i_{k+1} = \pi^i_k$ }
		\Else(){
			$\pi^i_{k+1} \sim (1 - e^i) \delta_{ \pi^i_k} + e^i \uniform ( \widehat{\Pi}^i ) 	$
			\BlankLine
		}
		
%
%
	}
	
	\caption{Subjective $\epsilon$-satisficing Policy Revision (for player $i \in \NN$)} \label{algo:oracle}
\end{algorithm}


Algorithm~\ref{algo:oracle} induces a time homogeneous Markov chain on $\widehat{\bPi} = \times_{i \in \NN} \widehat{\Pi}^i$. Policy $\bpi^{*} \in \widehat{\bPi}$ is absorbing for this Markov chain $\iff$ it is a $( \VV^{*}, \WW^{*})$-subjective $\epsilon$-equilibrium. At this point, the set $\widehat{\bPi}$ has not been explicitly defined, so the question of whether $\widehat{\bPi}$ admits a subjective equilibrium requires examination. Appropriate selection of $\widehat{\bPi}$ is discussed in the next subsection.  Moreover, even if $\widehat{\bPi}$ contains subjective equilibria, it is not clear \emph{a priori} that such subjective equilibria are accessible from particular non-equilibrium joint policies by a process of policy adjustment in which satisfied agents do not change their policies. This consideration leads to the following definitions. For the following definitions, let $\bG$ be a partially observed $n$-player mean-field game, let $i \in \NN$, let $\epsilon \geq 0$, and recall that $(\VV^{*}, \WW^{*})$ denotes the subjective function family for $\bG$ and was defined in Definition~\ref{def:subjective-function-family}.

\begin{definition} \label{def:subjective-satisficing-paths}
A sequence of policies $( \bpi_k )_{k \geq 0}$ in $\bPi_{S}$ is called a \emph{subjective $\epsilon$-satisficing path} if, for every $i \in \NN$ and $k \geq 0$, we have 
\[
\pi^i_k \in \SubjBR^i_{\epsilon} ( \bpi^{-i}_k , \VV^{*} , \WW^{*} ) \Rightarrow \pi^i_{k+1} = \pi^i_k .
\]
\end{definition}

Intuitively, when agents jointly update their policies along a subjective $\epsilon$-satisficing path, an agent only switches its policy when it (subjectively) deems the policy to be performing poorly. No further restrictions are placed on how an agent is allowed to switch (or not switch) its policy when it is subjectively unsatisfied.  

\begin{definition} 
Let $\widehat{\bPi} \subseteq \bPi_{S}$. The game $\bG$ is said to have the \emph{subjective $\epsilon$-satisficing paths property in $\widehat{\bPi} $} if the following holds: for every $\bpi \in \widehat{\bPi} $, there exists a subjective $\epsilon$-satisficing path $( \bpi_k )_{ k \geq 0}$ such that \textnormal{(i)} $\bpi_0 = \bpi$, \textnormal{(ii)} $\bpi_k \in \widehat{\bPi} $ for all $k \geq 0$, and \textnormal{(iii)} for some $K < \infty$, $\bpi_K \in \Subj_{\epsilon} ( \VV^{*} , \WW^{*})$. 
\end{definition}


This defines a subjective analog of the \emph{objective} satisficing paths theory first proposed in \cite[]{yongacoglu2023satisficing} for Markov games, also studied in \cite[]{yongacoglu2024generalizing} and \cite[]{yongacoglu2024paths}. We now describe how one may use \emph{quantization} to select the policy subset $\widehat{\bPi}$ such that the game $\bG$ has the subjective $\epsilon$-satisficing paths property within $\widehat{\bPi}$.

\subsection{Quantization of the Policy Space}  \label{sec:quantized-paths} 

In this section, we describe the quantization/discretization of policy sets. We argue that if the restricted set of policies $\widehat{\bPi}$ is obtained via a sufficiently fine quantization of the original set $\bPi_{S}$, then the performance loss for agent $i$ will be negligible if it optimizes over $\widehat{\Pi}^i$ instead of $\Pi^i_{S}$. Moreover, we argue that if the restricted subset of policies is suitably fine, symmetric, and soft (Definition~\ref{def:soft-policies}), then it will contain a subjective $\epsilon$-equilibrium and, further, satisficing dynamics will drive policies to such subjective equilibria from any initial joint policy. 

For the following definitions, let $\bG$ be a partially observed $n$-player mean-field game with mean-field observability, and let $i \in \NN$. Recall that $d^i$ is a metric on the set $\Pi^i_{S}$. 

\begin{definition}
Let $\xi > 0$ and $\tilde{\Pi}^i \subseteq \Pi^i_{S}$. A mapping $q^i : \tilde{\Pi}^i  \to \tilde{\Pi}^i $ is called a \emph{$\xi$-quantizer (on $\tilde{\Pi}^i$)} if \textnormal{(1)} $q^i ( \tilde{\Pi}^i  ) := \{ q^i ( \pi^i ) : \pi^i \in \tilde{\Pi}^i  \}$ is a finite set and \textnormal{(2)} $d^i ( \pi^i, q^i ( \pi^i )) < \xi$ for all $\pi^i \in \tilde{\Pi}^i $. 
\end{definition}

\begin{definition}
Let $\xi > 0$ and let $\tilde{\Pi}^i \subseteq \Pi^i_{S}$. A set of policies $\widehat{\Pi}^i \subseteq \tilde{\Pi}^i$ is called a $\xi$-quantization of $\tilde{\Pi}^i $ if $\widehat{\Pi}^i  = q^i ( \tilde{\Pi}^i )$, where $q^i$ is some $\xi$-quantizer on $\tilde{\Pi}^i $. 
\end{definition}

A set $\widehat{\Pi}^i \subseteq \Pi^i_{S}$ is called a quantization of $\Pi^i_{S}$ if it is a $\xi$-quantization of $\Pi^i_{S}$ for some $\xi > 0$. A quantization $\widehat{\Pi}^i$ is called \emph{soft} if each policy $\pi^i \in \widehat{\Pi}^i$ is soft in the sense of Definition~\ref{def:soft-policies}. The expression ``fine quantization'' will be used to reflect that a policy subset is a $\xi$-quantization for suitably small $\xi$. We extend these definitions and terminological conventions to also cover joint policies. For instance, $\widehat{\bPi} = \times_{i \in \NN} \widehat{\Pi}^i \subset \bPi_{S}$ is a $\xi$-quantization of $\bPi_{S}$ if each $\widehat{\Pi}^i$ is a $\xi$-quantization of $\Pi^i_{S}$, and so on.

\begin{definition}
Let $\widehat{\bPi} \subset \bPi_{S}$ be a quantization of $\bPi_{S}$. We say that $\widehat{\bPi}$ is \emph{symmetric} if $\widehat{\Pi}^i = \widehat{\Pi}^j$ for each $i, j \in \NN$. 
\end{definition}

\begin{lemma}  \label{lemma:mean-field-quantized-paths}
Let $\bG$ be an $n$-player mean-field game with mean-field observability (Assumption~\ref{ass:mean-field-state-observability}) and suppose Assumption~\ref{ass:state-visitation} holds. Let $\epsilon > 0$. There exists $\xi = \xi ( \epsilon ) > 0 $ such that if $\widehat{\bPi} \subset \bPi_{S}$ is any soft, symmetric $\xi$-quantization of $\bPi_{S}$, then we have 
\begin{enumerate}
	\item[1)] $\eq[\epsilon] \cap \widehat{\bPi} \not= \varnothing$ and $\Subj_{\epsilon} ( \VV^{*} , \WW^{*} ) \cap \widehat{\bPi} \not= \varnothing$, and
	\item[2)] $\bG$ has the $(\VV^{*} , \WW^{*} )$-subjective $\epsilon$-satisficing paths property in $\widehat{\bPi}$. 
\end{enumerate}
\end{lemma}

The first part can be seen using the proof of Lemma~\ref{lemma:mean-field-subjective-equilibrium}. The proof of the second part resembles that of \cite[Theorem 3.6]{yongacoglu2023satisficing}, which considered $n$-player symmetric stochastic games with full (global) state observability. One modification is needed: Corollary 2.9 of  \cite[]{yongacoglu2023satisficing} must be replaced by a partial observations analog, Lemma~\ref{lemma:mean-field-subjective-BR-iff}, whose statement and proof is given in Appendix~\ref{appendix:satisficing-results}

Lemma~\ref{lemma:mean-field-quantized-paths} guarantee that the game $\bG$ has the $(\VV^{*}, \WW^{*})$-subjective $\epsilon$-satisficing paths property within finely quantized subsets of policies. This has two desirable consequences for algorithm design purposes. First, players can restrict their policy search from an uncountable set (all stationary policies) to a finite subset of policies with only a small loss in performance. Second, since the game $\bG$ has the $(\VV^{*}, \WW^{*})$-subjective $\epsilon$-satisficing paths property within $\widehat{\bPi}$, play can be driven to (subjective) $\epsilon$-equilibrium by changing only the policies of those players that are ``$\epsilon$-unsatisfied," so to speak. We thus obtain a stopping condition, whereby player $i$ can settle on a policy whenever it is subjectively $\epsilon$-best-responding. 

Taken together, these points remove the need for \textit{coordinated} search of the joint policy space $\bPi_{S}$: since $(\VV^{*}, \WW^{*})$-subjective $\epsilon$-satisficing paths to equilibrium exist within $\widehat{\bPi}$ and $\widehat{\bPi}$ is finite, play can be driven to subjective $\epsilon$-equilibrium even by random policy updating by those players that are not subjectively $\epsilon$-best-responding. Moreover, this also removes the need for specialized policy update rules, such as inertial best-responding \cite[]{AY2017}, that take into account special structure in the game. 

The preceding remarks are formalized in Lemma~\ref{lemma:mean-field-oracle}, below.

\begin{lemma} \label{lemma:mean-field-oracle}
Let $\bG$ be a partially observed $n$-player mean-field game satisfying Assumptions~\ref{ass:mean-field-state-observability} and \ref{ass:state-visitation}, and let $\epsilon > 0$. Let $\widehat{\bPi}  \subset \bPi_{S}$ be a quantization of $\bPi_{S}$ and suppose $\widehat{\bPi} $ satisfies (1) $\widehat{\bPi}  \cap \Subj_{\epsilon} ( \VV^{*}, \WW^{*} ) \not= \varnothing$; (2) $\widehat{\Pi}^i = \widehat{\Pi}^j$ for all $i,j \in \NN$; and (3) every policy $\bpi \in \widehat{\bPi} $ is soft.  

Suppose that each agent $i \in \NN$ updates is policy sequence $\{ \pi^i_k \}_{ k \geq 0}$ according to Algorithm~\ref{algo:oracle} and that, for each $k \geq 0$, the policy updates for $\bpi_{k+1}$ are conditionally independent across agents given $\bpi_k$. Then, $\lim_{k \to \infty} \P ( \bpi_k \in \widehat{\bPi}  \cap \Subj_{\epsilon} ( \VV^{*}, \WW^{*} ) ) = 1. $
\end{lemma}

The proof of Lemma~\ref{lemma:mean-field-oracle} can be found in Appendix~\ref{appendix:satisficing-results}. The proof crucially leverages the subjective satisficing structure described in the second part of Lemma~\ref{lemma:mean-field-quantized-paths}.

\

\noindent \textbf{Remark:} The choice to update the policy according to $\pi^i_{k+1} \sim (1 - e^i) \delta_{\pi^i_k} + e^i \uniform (  \widehat{\Pi}^i)$, in Line 10 of Algorithm~\ref{algo:oracle}, was somewhat arbitrary. In particular, the choice to uniformly mix over $\widehat{\Pi}^i$ with probability $e^i > 0$ was made to ensure the paths to equilibrium exist in the Markov chain of Lemma~\ref{lemma:mean-field-oracle}. The choice to remain with one's old policy with probability $1-e^i$ was arbitrarily picked for ease of exposition, and can be replaced by any suitable distribution over $\widehat{\Pi}^i$ according to the taste of the system designer. Some choices may include gradient descent projected back onto $\widehat{\Pi}^i$ or selecting a best-response to $\bpi^{-i}_k$ within $\widehat{\Pi}^i$.  Such changes may result in a significant speed-up of convergence to equilibrium when they are well-suited to the underlying game, but the guarantee holds in any case due to the uniform randomization.

\subsection{An Independent Learning Algorithm for $n$-Player Mean-Field Games}  \label{ss:learning-algorithm}

We now present Algorithm~\ref{algo:main}, a win-stay, lose-shift algorithm that does not rely on non-local information and does not require the many agents of the system to use identical policies during learning. As such, Algorithm~\ref{algo:main} is suitable for online, decentralized learning applications in large-scale systems, where agents have a limited understanding of the system.

At a high level, Algorithm~\ref{algo:main} is a noise-perturbed, learning-based analog of Algorithm~\ref{algo:oracle}. Rather than receiving subjective function information from an oracle, here the subjective functions are learned using system feedback. The learned subjective functions are then used to estimate whether the agent is (subjectively) $\epsilon$-best-responding.

\begin{algorithm}[h] 
	\SetAlgoLined
	\DontPrintSemicolon
	\SetKw{Receive}{Receive}
	\SetKw{Reset}{Reset}
	\SetKw{parameters}{Set Parameters}
	\SetKw{initialize}{Initialize}
	\SetKwBlock{For}{for}{end}
	
	\parameters \;
	\Indp
	$\widehat{\Pi}^i \subset \Pi^i_{S}$ : a fine quantization of $\Pi^i_{S}$ \; 
	
	$\{ T_k \}_{k \geq 0}$: a sequence in $\nn$ of ``exploration phase" lengths \;
	\vspace{1pt}
	\Indp set $t_0 = 0$ and $t_{k+1} = t_k + T_k$ for all $k \geq 0.$ \;
	\vspace{1pt}
	\Indm $e^i \in (0,1)$: random policy updating probability \;
	$d^i\in(0,\infty)$: tolerance level for sub-optimality \;
	\Indm
	\BlankLine
	
	\initialize  $\pi_0^i \in \widehat{\Pi}^i $ (arbitrary), $\widehat{Q}_0^i = 0 \in \rr^{\yy \times \aa}$, $\widehat{J}^i_0 = 0 \in \rr^{\yy}$  \\

	\For($k \geq 0$ ($k^{th}$ exploration phase{)}){ 
		\For( $t = t_k, t_k +1, \dots, t_{k+1} - 1$)  
		{
			Observe $y^i_t = \varphi^i ( \bx_t )$ \;
			Select $a^i_t \sim 	\pi^i_k ( \cdot | y^i_t  )$  \;
			
			Observe $c^i_t := c( x^i_t, \upmu (  \bx_t), a^i_t )$ and $y^i_{t+1} $ \;
			Set $n_t^i = \sum_{\tau = t_k}^{t} \textbf{1} \{ ( y^i_{\tau} , a^i_{\tau} ) = ( y^i_t, a^i_t ) \}$ \;
			Set $m_t^i =  \sum_{\tau = t_k}^{t} \textbf{1} \{  y^i_{\tau} = y^i_t  \}$ \;
			\vspace{3pt}
			$\widehat{Q}_{t+1}^i( y^i_t , a_t^i)  =   \left(1- \frac{1}{ {n_t^i}} \right) \widehat{Q}_t^i(   y^i_t  , a_t^i) + \frac{1}{{n_t^i}}  \big[ c^i_t +  \gamma \min_{\tilde{a}^i} \widehat{Q}_t^i(  y^i_{t+1}  , \tilde{a} ^i) \big]$  \;
			\vspace{3pt}
			$\widehat{J}^i_{t+1} (  y^i_t ) = \left(1 - \frac{1}{{m^i_t}} \right) \widehat{J}^i_t ( y^i_t  ) + \frac{1}{ {m^i_t} } \left[  c^i_t + \gamma \widehat{J}^i_t ( y^i_{t+1} )	\right] $ \;
		}
		\vspace{6pt}

		\If { $\widehat{J}^i_{t_{k+1}} ( y ) \leq \min_{a^i} \widehat{Q}^i_{t_{k+1}} ( y , a^i )  + \epsilon + d^i$ $\forall y \in \yy$,}{ $\pi^i_{k+1} = \pi^i_k$ }
		\Else( ){$\pi^i_{k+1} \sim (1- e^i) \delta_{\pi^i_k} + e^i \uniform ( \widehat{\Pi}^i) $  }
		
		\vspace{3pt}
		\Reset $ \widehat{J}^i_{t_{k+1}} = 0 \in \rr^{\yy}$ and $ \widehat{Q}^i_{t_{k+1}} = 0 \in \rr^{\yy \times \aa}$ \;

	}
	
	\caption{Independent Learning} \label{algo:main}
\end{algorithm}

\subsubsection*{Learning with Mean-Field State Information }

We now present a result on the convergence behaviour of Algorithm~\ref{algo:main} under mean-field observability. Analogous results for other observation channels can be found in Appendix~\ref{appendix:learning-results-under-other-observation-channels}.

Under mean-field observability, policy iterates obtained from Algorithm~\ref{algo:main} drive play to $(\VV^{*}, \WW^{*})$-subjective $\epsilon$-equilibrium with high probability. In order to state this result formally in Theorem~\ref{theorem:mean-field-state-observability}, we now fix $\epsilon > 0$ and make the following assumptions on the various parameters of Algorithm~\ref{algo:main}.

\begin{assumption} \label{ass:mean-field-quantized-set}
Fix $\epsilon > 0$. Let $\widehat{\bPi}$ be a fine quantization of $\bPi_{S}$ satisfying: \textnormal{(1)}~$\widehat{\Pi}^i = \widehat{\Pi}^j$ for all $i,j \in \NN$; \textnormal{(2)} $\widehat{\bPi} \cap \Subj_{\epsilon} ( \VV^{*}, \WW^{*}) \not= \varnothing$; \textnormal{(3)} For any $\bpi \in \widehat{\bPi}$, $\bpi$ is soft.
\end{assumption}

Next we present a restriction on the parameters $\{ d^i \}_{i \in \NN}$. For each player $i \in \NN$, the tolerance parameter $d^i$ is taken to be positive, to account for noise in the learned estimates, but small, so that poorly performing policies are not mistaken for subjective $\epsilon$-best-responses. The bound $\bar{d}_{\mf}$ below is defined analogous to the term $\bar{d}$ in \cite{yongacoglu2023satisficing} and depends on both $\epsilon$ and $\widehat{\bPi}$. %

\begin{assumption} \label{ass:mean-field-delta}
For all $i \in \NN$, $d^i  \in (0, \bar{d}_{\mf} )$, where $\bar{d}_{\mf} = \bar{d}_{\mf} ( \epsilon, \widehat{\bPi})$ is specified in Appendix~\ref{appendix:proof-mean-field-state}. 
\end{assumption}

\begin{theorem} \label{theorem:mean-field-state-observability}
Let $\bG$ be a partially observed $n$-player mean-field game satisfying Assumptions~\ref{ass:mean-field-state-observability} and \ref{ass:state-visitation}, and let $\epsilon > 0$. Suppose the policy set $\widehat{\bPi}$ and the tolerance parameters $\{ d^i \}_{i \in \NN}$ satisfy Assumptions~\ref{ass:mean-field-quantized-set} and \ref{ass:mean-field-delta}, and suppose all players follow Algorithm~\ref{algo:main}. For any $\xi > 0$, there exists $\tilde{T} = \tilde{T} ( \xi , \epsilon, \widehat{\bPi} ,  \{ d^i \}_{i \in \NN} )$ such that if $T_k \geq \tilde{T}$ for all $k$, then 
\[
\P \left( \bpi_k \in \widehat{\bPi} \cap \Subj_{\epsilon} ( \VV^{*}, \WW^{*} ) \right) \geq 1 - \xi, 
\] 
for all sufficiently large $k$. 
\end{theorem}

\noindent A proof of Theorem~\ref{theorem:mean-field-state-observability} is available in Appendix~\ref{appendix:proof-mean-field-state}.  This result does not rule out the possibility that play settles at a subjective equilibrium that is not an objective equilibrium. That is, it is possible that $\left( \widehat{\bPi} \cap \Subj_{\epsilon} ( \VV^{*}, \WW^{*} ) \right)\setminus \eq[\epsilon] \not= \varnothing$, and that Algorithm~\ref{algo:main} drives play to such a policy.

\noindent \textbf{Remark:}  The guarantee of Theorem~\ref{theorem:mean-field-state-observability} continues to hold even if $\delta_{\pi^i_k}$ is replaced by any transition kernel in $\PP ( \widehat{\Pi}^i | \widehat{\Pi}^i \times \rr^{\yy} \times \rr^{\yy \times \aa} )$, where the distribution over $\widehat{\Pi}^i$ depends on $\pi^i_k$, $\widehat{J}^i_{t_{k+1}}$, and $\widehat{Q}^i_{t_{k+1}}$. Furthermore, each agent $i \in \NN$ may use a different transition kernel for this update, allowing for heterogeneity in the learning dynamics.

\subsubsection*{Learning Under Other Observation Channels}

Analogous convergence results can be stated under global observability (Theorem~\ref{theorem:global-state-observability}), compressed observability, or local observability. The latter two cases are described in Theorem~\ref{theorem:compressed-observability}, where the convergence result requires an additional assumption on the existence of subjective $\epsilon$-equilibrium.

\section{Simulation Study} \label{sec:simulation}

We ran Algorithm~\ref{algo:main} on a 20-player mean-field game with compressed observability, described below in \eqref{eq:simulation-game}. This game can be interpreted as a model for decision-making during an epidemic or as a model for vehicle use decisions in a traffic network. 

The game $\bG$ used for our simulation is given by the following list:
\begin{equation} \label{eq:simulation-game}
\bG = ( \NN , \xx , \yy , \aa , \{ \varphi^i \}_{i \in \NN}, P_{\loc} , c, \gamma , \nu_0,  ) .
\end{equation}

The components of $\bG$ are as follows.  $\NN = \{ 1, 2, \dots, 20 \}$ is a set of 20 players. The local state of an agent is interpreted as the agent's overall health, and takes values in $\xx = $ \{bad, medium, good\}. The agent selects between actions in  $\aa = $ \{go, wait, heal\}. The action ``go'' corresponds to proceeding with one's usual activities, ``wait'' corresponds to avoiding activities with high risk of disease transmission, and ``heal'' corresponds to seeking healthcare. Each agent receives a signal that the proportion of agents in bad condition is either ``low'' or ``high'', i.e. $\yy = \xx \times $\{low, high\}. For each $i \in \NN$ and $\bs \in \bX$, player $i$'s observation is given by $\varphi^i ( \bs ) = ( s^i , f ( \upmu ( \bs )) )$, where for any $\nu \in \Delta ( \xx )$, $f(\nu)$ is
		\[
		f ( \nu ) =	\begin{cases}
					\text{low} &\text{ if } \nu (\text{bad}) < 0.35 , \\
					\text{high} &\text{ if } \nu(\text{bad}) \geq 0.35. 
				\end{cases}
		\]

The stage cost function $c: \xx  \times  \Delta( \xx ) \times \aa \to \rr$  is given by 
		\begin{align*}
			c ( s, \nu , a )	:=& -  R_{\rm go } \cdot \textbf{1} \{ a = \text{go} \} + R_{\rm bad} \cdot  \textbf{1} \{ s = \text{bad} \} + R_{\rm heal} \cdot \textbf{1} \{ a = \text{heal} \}			\end{align*}
for all $( s, \nu , a ) \in \xx \times \Delta( \xx ) \times \aa,$ where $R_{\rm go} = 5$ is a reward for undertaking one's usual business, $R_{\rm bad} = 10$ is a penalty for being in bad condition, and $R_{\rm heal} = 3$ is the cost of seeking healthcare. We note that the stage cost $c(s,\nu,a)$ depends only on $(s,a)$, and the only strategic coupling in $\bG$ is through coupled state dynamics. 

The discount factor $\gamma = 0.8$, and $\nu_0 \in \Delta ( \bX )$ is the product uniform distribution: $x^i_0 \sim \uniform ( \xx )$ for each $i \in \NN$ and the random variables $\{ x^i_0 \}_{i \in \NN}$ are jointly independent.

We now describe the transition kernel $P_{\loc}$, which captures the state transition probabilities as a function of one's local state, the mean-field state, and one's local action.  First, for any $s \in \xx $ and $\nu \in \Delta ( \xx )$, we have $P_{\loc} ( \cdot | s, \nu,  \text{wait} ) = \delta_{ \{s \} }$. That is, when a player waits, its local state is left unchanged with probability 1.  For each $s \in  \xx $ and $\nu \in \Delta ( \xx )$, we summarize $P_{\loc} ( \cdot | s, \nu, \text{heal} )$ and $P_{\loc} ( \cdot | s, \nu, \text{go} )$  in the tables below:

\begin{center}
\begin{tabular}{ | c | c | c | c | } 
\hline
	& $P_{\loc} ( \text{bad} | s, \nu,  \text{heal} )$		& $P_{\loc} ( \text{medium} | s, \nu, \text{heal} )$		& $P_{\loc} ( \text{good} | s, \nu, \text{heal} )$ \\
\hline
$s = \text{bad}$		&0.25 & 0.5 & 0.25 \\ 
$s = \text{medium}$	&0.0 & 0.25 & 0.75 \\ 
$s = \text{good}$	&0.0 & 0.0 & 1.0 \\ 
\hline
\end{tabular}
\end{center}

\

To succinctly describe $P_{\loc} ( \cdot | s, \nu, \text{go} )$ for each $s \in \xx $ and $\nu \in \Delta ( \xx )$, we introduce a function to describe the likelihood of minor condition degradation (from good to medium or medium to bad) as a function of $\nu(\text{bad})$:
\[
m( \nu  ) := \frac{ \exp( 12 [ \nu(\text{bad}) - 0.3 ]  )   }{ 1 + \exp( 12 [ \nu(\text{bad}) - 0.3 ]  )  }
\]

We also encode a fixed 0.15 chance of severe condition deterioration playing ``go", independent of the mean-field term. This leads to the following transition probabilities: 

\begin{center}
\begin{tabular}{ | c | c | c | c | } 
\hline
$s $	& $P_{\loc} ( \text{bad} | s, \nu, \text{go} )$		& $P_{\loc} ( \text{medium} | s, \nu, \text{go} )$		& $P_{\loc} ( \text{good} | s,\nu, \text{go} )$ \\
\hline

bad		&1.0 & 0.0 & 0.0 \\ 

medium	&$0.15 + 0.85 \cdot m ( \nu ) $  &$1 - (0.15 + 0.85 \cdot  m ( \nu )   )$  &0.0 \\ 

good	&0.15 &$0.85 \cdot m ( \nu ) $ &$ 1 - (0.15 + 0.85 \cdot  m ( \nu )   )$  \\ 
\hline
\end{tabular}
\end{center}

As the number of agents in bad condition increases, the risk of degrading one's own condition when playing ``go'' also increases, slowly at first, then abruptly as the proportion of agents approaches and surpasses a threshold quantity.

The model data in \eqref{eq:simulation-game} has been chosen to encourage the following splitting behaviour: an agent in bad condition should always seek healthcare; an agent in good condition should always play ``go''; and an agent in medium condition should play ``wait'' when it observes that the proportion of agents in the bad state is high, while it should play ``go'' otherwise.

This behaviour is not a dominant strategy in the game theoretic sense: if, for some reason, a large number of agents play ``go'' when in bad condition, then an agent in medium condition does not gain anything by waiting, since the proportion of agents in bad condition will remain high. In this circumstance, the agent may prefer to either risk playing ``go'' or to simply play ``heal". As such, the dynamics to subjective equilibrium (if it exists) are non-trivial.

Using the game in \eqref{eq:simulation-game}, we ran 250 independent trials of self-play under Algorithm~\ref{algo:main}, were each trial consisted of 20 exploration phases and each exploration phase consisted for 25,000 stage games. Our chosen parameters were $\epsilon = 5$, $d^i \equiv 1.5$, and $e^i \equiv 0.25$. 

We define $\tilde{\Pi} \subset \PP ( \aa | \yy )$ to be the following finite subset of stationary policies: 
\[
\tilde{\Pi} :=  \{ \tilde{\pi} \in \PP ( \aa | \yy ) : 10^2 \cdot \tilde{\pi} ( a | y ) \in \zz \, , \; \forall y \in \yy, a \in \aa \} .
\]
That is, policies in $\tilde{\Pi}$ can be described using only two places after the decimal point in each component probability distribution. We then define a quantizer ${\rm quant} : \PP ( \aa | \yy ) \to \PP ( \aa | \yy ) $ as
\[
{\rm quant } ( \pi ) =	\begin{cases}
					\pi, &\text{ if } \pi \text{ is 0.025-soft} \\
					0.9 \cdot \pi + 0.1\cdot \pi_{\rm uniform} &\text{ otherwise,}
				\end{cases}
\]
where $\pi_{\rm uniform}$ is the policy such that $\pi_{\rm uniform} ( \cdot | y ) = \uniform( \aa )$ for each $y \in \yy$. In our simulation, players select policies from the finite set $\widehat{\Pi} := {\rm quant } (\tilde{\Pi} )$.

\begin{figure}[h]
\includegraphics[width=0.6\textwidth]{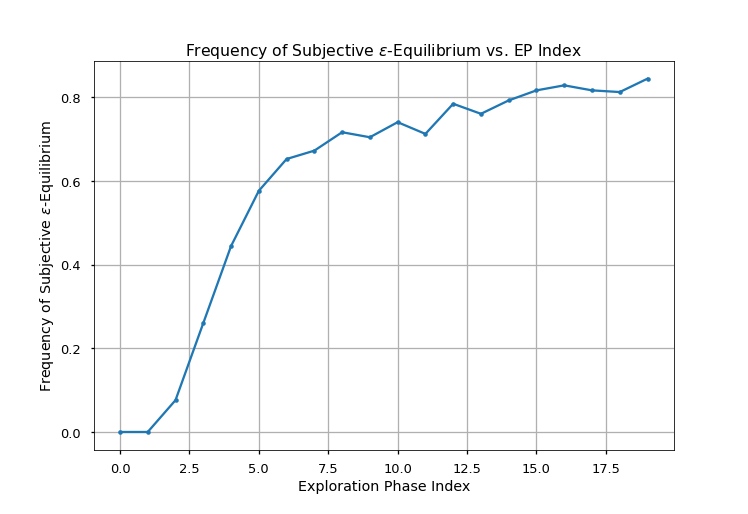}
\centering
\caption{Frequency of subjective $\epsilon$-equilibrium plotted against the exploration phase index, averaged over 250 trials.}\label{fig:equilibrium}
\end{figure}

\begin{figure}[h]
\includegraphics[width=0.6\textwidth]{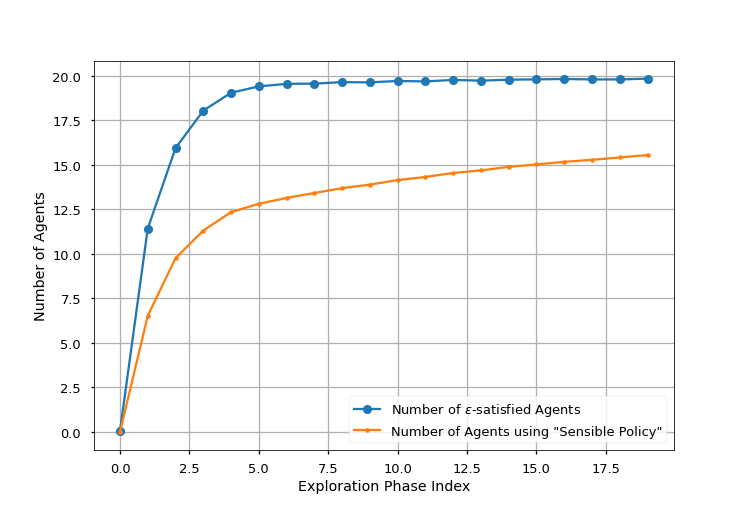}
\centering
\caption{Mean Number of players who are subjectively $\epsilon$-best-responding vs. exploration phase index, averaged over 250 trials.} \label{fig:mean-agents}
\end{figure}

Our results are summarized in Figures~\ref{fig:equilibrium} and \ref{fig:mean-agents} and in Table~\ref{table:results}. In Figure~\ref{fig:equilibrium}, we plot the frequency of subjective $\epsilon$-equilibrium against the exploration phase index. For our purposes, a policy at a given exploration phase is considered a subjective $\epsilon$-equilibrium if all 20 players perceive themselves to be $\epsilon$-best-responding, given their subjective state and action functions. We observe that the frequency steadily rises to over 84\% by the 20th exploration phase.  

\begin{table}
\centering \normalsize
\begin{tabular}{ c | c  } 
	Exploration Phase Index $k$	& $\frac{1}{250} \sum_{\tau = 1}^{250} \textbf{1} \{ \bpi_{k, \tau} \in \Subj_{\epsilon} (\VV^{*}, \WW^{*} )  \}    $ \\
\hline
	$k = 0$					& 0.0 \\
\hline
	$k=5$					&$0.576$ \\
\hline
	$k =10$					&$0.74$ \\
\hline
	$k =15$					&$0.816$ \\
\hline
	$k =19$					&$0.844$ \\
\hline
\end{tabular}
\caption{Frequency of Subjective $\epsilon$-equilibrium. $\bpi_{k, \tau}$ denotes the policy for EP $k$ during the $\tau^{th}$ trial.}
\label{table:results}
\end{table}

In Figure \ref{fig:mean-agents}, we plot two curves. The blue curve reports the mean number of agents, out of 20, that are subjectively $\epsilon$-best-responding at a given exploration phase index. This quantity quickly rises to, and remains at, 19.75, reflecting that a large majority of agents are $\epsilon$-satisfied even when the system is not at subjective $\epsilon$-equilibrium. Figure \ref{fig:mean-agents} also reports the number of agents using the quantized version of $\pi_{\rm sensible} \in \PP ( \aa | \yy )$, defined for each $(x, \sigma)$ pair as follows:  $   \pi_{\rm sensible} ( \cdot |  \text{good, low})    =   \pi_{\rm sensible} ( \cdot | \text{good, high})         = \delta_{\rm go} $ (i.e. uniformly playing ``go`` when in good condition); $  \pi_{\rm sensible} ( \cdot |  \text{bad, low}) =  \pi_{\rm sensible} ( \cdot |  \text{bad, high}) = \delta_{\rm heal}    $ (i.e. uniformly playing ``heal" when in bad condition); $ \pi_{\rm sensible} ( \cdot |  \text{medium, low})  = \delta_{\rm go}$ while $ \pi_{\rm sensible} ( \cdot |  \text{medium, high})  = \delta_{\rm wait}$. 

The orange curve in Figure~\ref{fig:mean-agents} reports the mean number of agents following the policy ${\rm quant} ( \pi_{\rm sensible} )$. Interestingly, this curve lags below the blue curve, reflecting that play often settled at an asymmetric joint policy that forms a subjective $\epsilon$-equilibrium.

\section{Discussion} \label{sec:discussion}

We now consider situations in which compressed observability (Assumption~\ref{ass:compressed-observability}) may be the most appropriate model for decentralized information. It is natural to begin by asking how a player's observations are actually obtained. (Note that the actual observation channel encountered need not be the same as the observation channel used in the model of the game.) We envision three (actual) observation channels as being the most plausible.

In the first scenario, agents obtain readings on the global state through local sensors. Such an (actual) observation channel is truly decentralized and would result in a limited view of the overall system, giving rise to compressed observability. In this case, the actual observation channel may be used as the model's observation channel. In a related second scenario, agents supplement local sensor readings by communicating with neighbours. Here, too, one may naturally take the model's observation channel to be the same as the actual observation channel, and Assumption~\ref{ass:compressed-observability} may be more appropriate than Assumption~\ref{ass:mean-field-state-observability}. 
	
In a third scenario, a centralized entity monitors the global state and broadcasts a compressed signal about the mean-field to the agents. This is the case, for example, in vehicle routing. Here, local states are locations in a traffic network, and action selection corresponds to selection of a path to one's destination. Agents may rely on a satellite navigation system to locate themselves in the network and to identify which paths are congested.  If the navigation system reports congestion using a tiered system of low-, moderate-, and high-congestion roads, then compressed observability is the relevant observation channel. 

In addition to applications where the actual observation channel results in a limited view of the overall system, compressed observability may also arise in systems with rich 
(actual) observation channels if players voluntarily choose to discard information in their learning process. When the number of players, $n$, and the number of local states, $| \xx |$, are both moderately large, the set of empirical measures $\Emp_n = \{ \upmu (  \bs ) : \bs \in \bX \}$ becomes unwieldy for the purposes of (tabular) learning. As such, it is reasonable to expect that naive independent learning agents will employ some form of function approximation, with quantization of $\Emp_n$ offering the simplest form of function approximation. Moreover, a common compression scheme may be shared by all agents in such an application, perhaps as the result of some shared ``conventional wisdom'' about the system. Building on the ideas of \cite[]{patil2024learning}, the question of whether non-tabular algorithms can learn well-performing, history-dependent policies involving compressed mean-field observations is an interesting direction for future research.

\section{Conclusion} \label{sec:conclusion}

In this paper, we considered partially observed $n$-player mean-field games from the point of view of decentralized independent learners. Independent learning is characterized by ignoring the presence of other strategic agents in the system, treating one's environment as if it were a single-agent MDP, and naively running single-agent learning algorithms to select one's policy. We studied the convergence of naive single-agent learning iterates in the game setting, proving almost sure convergence under mild assumptions. By analogy to near-optimality criteria for MDPs, we used the limiting values of these learning iterates to develop a notion of subjective $\epsilon$-equilibrium. After establishing the existence of objective perfect equilibrium as well as subjective $\epsilon$-equilibrium under mean-field observability, we extended the notion of (objective) $\epsilon$-satisficing paths of \cite{yongacoglu2023satisficing} to subjective value functions. In this framework, we studied the structural properties of $n$-player mean-field games and showed that subjective $\epsilon$-satisficing paths to subjective $\epsilon$-equilibrium exist under various information structures for  partially observed $n$-player mean-field games.

Apart from the structural and conceptual contributions described above, we have also presented Algorithm~\ref{algo:main}, a decentralized independent learner for playing partially observed $n$-player mean-field games, and we have argued that Algorithm~\ref{algo:main} drives policies to subjective $\epsilon$-equilibrium under self-play. Unlike the bulk of results on learning in mean-field games, the convergence guarantees of Algorithm~\ref{algo:main} do not mandate that players use the same policy at a given time or that they use the same policy update rule to switch policies at the end of an exploration phase. As such, our algorithm is capable of describing learning dynamics for a population of homogeneous agents that may arrive at a joint policy consisting of heterogeneous policies. The learning dynamics presented here result in system stability, in that policies settle to a particular joint policy, but the emergent behaviour need not be an objective equilibrium. Convergence to non-equilibrium policies may arise in real-world strategic environments, and our notion of subjective equilibrium may present an interpretation for such real-world stability in some instances. 

This paper leaves open the question of whether subjective $\epsilon$-equilibrium always exist in mean-field games with compressed or local observability. Determining sufficient conditions for the existence of subjective equilibrium is an interesting topic for future research.

\bibliographystyle{ieeetr}
\bibliography{JMLR_2025_updated}

\begin{thebibliography}{96}
\providecommand{\natexlab}[1]{#1}
\providecommand{\url}[1]{\texttt{#1}}
\expandafter\ifx\csname urlstyle\endcsname\relax
  \providecommand{\doi}[1]{doi: #1}\else
  \providecommand{\doi}{doi: \begingroup \urlstyle{rm}\Url}\fi

\bibitem[Adlakha et~al.(2015)Adlakha, Johari, and
  Weintraub]{adlakha2015equilibria}
Sachin Adlakha, Ramesh Johari, and Gabriel~Y. Weintraub.
\newblock Equilibria of dynamic games with many players: Existence,
  approximation, and market structure.
\newblock \emph{Journal of Economic Theory}, 156:\penalty0 269--316, 2015.

\bibitem[Anahtarci et~al.(2020)Anahtarci, Kariksiz, and
  Saldi]{anahtarci2020value}
Berkay Anahtarci, Can~Deha Kariksiz, and Naci Saldi.
\newblock Value iteration algorithm for mean-field games.
\newblock \emph{Systems \& Control Letters}, 143:\penalty0 104744, 2020.

\bibitem[Anahtarci et~al.(2023{\natexlab{a}})Anahtarci, Kariksiz, and
  Saldi]{anahtarci2023learning}
Berkay Anahtarci, Can~Deha Kariksiz, and Naci Saldi.
\newblock Learning mean-field games with discounted and average costs.
\newblock \emph{Journal of Machine Learning Research}, 24\penalty0
  (17):\penalty0 1--59, 2023{\natexlab{a}}.

\bibitem[Anahtarci et~al.(2023{\natexlab{b}})Anahtarci, Kariksiz, and
  Saldi]{anahtarci2023q}
Berkay Anahtarci, Can~Deha Kariksiz, and Naci Saldi.
\newblock Q-learning in regularized mean-field games.
\newblock \emph{Dynamic Games and Applications}, 13\penalty0 (1):\penalty0
  89--117, 2023{\natexlab{b}}.

\bibitem[Angiuli et~al.(2022)Angiuli, Fouque, and
  Lauri{\`e}re]{angiuli2022unified}
Andrea Angiuli, Jean-Pierre Fouque, and Mathieu Lauri{\`e}re.
\newblock Unified reinforcement {Q}-learning for mean field game and control
  problems.
\newblock \emph{Mathematics of Control, Signals, and Systems}, 34\penalty0
  (2):\penalty0 217--271, 2022.

\bibitem[Arapostathis et~al.(2017)Arapostathis, Biswas, and
  Carroll]{arapostathis2017solutions}
Ari Arapostathis, Anup Biswas, and Johnson Carroll.
\newblock On solutions of mean field games with ergodic cost.
\newblock \emph{Journal de Math{\'e}matiques Pures et Appliqu{\'e}es},
  107\penalty0 (2):\penalty0 205--251, 2017.

\bibitem[Arslan and Y\"{u}ksel(2017)]{AY2017}
G{\"{u}}rdal Arslan and Serdar Y\"{u}ksel.
\newblock Decentralized {Q}-learning for stochastic teams and games.
\newblock \emph{IEEE Transactions on Automatic Control}, 62\penalty0
  (4):\penalty0 1545--1558, 2017.

\bibitem[Arslan and Y{\"u}ksel(2023)]{arslan2023subjective}
G{\"u}rdal Arslan and Serdar Y{\"u}ksel.
\newblock Subjective equilibria under beliefs of exogenous uncertainty for
  dynamic games.
\newblock \emph{SIAM Journal on Control and Optimization}, 61\penalty0
  (3):\penalty0 1038--1062, 2023.

\bibitem[Bauso et~al.(2012)Bauso, Tembine, and Ba{\c{s}}ar]{bauso2012robust}
Dario Bauso, Hamidou Tembine, and Tamer Ba{\c{s}}ar.
\newblock Robust mean field games with application to production of an
  exhaustible resource.
\newblock \emph{IFAC Proceedings Volumes}, 45\penalty0 (13):\penalty0 454--459,
  2012.

\bibitem[Biswas(2015)]{biswas2015mean}
Anup Biswas.
\newblock Mean field games with ergodic cost for discrete time {M}arkov
  processes.
\newblock \emph{arXiv preprint arXiv:1510.08968}, 2015.

\bibitem[Cabannes et~al.(2022)Cabannes, Lauri\`{e}re, Perolat, Marinier,
  Girgin, Perrin, Pietquin, Bayen, Goubault, and Elie]{cabannes2022solving}
Theophile Cabannes, Mathieu Lauri\`{e}re, Julien Perolat, Raphael Marinier,
  Sertan Girgin, Sarah Perrin, Olivier Pietquin, Alexandre~M. Bayen, Eric
  Goubault, and Romuald Elie.
\newblock Solving {N}-player dynamic routing games with congestion: {A}
  mean-field approach.
\newblock In \emph{Proceedings of the 21st International Conference on
  Autonomous Agents and Multiagent Systems}, pages 1557--1559, 2022.

\bibitem[Campi and Fischer(2022)]{campi2022correlated}
Luciano Campi and Markus Fischer.
\newblock Correlated equilibria and mean field games: {A} simple model.
\newblock \emph{Mathematics of Operations Research}, 47\penalty0 (3):\penalty0
  2240--2259, 2022.

\bibitem[Candogan et~al.(2013)Candogan, Ozdaglar, and
  Parrilo]{candogan2013near}
Ozan Candogan, Asuman Ozdaglar, and Pablo~A. Parrilo.
\newblock Near-potential games: Geometry and dynamics.
\newblock \emph{ACM Transactions on Economics and Computation (TEAC)},
  1\penalty0 (2):\penalty0 1--32, 2013.

\bibitem[Chandak et~al.(2024)Chandak, Shah, Borkar, and
  Dodhia]{chandak2024reinforcement}
Siddharth Chandak, Pratik Shah, Vivek~S. Borkar, and Parth Dodhia.
\newblock Reinforcement learning in non-{M}arkovian environments.
\newblock \emph{Systems \& Control Letters}, 185:\penalty0 105751, 2024.

\bibitem[Chasparis et~al.(2013)Chasparis, Arapostathis, and
  Shamma]{Chasparis2013aspiration}
Georgios~C. Chasparis, Ari Arapostathis, and Jeff~S. Shamma.
\newblock Aspiration learning in coordination games.
\newblock \emph{{SIAM} J. Control and Optimization}, 51\penalty0 (1):\penalty0
  465--490, 2013.

\bibitem[Chevalier et~al.(2015)Chevalier, Le~Ny, and
  Malham{\'e}]{chevalier2015micro}
Geoffroy Chevalier, Jerome Le~Ny, and Roland Malham{\'e}.
\newblock A micro-macro traffic model based on mean-field games.
\newblock In \emph{2015 American Control Conference (ACC)}, pages 1983--1988.
  IEEE, 2015.

\bibitem[Chien and Sinclair(2011)]{chien2011convergence}
Steve Chien and Alistair Sinclair.
\newblock Convergence to approximate {N}ash equilibria in congestion games.
\newblock \emph{Games and Economic Behavior}, 71\penalty0 (2):\penalty0
  315--327, 2011.

\bibitem[Claus and Boutilier(1998)]{Claus1998}
Caroline Claus and Craig Boutilier.
\newblock The dynamics of reinforcement learning in cooperative multiagent
  systems.
\newblock In \emph{Proceedings of the Tenth Innovative Applications of
  Artificial Intelligence Conference, Madison, Wisconsin}, pages 746--752,
  1998.

\bibitem[Condon(1990)]{condon1990}
Anne Condon.
\newblock On algorithms for simple stochastic games.
\newblock \emph{Advances in Computational Complexity Theory}, 13:\penalty0
  51--72, 1990.

\bibitem[Cui and Koeppl(2021)]{cui2021approximately}
Kai Cui and Heinz Koeppl.
\newblock Approximately solving mean field games via entropy-regularized deep
  reinforcement learning.
\newblock In \emph{International Conference on Artificial Intelligence and
  Statistics}, pages 1909--1917. PMLR, 2021.

\bibitem[Daskalakis et~al.(2020)Daskalakis, Foster, and
  Golowich]{daskalakis2020independent}
Constantinos Daskalakis, Dylan~J. Foster, and Noah Golowich.
\newblock Independent policy gradient methods for competitive reinforcement
  learning.
\newblock \emph{Advances in Neural Information Processing Systems},
  33:\penalty0 5527--5540, 2020.

\bibitem[Ding et~al.(2022)Ding, Wei, Zhang, and Jovanovic]{ding2022independent}
Dongsheng Ding, Chen-Yu Wei, Kaiqing Zhang, and Mihailo Jovanovic.
\newblock Independent policy gradient for large-scale {M}arkov potential games:
  Sharper rates, function approximation, and game-agnostic convergence.
\newblock In \emph{International Conference on Machine Learning}, pages
  5166--5220. PMLR, 2022.

\bibitem[Dong et~al.(2022)Dong, Van~Roy, and Zhou]{dong2022simple}
Shi Dong, Benjamin Van~Roy, and Zhengyuan Zhou.
\newblock Simple agent, complex environment: Efficient reinforcement learning
  with agent states.
\newblock \emph{Journal of Machine Learning Research}, 23\penalty0
  (1):\penalty0 11627--11680, 2022.

\bibitem[Elie et~al.(2020)Elie, Perolat, Lauri{\`e}re, Geist, and
  Pietquin]{elie2020convergence}
Romuald Elie, Julien Perolat, Mathieu Lauri{\`e}re, Matthieu Geist, and Olivier
  Pietquin.
\newblock On the convergence of model free learning in mean field games.
\newblock In \emph{Proceedings of the AAAI Conference on Artificial
  Intelligence}, volume~34, pages 7143--7150, 2020.

\bibitem[Fink(1964)]{fink1964equilibrium}
Arlington~M. Fink.
\newblock Equilibrium in a stochastic $n$-person game.
\newblock \emph{Journal of Science of the Hiroshima University, Series AI
  (Mathematics)}, 28\penalty0 (1):\penalty0 89--93, 1964.

\bibitem[Fischer(2017)]{fischer2017connection}
Markus Fischer.
\newblock On the connection between symmetric $n$-player games and mean field
  games.
\newblock \emph{The Annals of Applied Probability}, 27\penalty0 (2):\penalty0
  757--810, 2017.

\bibitem[Foerster et~al.(2018)Foerster, Farquhar, Afouras, Nardelli, and
  Whiteson]{foerster2018counterfactual}
Jakob Foerster, Gregory Farquhar, Triantafyllos Afouras, Nantas Nardelli, and
  Shimon Whiteson.
\newblock Counterfactual multi-agent policy gradients.
\newblock In \emph{Proceedings of the AAAI Conference on Artificial
  Intelligence}, volume~32, 2018.

\bibitem[Foster and Young(2006)]{foster2006regret}
Dean Foster and H.~Peyton Young.
\newblock Regret testing: Learning to play {N}ash equilibrium without knowing
  you have an opponent.
\newblock \emph{Theoretical Economics}, 1:\penalty0 341--367, 2006.

\bibitem[Fox et~al.(2022)Fox, McAleer, Overman, and
  Panageas]{fox2022independent}
Roy Fox, Stephen~M. McAleer, Will Overman, and Ioannis Panageas.
\newblock Independent natural policy gradient always converges in {M}arkov
  potential games.
\newblock In \emph{International Conference on Artificial Intelligence and
  Statistics}, pages 4414--4425. PMLR, 2022.

\bibitem[Gomes and Sa{\'u}de(2014)]{gomes2014mean}
Diogo~A. Gomes and Jo{\~a}o Sa{\'u}de.
\newblock Mean field games models: {A} brief survey.
\newblock \emph{Dynamic Games and Applications}, 4:\penalty0 110--154, 2014.

\bibitem[Gomes et~al.(2014)Gomes, Velho, and Wolfram]{gomes2014socio}
Diogo~A. Gomes, Roberto~M. Velho, and Marie-Therese Wolfram.
\newblock Socio-economic applications of finite state mean field games.
\newblock \emph{Philosophical Transactions of the Royal Society A:
  Mathematical, Physical and Engineering Sciences}, 372\penalty0 (2028), 2014.

\bibitem[Guo et~al.(2019)Guo, Hu, Xu, and Zhang]{guo2019learning}
Xin Guo, Anran Hu, Renyuan Xu, and Junzi Zhang.
\newblock Learning mean-field games.
\newblock \emph{Advances in Neural Information Processing Systems}, 32, 2019.

\bibitem[Hu and Wellman(2003)]{Hu2003}
Junling Hu and Michael~P. Wellman.
\newblock Nash {Q}-learning for general-sum stochastic games.
\newblock \emph{Journal of Machine Learning Research}, 4\penalty0
  (Nov):\penalty0 1039--1069, 2003.

\bibitem[Huang et~al.(2006)Huang, Malham{\'e}, and Caines]{huang2006large}
Minyi Huang, Roland~P. Malham{\'e}, and Peter~E. Caines.
\newblock Large population stochastic dynamic games: {C}losed-loop
  {M}c{K}ean-{V}lasov systems and the {N}ash certainty equivalence principle.
\newblock \emph{Communications in Information \& Systems}, 6\penalty0
  (3):\penalty0 221--252, 2006.

\bibitem[Huang et~al.(2007)Huang, Caines, and Malham{\'e}]{huang2007large}
Minyi Huang, Peter~E. Caines, and Roland~P. Malham{\'e}.
\newblock Large-population cost-coupled {LQG} problems with nonuniform agents:
  {I}ndividual-mass behavior and decentralized $\epsilon$-{N}ash equilibria.
\newblock \emph{IEEE Transactions on Automatic Control}, 52\penalty0
  (9):\penalty0 1560--1571, 2007.

\bibitem[Iyer et~al.(2014)Iyer, Johari, and Sundararajan]{iyer2014mean}
Krishnamurthy Iyer, Ramesh Johari, and Mukund Sundararajan.
\newblock Mean field equilibria of dynamic auctions with learning.
\newblock \emph{Management Science}, 60\penalty0 (12):\penalty0 2949--2970,
  2014.

\bibitem[Kalai and Lehrer(1993)]{kalai1993rational}
Ehud Kalai and Ehud Lehrer.
\newblock Rational learning leads to {N}ash equilibrium.
\newblock \emph{Econometrica}, pages 1019--1045, 1993.

\bibitem[Kalai and Lehrer(1995)]{kalai1995subjective}
Ehud Kalai and Ehud Lehrer.
\newblock Subjective games and equilibria.
\newblock \emph{Games and Economic Behavior}, 8\penalty0 (1):\penalty0
  123--163, 1995.

\bibitem[Kara and Y{\"u}ksel(2022)]{kara2022convergence}
Ali~D. Kara and Serdar Y{\"u}ksel.
\newblock Convergence of finite memory {Q}-learning for {POMDP}s and near
  optimality of learned policies under filter stability.
\newblock \emph{Mathematics of Operations Research}, 48\penalty0 (4):\penalty0
  2066–2093, 2022.

\bibitem[Kara and Y\"{u}ksel(2024)]{kara2024qlearning}
Ali~D. Kara and Serdar Y\"{u}ksel.
\newblock Q-learning for stochastic control under general information
  structures and non-{M}arkovian environments.
\newblock \emph{Transactions on Machine Learning Research}, 2024.
\newblock ISSN 2835-8856.

\bibitem[Lacker(2015)]{lacker2015mean}
Daniel Lacker.
\newblock Mean field games via controlled martingale problems: {E}xistence of
  {M}arkovian equilibria.
\newblock \emph{Stochastic Processes and their Applications}, 125\penalty0
  (7):\penalty0 2856--2894, 2015.

\bibitem[Lacker(2020)]{lacker2020convergence}
Daniel Lacker.
\newblock {On the convergence of closed-loop {N}ash equilibria to the mean
  field game limit}.
\newblock \emph{The Annals of Applied Probability}, 30\penalty0 (4):\penalty0
  1693 -- 1761, 2020.

\bibitem[Lasry and Lions(2007)]{lasry2007mean}
Jean-Michel Lasry and Pierre-Louis Lions.
\newblock Mean field games.
\newblock \emph{Japanese Journal of Mathematics}, 2\penalty0 (1):\penalty0
  229--260, 2007.

\bibitem[Lauri{\`e}re et~al.(2022)Lauri{\`e}re, Perrin, Geist, and
  Pietquin]{lauriere2022learning}
Mathieu Lauri{\`e}re, Sarah Perrin, Matthieu Geist, and Olivier Pietquin.
\newblock Learning mean field games: A survey.
\newblock \emph{arXiv preprint arXiv:2205.12944}, 2022.

\bibitem[Lee et~al.(2021)Lee, Rengarajan, Kalathil, and
  Shakkottai]{lee2021reinforcement}
Kiyeob Lee, Desik Rengarajan, Dileep Kalathil, and Srinivas Shakkottai.
\newblock Reinforcement learning for mean field games with strategic
  complementarities.
\newblock In \emph{International Conference on Artificial Intelligence and
  Statistics}, pages 2458--2466. PMLR, 2021.

\bibitem[Leonardos et~al.(2022)Leonardos, Overman, Panageas, and
  Piliouras]{leonardos2022global}
Stefanos Leonardos, Will Overman, Ioannis Panageas, and Georgios Piliouras.
\newblock Global convergence of multi-agent policy gradient in {M}arkov
  potential games.
\newblock In \emph{International Conference on Learning Representations}, 2022.

\bibitem[Li et~al.(2016)Li, Bhattacharyya, Paul, Shakkottai, and
  Subramanian]{li2016incentivizing}
Jian Li, Rajarshi Bhattacharyya, Suman Paul, Srinivas Shakkottai, and Vijay
  Subramanian.
\newblock Incentivizing sharing in realtime d2d streaming networks: A mean
  field game perspective.
\newblock \emph{IEEE/ACM Transactions on Networking}, 25\penalty0 (1):\penalty0
  3--17, 2016.

\bibitem[Li et~al.(2018)Li, Xia, Geng, Ming, Shakkottai, Subramanian, and
  Xie]{li2018mean}
Jian Li, Bainan Xia, Xinbo Geng, Hao Ming, Srinivas Shakkottai, Vijay
  Subramanian, and Le~Xie.
\newblock Mean field games in nudge systems for societal networks.
\newblock \emph{ACM Transactions on Modeling and Performance Evaluation of
  Computing Systems (TOMPECS)}, 3\penalty0 (4):\penalty0 1--31, 2018.

\bibitem[Li et~al.(2019)Li, Qin, Jiao, Yang, Wang, Wang, Wu, and
  Ye]{li2019efficient}
Minne Li, Zhiwei Qin, Yan Jiao, Yaodong Yang, Jun Wang, Chenxi Wang, Guobin Wu,
  and Jieping Ye.
\newblock Efficient ridesharing order dispatching with mean field multi-agent
  reinforcement learning.
\newblock In \emph{The World Wide Web Conference}, pages 983--994, 2019.

\bibitem[Littman(2001)]{Littman2001ffq}
Michael~L. Littman.
\newblock Friend-or-foe {Q}-learning in general-sum games.
\newblock In \emph{International Conference on Machine Learning}, volume~1,
  pages 322--328, 2001.

\bibitem[Littman and Szepesv{\'a}ri(1996)]{littman1996generalized}
Michael~L. Littman and Csaba Szepesv{\'a}ri.
\newblock A generalized reinforcement-learning model: Convergence and
  applications.
\newblock In \emph{International Conference on Machine Learning}, volume~96,
  pages 310--318, 1996.

\bibitem[Lowe et~al.(2017)Lowe, Wu, Tamar, Harb, Abbeel, and
  Mordatch]{lowe2017multi}
Ryan Lowe, Yi~Wu, Aviv Tamar, Jean Harb, Pieter Abbeel, and Igor Mordatch.
\newblock Multi-agent actor-critic for mixed cooperative-competitive
  environments.
\newblock \emph{Advances in Neural Information Processing Systems}, 30, 2017.

\bibitem[Manjrekar et~al.(2019)Manjrekar, Ramaswamy, Raja, and
  Shakkottai]{manjrekar2019mean}
Mayank Manjrekar, Vinod Ramaswamy, Vamseedhar~Reddyvari Raja, and Srinivas
  Shakkottai.
\newblock A mean field game approach to scheduling in cellular systems.
\newblock \emph{IEEE Transactions on Control of Network Systems}, 7\penalty0
  (2):\penalty0 568--578, 2019.

\bibitem[Matignon et~al.(2009)Matignon, Laurent, and
  Le~Fort-Piat]{matignon2009coordination}
Laetitia Matignon, Guillaume~J. Laurent, and Nadine Le~Fort-Piat.
\newblock Coordination of independent learners in cooperative {M}arkov games.
\newblock \emph{HAL preprint hal-00370889}, 2009.

\bibitem[Matignon et~al.(2012)Matignon, Laurent, and
  Le~Fort-Piat]{matignon2012survey}
Laetitia Matignon, Guillaume~J. Laurent, and Nadine Le~Fort-Piat.
\newblock Independent reinforcement learners in cooperative {M}arkov games: {A}
  survey regarding coordination problems.
\newblock \emph{Knowledge Engineering Review}, 27\penalty0 (1):\penalty0 1--31,
  2012.

\bibitem[Mguni et~al.(2018)Mguni, Jennings, and
  de~Cote]{mguni2018decentralised}
David Mguni, Joel Jennings, and Enrique~Munoz de~Cote.
\newblock Decentralised learning in systems with many, many strategic agents.
\newblock In \emph{Proceedings of the AAAI Conference on Artificial
  Intelligence}, volume~32, 2018.

\bibitem[Mguni et~al.(2021)Mguni, Wu, Du, Yang, Wang, Li, Wen, Jennings, and
  Wang]{mguni2021learning}
David Mguni, Yutong Wu, Yali Du, Yaodong Yang, Ziyi Wang, Minne Li, Ying Wen,
  Joel Jennings, and Jun Wang.
\newblock Learning in nonzero-sum stochastic games with potentials.
\newblock In \emph{International Conference on Machine Learning}, pages
  7688--7699. PMLR, 2021.

\bibitem[Mishra et~al.(2023)Mishra, Vishwanath, and Vasal]{mishra2023model}
Rajesh Mishra, Sriram Vishwanath, and Deepanshu Vasal.
\newblock Model-free reinforcement learning for mean field games.
\newblock \emph{IEEE Transactions on Control of Network Systems}, 2023.

\bibitem[Muller et~al.(2022)Muller, Elie, Rowland, Lauri{\`e}re, Perolat,
  Perrin, Geist, Piliouras, Pietquin, and Tuyls]{muller2022learning}
Paul Muller, Romuald Elie, Mark Rowland, Mathieu Lauri{\`e}re, Julien Perolat,
  Sarah Perrin, Matthieu Geist, Georgios Piliouras, Olivier Pietquin, and Karl
  Tuyls.
\newblock Learning correlated equilibria in mean-field games.
\newblock \emph{arXiv preprint arXiv:2208.10138}, 2022.

\bibitem[Nachbar(2005)]{nachbar2005beliefs}
John~H. Nachbar.
\newblock Beliefs in repeated games.
\newblock \emph{Econometrica}, 73\penalty0 (2):\penalty0 459--480, 2005.

\bibitem[Patil et~al.(2024)Patil, Mahajan, and Precup]{patil2024learning}
Gandharv Patil, Aditya Mahajan, and Doina Precup.
\newblock On learning history-based policies for controlling {M}arkov decision
  processes.
\newblock In \emph{International Conference on Artificial Intelligence and
  Statistics}, pages 3511--3519. PMLR, 2024.

\bibitem[Perrin et~al.(2022)Perrin, Lauri{\`e}re, P{\'e}rolat, {\'E}lie, Geist,
  and Pietquin]{perrin2022generalization}
Sarah Perrin, Mathieu Lauri{\`e}re, Julien P{\'e}rolat, Romuald {\'E}lie,
  Matthieu Geist, and Olivier Pietquin.
\newblock Generalization in mean field games by learning master policies.
\newblock In \emph{Proceedings of the AAAI Conference on Artificial
  Intelligence}, volume~36, pages 9413--9421, 2022.

\bibitem[Posch(1999)]{posch1999win}
Martin Posch.
\newblock Win--stay, lose--shift strategies for repeated games---{M}emory
  length, aspiration levels and noise.
\newblock \emph{Journal of Theoretical Biology}, 198\penalty0 (2):\penalty0
  183--195, 1999.

\bibitem[Saldi et~al.(2018)Saldi, Ba\c{s}ar, and Raginsky]{saldi2018markov}
Naci Saldi, Tamer Ba\c{s}ar, and Maxim Raginsky.
\newblock Markov--{N}ash equilibria in mean-field games with discounted cost.
\newblock \emph{SIAM Journal on Control and Optimization}, 56\penalty0
  (6):\penalty0 4256--4287, 2018.

\bibitem[Salhab et~al.(2018)Salhab, Le~Ny, and Malham{\'e}]{salhab2018mean}
Rabih Salhab, Jerome Le~Ny, and Roland~P Malham{\'e}.
\newblock A mean field route choice game model.
\newblock In \emph{2018 IEEE Conference on Decision and Control (CDC)}, pages
  1005--1010. IEEE, 2018.

\bibitem[Sanjari et~al.(2022)Sanjari, Saldi, and
  Y{\"u}ksel]{sanjari2022optimality}
Sina Sanjari, Naci Saldi, and Serdar Y{\"u}ksel.
\newblock Optimality of independently randomized symmetric policies for
  exchangeable stochastic teams with infinitely many decision makers.
\newblock \emph{Mathematics of Operations Research}, 2022.

\bibitem[Sayin et~al.(2021)Sayin, Zhang, Leslie, Ba\c{s}ar, and
  Ozdaglar]{sayin2021decentralized}
Muhammed~O. Sayin, Kaiqing Zhang, David Leslie, Tamer Ba\c{s}ar, and Asuman
  Ozdaglar.
\newblock Decentralized {Q}-learning in zero-sum {M}arkov games.
\newblock \emph{Advances in Neural Information Processing Systems},
  34:\penalty0 18320--18334, 2021.

\bibitem[Sayin et~al.(2022)Sayin, Parise, and Ozdaglar]{sayin2022fictitious}
Muhammed~O. Sayin, Francesca Parise, and Asuman Ozdaglar.
\newblock Fictitious play in zero-sum stochastic games.
\newblock \emph{SIAM Journal on Control and Optimization}, 60\penalty0
  (4):\penalty0 2095--2114, 2022.

\bibitem[Sen et~al.(1994)Sen, Sekaran, and Hale]{Sen94}
Sandip Sen, Mahendra Sekaran, and John Hale.
\newblock Learning to coordinate without sharing information.
\newblock In \emph{Proceedings of the 12th National Conference on Artificial
  Intelligence}, pages 426--431, 1994.

\bibitem[Simon(1956)]{simon1956rational}
Herbert~A. Simon.
\newblock Rational choice and the structure of the environment.
\newblock \emph{Psychological Review}, 63\penalty0 (2):\penalty0 129, 1956.

\bibitem[Sinha et~al.(2024)Sinha, Geist, and Mahajan]{sinha2024periodic}
Amit Sinha, Matthieu Geist, and Aditya Mahajan.
\newblock Periodic agent-state based {Q}-learning for {POMDP}s.
\newblock \emph{Advances in Neural Information Processing Systems}, 2024.

\bibitem[Stella et~al.(2013)Stella, Bagagiolo, Bauso, and
  Como]{stella2013opinion}
Leonardo Stella, Fabio Bagagiolo, Dario Bauso, and Giacomo Como.
\newblock Opinion dynamics and stubbornness through mean-field games.
\newblock In \emph{52nd IEEE Conference on Decision and Control}, pages
  2519--2524. IEEE, 2013.

\bibitem[Subramanian and Mahajan(2019)]{subramanian2019reinforcement}
Jayakumar Subramanian and Aditya Mahajan.
\newblock Reinforcement learning in stationary mean-field games.
\newblock In \emph{Proceedings of the 18th International Conference on
  Autonomous Agents and MultiAgent Systems}, pages 251--259, 2019.

\bibitem[Subramanian et~al.(2022{\natexlab{a}})Subramanian, Sinha, Seraj, and
  Mahajan]{subramanian2022approximate}
Jayakumar Subramanian, Amit Sinha, Raihan Seraj, and Aditya Mahajan.
\newblock Approximate information state for approximate planning and
  reinforcement learning in partially observed systems.
\newblock \emph{Journal of Machine Learning Research}, 23\penalty0
  (1):\penalty0 483--565, 2022{\natexlab{a}}.

\bibitem[Subramanian et~al.(2022{\natexlab{b}})Subramanian, Taylor, Crowley,
  and Poupart]{subramanian2022decentralized}
Sriram~G. Subramanian, Matthew~E. Taylor, Mark Crowley, and Pascal Poupart.
\newblock Decentralized mean field games.
\newblock In \emph{Proceedings of the AAAI Conference on Artificial
  Intelligence}, volume~36, pages 9439--9447, 2022{\natexlab{b}}.

\bibitem[Tan(1993)]{tan1993multi}
Ming Tan.
\newblock Multi-agent reinforcement learning: Independent vs. cooperative
  agents.
\newblock In \emph{Proceedings of the Tenth International Conference on Machine
  Learning}, pages 330--337, 1993.

\bibitem[Tsitsiklis(1994)]{tsitsiklis1994asynchronous}
J.~N. Tsitsiklis.
\newblock Asynchronous stochastic approximation and {Q}-learning.
\newblock \emph{Machine Learning}, 16\penalty0 (3):\penalty0 185--202, 1994.

\bibitem[Unlu and Sayin(2023)]{unlu2023episodic}
Onur Unlu and Muhammed~O. Sayin.
\newblock Episodic {L}ogit-{Q} dynamics for efficient learning in stochastic
  teams.
\newblock In \emph{2023 62nd IEEE Conference on Decision and Control (CDC)},
  pages 1985--1990. IEEE, 2023.

\bibitem[Vasal(2023)]{vasal2023sequential}
Deepanshu Vasal.
\newblock Sequential decomposition of discrete-time mean-field games.
\newblock \emph{Dynamic Games and Applications}, pages 1--19, 2023.

\bibitem[Wang et~al.(2020)Wang, Yang, and Wang]{wang2020breaking}
Lingxiao Wang, Zhuoran Yang, and Zhaoran Wang.
\newblock Breaking the curse of many agents: Provable mean embedding
  {Q}-iteration for mean-field reinforcement learning.
\newblock In \emph{International Conference on Machine Learning}, pages
  10092--10103. PMLR, 2020.

\bibitem[Watkins(1989)]{Watkins89}
Christopher Watkins.
\newblock \emph{Learning from Delayed Rewards}.
\newblock PhD thesis, Cambridge University, 1989.

\bibitem[Wei and Luke(2016)]{wei2016lenient}
Ermo Wei and Sean Luke.
\newblock Lenient learning in independent-learner stochastic cooperative games.
\newblock \emph{Journal of Machine Learning Research}, 17\penalty0
  (1):\penalty0 2914--2955, 2016.

\bibitem[Weintraub et~al.(2005)Weintraub, Benkard, and
  Van~Roy]{weintraub2005oblivious}
Gabriel~Y. Weintraub, C.~Lanier Benkard, and Benjamin Van~Roy.
\newblock Oblivious equilibrium: A mean field approximation for large-scale
  dynamic games.
\newblock \emph{Advances in Neural Information Processing Systems}, 18, 2005.

\bibitem[Weintraub et~al.(2008)Weintraub, Benkard, and
  Van~Roy]{weintraub2008markov}
Gabriel~Y. Weintraub, C.~Lanier Benkard, and Benjamin Van~Roy.
\newblock Markov perfect industry dynamics with many firms.
\newblock \emph{Econometrica}, 76\penalty0 (6):\penalty0 1375--1411, 2008.

\bibitem[Xia et~al.(2019)Xia, Shakkottai, and Subramanian]{xia2019small}
Bainan Xia, Srinivas Shakkottai, and Vijay Subramanian.
\newblock Small-scale markets for a bilateral energy sharing economy.
\newblock \emph{IEEE Transactions on Control of Network Systems}, 6\penalty0
  (3):\penalty0 1026--1037, 2019.

\bibitem[Xie et~al.(2021)Xie, Yang, Wang, and Minca]{xie2021learning}
Qiaomin Xie, Zhuoran Yang, Zhaoran Wang, and Andreea Minca.
\newblock Learning while playing in mean-field games: Convergence and
  optimality.
\newblock In \emph{International Conference on Machine Learning}, pages
  11436--11447. PMLR, 2021.

\bibitem[Yardim et~al.(2023)Yardim, Cayci, Geist, and He]{yardim2023policy}
Batuhan Yardim, Semih Cayci, Matthieu Geist, and Niao He.
\newblock Policy mirror ascent for efficient and independent learning in mean
  field games.
\newblock In \emph{International Conference on Machine Learning}, pages
  39722--39754. PMLR, 2023.

\bibitem[Yongacoglu et~al.(2022)Yongacoglu, Arslan, and Y{\"u}ksel]{YAY-TAC}
Bora Yongacoglu, G{\"u}rdal Arslan, and Serdar Y{\"u}ksel.
\newblock Decentralized learning for optimality in stochastic dynamic teams and
  games with local control and global state information.
\newblock \emph{IEEE Transactions on Automatic Control}, 67\penalty0
  (10):\penalty0 5230--5245, 2022.

\bibitem[Yongacoglu et~al.(2023)Yongacoglu, Arslan, and
  Y{\"u}ksel]{yongacoglu2023satisficing}
Bora Yongacoglu, G{\"u}rdal Arslan, and Serdar Y{\"u}ksel.
\newblock Satisficing paths and independent multiagent reinforcement learning
  in stochastic games.
\newblock \emph{SIAM Journal on Mathematics of Data Science}, 5\penalty0
  (3):\penalty0 745--773, 2023.

\bibitem[Yongacoglu et~al.(2024{\natexlab{a}})Yongacoglu, Arslan, Pavel, and
  Y{\"u}ksel]{yongacoglu2024generalizing}
Bora Yongacoglu, G{\"u}rdal Arslan, Lacra Pavel, and Serdar Y{\"u}ksel.
\newblock Generalizing better response paths and weakly acyclic games.
\newblock \emph{2024 IEEE 63rd Conference on Decision and Control (CDC)},
  2024{\natexlab{a}}.

\bibitem[Yongacoglu et~al.(2024{\natexlab{b}})Yongacoglu, Arslan, Pavel, and
  Yuksel]{yongacoglu2024paths}
Bora Yongacoglu, Gurdal Arslan, Lacra Pavel, and Serdar Yuksel.
\newblock Paths to equilibrium in games.
\newblock \emph{Advances in Neural Information Processing Systems},
  2024{\natexlab{b}}.

\bibitem[Zaman et~al.(2020{\natexlab{a}})Zaman, Zhang, Miehling, and
  Ba\c{s}ar]{zaman2020reinforcement}
Muhammad Aneeq~uz Zaman, Kaiqing Zhang, Erik Miehling, and Tamer Ba\c{s}ar.
\newblock Reinforcement learning in non-stationary discrete-time
  linear-quadratic mean-field games.
\newblock In \emph{2020 59th IEEE Conference on Decision and Control (CDC)},
  pages 2278--2284. IEEE, 2020{\natexlab{a}}.

\bibitem[Zaman et~al.(2020{\natexlab{b}})Zaman, Zhang, Miehling, and
  Ba{\c{s}}ar]{zaman2020approximate}
Muhammad Aneeq~uz Zaman, Kaiqing Zhang, Erik Miehling, and Tamer Ba{\c{s}}ar.
\newblock Approximate equilibrium computation for discrete-time
  linear-quadratic mean-field games.
\newblock In \emph{2020 American Control Conference (ACC)}, pages 333--339.
  IEEE, 2020{\natexlab{b}}.

\bibitem[Zaman et~al.(2023)Zaman, Koppel, Bhatt, and
  Ba\c{s}ar]{zaman2023oracle}
Muhammad Aneeq~Uz Zaman, Alec Koppel, Sujay Bhatt, and Tamer Ba\c{s}ar.
\newblock Oracle-free reinforcement learning in mean-field games along a single
  sample path.
\newblock In \emph{International Conference on Artificial Intelligence and
  Statistics}, pages 10178--10206. PMLR, 2023.

\bibitem[Zhang et~al.(2021)Zhang, Yang, and Ba{\c{s}}ar]{zhang2021multi}
Kaiqing Zhang, Zhuoran Yang, and Tamer Ba{\c{s}}ar.
\newblock Multi-agent reinforcement learning: A selective overview of theories
  and algorithms.
\newblock \emph{Handbook of Reinforcement Learning and Control}, pages
  321--384, 2021.

\bibitem[Zhang et~al.(2022)Zhang, Mei, Dai, Schuurmans, and
  Li]{zhang2022global}
Runyu Zhang, Jincheng Mei, Bo~Dai, Dale Schuurmans, and Na~Li.
\newblock On the global convergence rates of decentralized softmax gradient
  play in {M}arkov potential games.
\newblock \emph{Advances in Neural Information Processing Systems},
  35:\penalty0 1923--1935, 2022.

\end{thebibliography}

\newpage 
\appendix

\section*{Appendices}
\section{Fully and Partially Observed Markov Decision Problems} \label{ss:POMDPs}

A finite, partially observed Markov decision problem (POMDP) with the discounted cost criterion is given by a list $\MM$:
\begin{equation}   \label{eq:POMDP}
\MM = \left( 	\XX, \YY, \AA, C , P_{\MM}  , \upphi , 	\upgamma , \upnu_0	\right) .
\end{equation}

At time $t \in \zz_{\geq 0}$, the system's state is denoted $X_t$ and takes values in the finite set $\XX$, with $X_0 \sim \upnu_0$. An observation $Y_t$ taking values in the finite set $\YY$ is generated according to a noisy reading of $X_t$ through the observation channel $\upphi \in \PP ( \YY | \XX )$, as  $Y_t \sim \upphi ( \cdot | X_t )$. The agent uses its observable history variable, $\mathcal{h}^{\rm ob}_t$, to be defined shortly, to select its action $A_t$ from the finite set $\AA$. The agent then incurs a stage cost $C_t$, given by the cost function $C$ as $C_t :=  C ( X_t, A_t)$. The system's state subsequently evolves according to $X_{t+1} \sim P_{\MM} ( \cdot | X_t , A_t)$, where $P_{\MM} \in \PP ( \XX | \XX \times \AA )$. A discount factor $\upgamma \in (0,1)$ is used to discount the sequence of costs incurred by the agent.

We define the \emph{system history sets} $\{ \HH_t \}_{t \in \zz_{\geq 0}}$ as follows: 
\begin{align*}
 \HH_t	:=  \left( \XX \times \YY \times \AA  \right)^{  t   } \times \XX \times \YY , \quad \forall t \geq 0. 
\end{align*}

Elements of $\HH_t$ are called \textit{system histories of length $t$}. We define a random quantity $\mathcal{h}_t$, taking values in $\HH_t$ and given by 
\[
\mathcal{h}_t = ( X_0, Y_0,  A_0, \cdots, X_{t-1}, Y_{t-1}, A_{t-1}, X_t,  Y_t),
\]
and we use $\mathcal{h}_t$ to denote the $t^{\rm th}$ \textit{system history variable}. To capture the information actually observed by the agent controlling the system, we define \emph{observable history sets} $\{  \HH^{\rm ob}_{t}   \}_{t \geq 0}$ as 
\begin{align*}
\HH^{\rm ob}_{t}		:=  \Delta( \XX ) \times \left( \YY \times \AA  \times \rr \right)^{  t    } \times \YY , \quad \forall t \geq 0. 
\end{align*}

Elements of $\HH^{\rm ob}_{t}$ are called \emph{observable histories of length $t$}, and we let 
\[
\mathcal{h}^{\rm ob}_t = ( \upnu_0, Y_0, A_0, C_0, \cdots, C_{t-1},  Y_t),
\]
denote the $\HH^{\rm ob}_t$-valued random quantity representing the $t^{\rm th}$ \emph{observable history variable}.

\begin{definition}
A sequence $\pi = ( \pi_t )_{t \geq 0}$ such that $\pi_t \in \PP ( \AA | \HH^{\rm ob}_{t} )$ for each $t$ is called a \emph{policy} for the POMDP $\MM$.
\end{definition}

We denote the set of all policies for the POMDP $\MM$ by $\Pi_{\MM}$. Fixing a policy $\pi \in \Pi_{\MM}$ and an initial measure $\upnu \in \Delta(  \XX )$ induces a unique probability measure $\P^{\pi}_{\upnu}$ on the trajectories of play, i.e. on sequences $( X_t, Y_t, A_t , C_t )_{t = 0}^{\infty}$. We let $E^{\pi}_{\upnu}$  denote the expectation associated with $\P^{\pi}_{\upnu}$ by define the agent's objective function as 
\[
\mathcal{J}_{\MM} (  \pi  , \upnu) := E^{\pi}_{\upnu} \left[ \sum_{t \geq 0} \upgamma^t C ( X_t,  A_t ) \right], \quad \forall \pi \in \Pi_{\MM}, \upnu \in \Delta ( \XX ). 
\]

\noindent In the special case that $\upnu = \delta_{s}$ for some state $s \in \XX$, we simply write $\mathcal{J}_{\MM} (\pi , s)$.%

\begin{definition}[Optimal Policy]
For $\epsilon \geq 0$, a policy $\pi^{*} \in \Pi_{\MM}$ is called \emph{$\epsilon$-optimal with respect to $\upnu \in \Delta(\XX)$} if 
\[
\JJ_{\MM} ( \pi^{*} , \upnu ) \leq \inf_{ \pi \in \Pi_{\MM} } \JJ_{\MM} (\pi , \upnu ) + \epsilon. 
\] 
If $\pi^{*}$ is $\epsilon$-optimal with respect to every $\upnu \in \Delta(\XX)$, then $\pi^{*}$ is called \emph{uniformly $\epsilon$-optimal.} 
\end{definition}

\noindent If $\epsilon = 0$ in the preceding definition, we simply refer to $\pi^{*}$ as being optimal, either uniformly or with respect to a given initial distribution $\upnu$.

\begin{definition}[Stationary Policies]
A policy $\pi \in \Pi_{\MM} $ is called \emph{(memoryless) stationary (or simply stationary)} if, for some $g \in \PP ( \AA | \YY )$, the following holds: for any $ t \geq 0$ and $\tilde{\mathcal{h}}^{\rm ob}_t = ( \tilde{\nu}, \tilde{Y}_0, \dots, \tilde{Y}_t )\in \HH^{\rm ob}_t$, we have 
\[
\pi_t ( \cdot | \tilde{\mathcal{h}}^{\rm ob}_t ) = g ( \cdot | \tilde{Y}_t ).
\]
We let $\Pi_{\MM,S}$ denote the set of stationary policies for the POMDP $\MM$. 
\end{definition}

\begin{definition}[Soft Policies]   \label{def:soft-policies}
For $\xi > 0$, a policy $\pi \in \Pi_{\MM}$ is called \emph{$\xi$-soft} if, for any $t \geq 0$ and $\tilde{\mathcal{h}}^{\rm ob}_t \in \HH^{\rm ob}_t$, we have $\pi ( a | \tilde{\mathcal{h}}^{\rm ob}_t ) \geq \xi$ for all $a \in \AA$. A policy $\pi \in \Pi_{\MM}$ is called \emph{soft} if it is $\xi$-soft for some $\xi > 0$. 
\end{definition}

The goal for an agent controlling the POMDP $\MM$ is to find an optimal policy.  

\

\noindent \textbf{Remark:} We have chosen to present a model in which costs are measurable functions of the state and action. One can also allow for costs to be random variables with expectation given by the cost function $C$. Since our objective involves an expectation, this does not change the ensuing analysis, and we opt for the simpler model.

One can also generalize the POMDP model above (or the MDP model below) to allow the transition/observation probabilities and cost function to additionally depend on time. Such models may be called \emph{time inhomogeneous}. By contrast, the models presented here are \emph{time homogeneous}, as the transition/observation probabilities and cost function do not vary across time. Any reference to a POMDP (or fully observed MDP) in this article shall be understood to refer to a time homogeneous POMDP (or MDP). 

\subsection{Fully Observed Markov Decision Problems}  \label{ss:MDPs}

\begin{definition}[MDP]
A fully observed Markov decision problem (or simply an MDP) is a POMDP for which $\XX = \YY$ and $\upphi ( \cdot | s ) = \delta_{s}$ for each state $s \in \XX$. 
\end{definition}

It is well-known that if $\MM$ is an MDP, then there exists a (uniformly) optimal policy $\pi^{*} \in \Pi_{\MM,S}$, and furthermore it suffices to check for optimality along Dirac distributions $\{ \delta_s : s \in \XX \}$. Using the existence of an optimal policy $\pi^{*} \in \Pi_{\MM,S}$, we define the (optimal) Q-function for the MDP $\MM$ as follows:  $Q^{*}_{\MM} : \XX \times \AA \to \rr$ is given by 
\[
Q^{*}_{\MM} ( s, a ) := E^{\pi^{*}}_{\upnu} \left[ \sum_{t = 0}^{\infty} \upgamma^t C ( X_t,  A_t ) \middle| X_0 = s , A_0 = a \right], \quad \forall (s,a) \in \XX \times \AA,
\]
where $\pi^{*} \in \Pi_{\MM,S}$ is an optimal policy for $\MM$ and $\upnu \in \Delta( \XX )$ is any initial state distribution.\footnote{The function $Q^{*}_{\MM}$ is also called the action-value function for $\MM$ and is usually defined as the fixed point of a Bellman operator. The brief presentation above invokes some well-known properties of MDPs.} One can show that, for any $s \in \XX$, we have $\JJ_{\MM} ( \pi^{*} , s ) = \min_{a \in \AA } Q^{*}_{\MM} ( s, a )$. 

\begin{lemma} \label{lemma:stopping-condition}
Let $\pi \in \Pi_{\MM,S}$, $\epsilon \geq 0$. Then, $\pi$ is $\epsilon$-optimal for the MDP $\MM$ if and only if 
\begin{equation} \label{eq:stopping-condition}
\JJ_{\MM} ( \pi , s ) \leq \min_{a \in \AA } Q^{*}_{\MM} ( s, a ) + \epsilon , \quad \forall s \in \XX . 
\end{equation}
\end{lemma}

Under mild conditions on the MDP $\MM$, the action value function $Q^{*}_{\MM}$ can be learned iteratively using the Q-learning algorithm \cite[]{Watkins89}. Similarly, for stationary policies $\pi \in \Pi_{\MM,S}$, the value function $\JJ_{\MM} ( \pi, \cdot )$ can be learned iteratively. That is, the agent can produce sequences of iterates $\{ \widehat{J}_t, \widehat{Q}_t \}_{t \geq 0}$, with $\widehat{J}_t \in \rr^{\XX}$ and $\widehat{Q}_t \in \rr^{\XX \times \AA}$ for each $t$, such that, almost surely, 
\[
\widehat{Q}_t \to Q^{*}_{\MM} \quad \text{and} \quad \widehat{J}_t (s) \to \JJ_{\MM} ( \pi, s ),  \quad \forall s \in \XX.
\]

Thus, an agent controlling the MDP $\MM$ using a stationary policy $\pi \in \Pi_{\MM, S}$ may use an estimated surrogate of the inequality of \eqref{eq:stopping-condition}---involving stochastic estimates of $Q^{*}_{\MM}$ and $\JJ_{\MM} ( \pi, \cdot )$---as a stopping condition when searching for an $\epsilon$-optimal policy. This idea will feature heavily in the subsequent sections. In particular, we will use an analogous condition for our definition of subjective best-responding. As a preview of forthcoming definitions, we now state a definition that is motivated by analogy to \eqref{eq:stopping-condition}.

\begin{definition} \label{def:subjective-optimality}
Let $\widehat{J} \in \rr^{\XX}$ and  $\widehat{Q} \in \rr^{\XX \times \AA}$. The function $\widehat{J}$ is said to be \emph{subjectively $\epsilon$-optimal with respect to $\widehat{Q}$} if
\[
\widehat{J} ( s ) \leq \min_{a \in \AA} \widehat{Q} ( s, a ) + \epsilon, \quad \forall s \in \XX . 
\]
\end{definition}

The intuition underlying Definition~\ref{def:subjective-optimality} is that $\widehat{J} (s)$ plays the role of $\JJ_{\MM} ( \pi, s )$ while $\widehat{Q}(s,a)$ plays the role of $Q^{*}_{\MM} (s,a)$ in \eqref{eq:stopping-condition}, with $\widehat{J}$ and $\widehat{Q}$ arising from a learning process. In employing such a comparison to test for $\epsilon$-optimality of $\pi$, the agent replaces objective quantities with learned estimates that represent subjective knowledge obtained through system interaction and learning.

\section{Implied MDPs and Subjective Functions}  \label{appendix:implied-MDPs}

In this section, we explicitly characterize the subjective value functions of Theorem~\ref{theorem:naive-learning} in Section~\ref{sec:naive-learning}. These subjective value functions arise as the limit points of an independent learning process, in which each agent used a stationary policy during the learning process and obtained a sequence of stochastic iterates meant to approximate action values and state values of an assumed MDP. However, as described in Lemma~\ref{lemma:POMDP}, each agent does not face a fully observed MDP but rather a POMDP. To understand the limiting values of the learning process in Algorithm~\ref{algo:Q-learning}, we introduce \emph{implied MDPs} associated with POMDPs. We then use the connection between $n$-player mean-field games and POMDPs to define the subjective functions of Theorem~\ref{theorem:naive-learning} using the language of implied MDPs.

\subsection{Implied MDPs and Subjective Functions for Ergodic POMDPs}

To begin, let $\MM = \left( 	\XX, \YY, \AA, C , P_{\MM}  , \upphi , \upgamma , \upnu_0	\right)$ be a POMDP as described in \eqref{eq:POMDP}. Recall that $\Pi_{\MM, S}$ denotes the set of stationary policies for the POMDP $\MM$, and is identified with the set $\PP ( \AA | \YY)$ of transition kernels on $\AA$ given $\YY$.  We now define several objects relevant to the analysis of learning algorithms for POMDPs.

\begin{definition}
For $\lambda \in \Delta ( \XX )$, the \emph{backward channel} $B^{\lambda} \in \PP ( \XX | \YY )$ is a transition kernel on $\XX$ given $\YY$ defined by
\[
B^{\lambda} ( x | y ) := \frac{ \upphi( y | x ) \lambda ( x ) 	} { \sum_{s \in \XX } \upphi ( y | s )  \lambda (s )	} 	, \quad \forall x \in \XX , y \in \YY . 
\]
\end{definition}

One may interpret the quantity $B^{\lambda} ( x | y )$ as the posterior probability that $x_0 = x$ given that $x_0 \sim \lambda$ and the first observation symbol was $y_0 = y$.

\begin{definition}
For $\lambda \in \Delta ( \XX )$, the \emph{implied transition kernel} $\TT_{\lambda} \in \PP ( \YY | \YY \times \AA )$ is given by 
\[
\TT_{\lambda} ( y^{\prime} | y, a ) := \sum_{ s^{\prime} \in \XX } \upphi ( y^{\prime} | s^{\prime} ) \sum_{s \in \XX } \TT ( s^{\prime} | s, a ) B^{\lambda} ( s | y ) , \quad \forall y , y^{\prime} \in \YY, a \in \AA . 
\]
\end{definition}

The quantity $\TT_{\lambda} ( y^{\prime} | y, a ) $ represents the conditional probability that the observation $y_1 = y^{\prime}$ given $y_0 = y, u_0 = u$ and $x_0 \sim \lambda$.

\begin{definition}
For $\lambda \in \Delta( \XX )$, the \emph{implied cost function} $C_{\lambda} : \YY \times \AA \to \rr$ is given by 
\[
C_{\lambda} ( y, a ) := \sum_{x \in \XX } C ( x, a ) B^{\lambda} ( x | y  ) , \quad \forall y \in \YY  , a \in \AA .
\]
\end{definition}

We will employ the following standing assumption for the remainder of this section.

\begin{assumption} \label{ass:ergodic-POMDP}
For any policy $\pi \in \Pi_{\MM, S}$, there exists unique $\lambda_{\pi} \in \Delta ( \XX )$ such that
\[
\lim_{t \to \infty} \P^{\pi}_{\lambda_0} \left( X_t \in \cdot \right) = \lambda_{\pi} ( \cdot ) , \quad \forall \lambda_0 \in \Delta(  \XX ) . 
\]
\end{assumption}

In words, this assumption says that the law of the hidden state at time $t$, $X_t$, converges, as $t \to \infty$, to some invariant ergodic distribution $\lambda_{\pi} \in \Delta ( \XX )$. The invariant distribution will depend on the policy $\pi$, but convergence to this distribution holds for any distribution $\lambda_0 \in \Delta ( \XX ) $ of the initial state $X_0$.

\vspace{7.5pt}

\noindent \textbf{Remark:} Assumption~\ref{ass:ergodic-POMDP} is made for clarity of presentation, and in fact can be relaxed. For the results below, one can alternatively assume the following.

\vspace{7.5pt} 
\noindent\textbf{Assumption~\ref{ass:ergodic-POMDP}$^{\star}$} \textit{For any policy $\pi \in \Pi_{\MM,S}$, there exists unique  $\lambda_{\pi} \in \Delta ( \XX )$ such that}
\[
\lim_{T \to \infty} \frac{1}{T} \sum_{t = 0}^{T-1}  {  \P^{\pi}_{\lambda_0} }  \left( X_t \in \cdot \right) = \lambda_{\pi} ( \cdot ) , \quad \forall \lambda_0 \in \Delta ( \XX ) . 
\]

\vspace{7.5pt}

Under either Assumption~\ref{ass:ergodic-POMDP} or Assumption~\ref{ass:ergodic-POMDP}$^{\star}$, each stationary policy is assigned a corresponding backward channel, implied transition kernel, and implied cost function: for each $\pi \in \Pi_{\MM,S}$, we say that $B^{\lambda_{\pi}}$ is the backward channel associated with $\pi$, $\TT_{\lambda_{\pi}}$ is the implied transition kernel associated with $\pi$, and $C_{\lambda_{\pi}}$ is the implied cost unction associated with $\pi$.

For each $\pi \in \Pi_{\MM,S}$, we define an operator $f_{\pi} : \rr^{\YY \times \AA} \to \rr^{\YY \times \AA}$, whose $(y,a)$-component mapping is given by 
\[
\left[ f_{\pi} Q \right] ( y, a ) := C_{\lambda_{\pi}} ( y, a ) + \gamma \sum_{y^{\prime} \in \YY } \min_{ a^{\prime} \in \AA } Q ( y^{\prime}, a^{\prime} )  \TT_{\lambda_{\pi}} ( y^{\prime} | y, a ) ,  \quad \forall \: Q \in \rr^{\YY \times \AA } . 
\]

It is easily verified that $f_{\pi}$ is a $\gamma$-contraction on $\rr^{\YY \times \AA  }$ and therefore admits a unique fixed point, which we denote by {$\QQ^{*}_{\MM,\pi} \in \rr^{\YY \times \AA }$.}

\begin{definition}
The \emph{implied Q-function} associated with $\pi$  is defined as {$\QQ^{*}_{\MM,\pi}$. }
\end{definition}

Structurally, one can see that the implied Q-function $\QQ^{*}_{\MM,\pi}$ is the optimal Q-function for a suitably defined \emph{implied MDP}, namely that with state space $\YY$, action space $\AA$, stage cost function $C_{\lambda_{\pi}}$, and transition kernel $\TT_{\lambda_{\pi}}$. 

\vspace{5pt}

\noindent \textbf{Remark:} To be clear, we do not claim that using the greedy policy with respect to $\QQ^{*}_{\MM,\pi}$ will yield meaningful performance guarantees for the POMDP $\MM$.

\

In addition to the operator $f_{\pi}$, we define an operator $g_{\pi} : \rr^{\YY} \to \rr^{\YY}$, whose $y$-component mapping is given by 
\[ 
\left[ g_{\pi} J \right] ( y ) := \sum_{ a \in \AA } \pi ( a | y ) \left\{  C_{\lambda_{\pi}} ( y, a ) + \gamma \sum_{ y^{\prime} \in \YY } J ( y^{\prime} ) \TT_{\lambda_{\pi}} ( y^{\prime} | y, a )		\right\}, \quad \forall J \in \rr^{  \YY } .
\]

As with $f_{\pi}$, one can verify that $g_{\pi}$ is a contraction and admits a unique fixed point, which we denote by {$\JJ^{*}_{\MM, \pi} \in \rr^{\YY}$.}

\begin{definition}
The \emph{implied value function} associated with $\pi$ is defined as {$\JJ^{*}_{\MM, \pi}$.}
\end{definition}

One can interpret the function $\JJ^{*}_{\MM, \pi}$ as the state value function of the policy $\pi \in \Pi_{\MM, S}$ in the same implied MDP as before, with state space $\YY$, action space $\AA$, stage cost function $C_{\lambda_{\pi}}$, and transition kernel $\TT_{\lambda_{\pi}}$.

\subsection{Implied MDPs and Subjective Functions in Partially Observed $n$-Player Mean-Field Games}

Let $\bG =  ( \NN, \xx , \yy,  \aa,   \{ \varphi^i \}_{i \in \NN}, c , \gamma, P_{\loc}, \nu_0 )$ be a partially observed $n$-player mean-field game, with the symbols inheriting their meanings from \eqref{eq:mean-field-game}. By Lemma~\ref{lemma:POMDP}, we have that if $\bpi^{-i} \in \bPi^{-i}_{S}$ is a stationary policy for the remaining players, then player $i$ faces a POMDP $\MM_{\bpi^{-i}}$, with partially observed state process $\{ \bx_t \}_{t \geq 0}$ and observation process $\{ y^i_t = \varphi^i ( \bx_t ) \}_{t \geq 0}$.

Under Assumption~\ref{ass:state-visitation}, if $(\pi^i, \bpi^{-i} ) \in \Pi^i_{S} \times \bPi^{-i}_{S}$ is stationary, then the ergodicity condition of Assumption~\ref{ass:ergodic-POMDP} holds for the POMDP $\MM_{\bpi^{-i}}$. Thus, for any stationary policy $\pi^i \in \Pi^i_{S}$, we may define the objects of the preceding section: that is, we may define backward channels, implied transition kernels, implied cost functions, implied Q-functions, and implied value functions associated with $\pi$ in the POMDP $\MM_{\bpi^{-i}}$.

\begin{definition}
Let $\bG$ be a partially observed $n$-player mean-field game satisfying Assumption~\ref{ass:state-visitation}. For each player $i \in \NN$ and policy $\bpi^{-i} \in \bPi^{-i}_{S}$, let $\MM_{\bpi^{-i}}$ denote the POMDP faced by player $i$.

For any $\bpi = (\pi^i, \bpi^{-i}) \in \Pi^i_{S} \times \bPi^{-i}_{S}$, the \emph{subjective value function $V^{*i}_{\bpi}$} is defined as the implied value function associated with $\pi^i$ in the POMDP $\MM_{\bpi^{-i}}$. That is, $V^{*i}_{\bpi} := {\JJ^{\star}_{\MM_{\bpi^{-i}}, \pi^i }}$.

For any $\bpi = (\pi^i, \bpi^{-i}) \in \Pi^i_{S} \times \bPi^{-i}_{S}$, the \emph{subjective Q-function $W^{*i}_{\bpi}$} is defined as the implied Q-function associated with $\pi^i$ in the POMDP $\MM_{\bpi^{-i}}$. That is, $W^{*i}_{\bpi} := {\QQ^{\star}_{\MM_{\bpi^{-i}}, \pi^i }}$.
\end{definition}

\section{Proof of Theorem~\ref{theorem:naive-learning} } \label{appendix:subjective-iterates}

By Lemma~\ref{lemma:POMDP}, player $i$ faces a POMDP with observation sequence $\{ y^i_t \}_{t \geq 0}$ and underlying state process $\{ \bx_t \}_{t \geq 0}$. By Assumption~\ref{ass:state-visitation} and the assumed softness of $\bpi$, we have that all pairs $(\bs, a^i ) \in \bX \times \aa$ are visited infinitely often $\P^{\bpi}_{\nu}$-almost surely for any $\nu \in \Delta ( \bX)$. We can therefore invoke {\cite[Theorem 4(i)]{kara2022convergence}} to establish the almost sure convergence of Q-factor iterates $\{ \bar{Q}^i_t \}_{t \geq 0}$. The same analysis can be used to establish the convergence of the value function iterates $\{ \bar{J}^i_t \}_{ t\geq 0}$. This proves the first part.

Under Assumption~\ref{ass:global-state-observability}, player $i$ faces an MDP with state process $\{ \bx_t \}_{t \geq 0}$. Again by the softness of $\bpi$ and Assumption~\ref{ass:state-visitation}, each $(\bs, a^i ) \in \bX \times \aa$ is visited infinitely often almost surely. The convergence of Q-factors under these conditions is well-known; see for instance \cite{tsitsiklis1994asynchronous}. The same analysis  can be used to prove that $\lim_{t \to \infty} \bar{J}^i_t ( \bs ) = J^i (\bpi ,\bs )$ almost surely for each $\bs \in \bX$, proving the second part.

The proof of the third part parallels that of the second part,  replacing $\bx_t$ by $y^i_t = ( x^i_t, \upmu ( \bx_t ))$ and using Theorem~\ref{theorem:mean-field-MDP} to see that player $i$ faces an MDP in $\{ y^i_t \}_{t \geq 0}$.

\section{Proof of Lemma \ref{lemma:mean-field-subjective-equilibrium}} \label{appendix:mean-field-subjective-equilibrium}

For any $i \in \NN$, the sets $\bPi_{S, \sym}$ and $\bPi^{-i}_{S, \sym}$ are compact under the topologies induced by the metrics $\bd_{\sym}$ and $\bd^{-i}_{\sym}$, respectively. By compactness, Lemma~\ref{lemma:continuous-cost}, and Lemma~\ref{lemma:mean-field-continuous-min-Q-factors}, we see that the functions of Lemmas~\ref{lemma:continuous-cost} and \ref{lemma:mean-field-continuous-min-Q-factors} are in fact uniformly continuous on $\bPi_{S, \sym}$ and $\bPi^{-i}_{S, \sym}$. It follows that there exists $\xi > 0 $ such that if two joint policies $\bpi, \bpi' \in \bPi_{S, \sym}$ satisfy $\bd_{\sym} ( \bpi, \bpi' ) < \xi$, then 
\begin{equation} \label{eq:close-Q-and-J}
\left| J^i ( \bpi , \bs ) - J^i ( \bpi' , \bs ) \right| < \frac{\epsilon}{2} \quad \text{and} \quad \left| \min_{a^i \in \aa } Q^{*i}_{\bpi^{-i}} ( \varphi^i ( \bs ) , a^i )   -    \min_{a^i \in \aa } Q^{*i}_{\bpi'^{-i}} ( \varphi^i ( \bs) , a^i ) \right| < \frac{\epsilon}{2},
\end{equation}
for any $\bs \in \bX$, where $\varphi^i ( \bs ) = (s^i, \upmu ( \bs ))$ due to Assumption~\ref{ass:mean-field-state-observability}. %
We fix $\bpi^{*} \in \eq[0]_{S} \cap \bPi_{\sym}$ to be a symmetric perfect equilibrium policy, which exists by Theorem~\ref{theorem:mean-field-equilibrium}. (That a \emph{symmetric} perfect equilibrium exists can be seen from the proof of Theorem~\ref{theorem:mean-field-equilibrium}.) Let $\bpi_{\rm soft} \in \bPi_{S, \sym}$ be a soft, symmetric joint policy satisfying $\bd_{\sym} ( \bpi^{*} , \bpi_{\rm soft} ) < \xi$. 

By  part 3 of Theorem~\ref{theorem:naive-learning}, since $\bpi_{\rm soft}$ is soft, we have that 
\[
W^{*i}_{\bpi_{\rm soft}}  = Q^{*i}_{\bpi^{-i}_{\rm soft}}  \quad \text{and} \quad V^{*i}_{\bpi_{\rm soft}} (    \varphi^i ( \bs ) )  = J^i ( \bpi_{\rm soft} ,  \bs ), \quad \forall i \in \NN,  \bs \in \bX .
\]
Combining this with \eqref{eq:close-Q-and-J}, it follows that, for any $\bs \in \bX$ and $i \in \NN$, we have $V^{*i}_{\bpi_{\rm soft}} ( \varphi^i( \bs )) \leq \min_{a^i \in \aa} W^{*i}_{\bpi_{\rm soft}} ( \varphi^i ( \bs ), a^i ) + \epsilon$, which shows that $\bpi_{\rm soft} \in \Subj_{\epsilon} ( \VV^{*}, \WW^{*} )$. \hfill $\square$

\section{Proof of Lemma \ref{lemma:mean-field-oracle} } \label{appendix:satisficing-results}

To prove Lemma \ref{lemma:mean-field-oracle}, we first need the following auxiliary result, which also appears in the proof of Lemma~\ref{lemma:mean-field-quantized-paths}.

\begin{lemma} \label{lemma:mean-field-subjective-BR-iff}
Let $\bG$ be a partially observed $n$-player mean-field game for which Assumptions~\ref{ass:mean-field-state-observability} and \ref{ass:state-visitation} hold, and let $\epsilon \geq 0$. Let $\bpi \in \bPi_{S}$ be soft. For $i, j \in \NN$, suppose $\pi^i$ and $\pi^j$ are symmetric. Then, we have 
\[
\pi^i \in \SubjBR^{i}_{\epsilon} ( \bpi^{-i} , \VV^{*}, \WW^{*} ) \iff \pi^j \in \SubjBR^{j}_{\epsilon} ( \bpi^{-j} , \VV^{*}, \WW^{*} ).
\]
\end{lemma}

\begin{proof}   %
To prove this result, we argue that the subjective functions of players $i$ and $j$ satisfy $V^{*i}_{\bpi}= V^{*j}_{\bpi}$ and $W^{*i}_{\bpi} = W^{*j}_{\bpi}$ whenever $\pi^i$ and $\pi^j$ are symmetric and soft. In Section~\ref{sec:naive-learning}, we observed that, under Assumptions~\ref{ass:mean-field-state-observability} and \ref{ass:state-visitation}, the learned values $V^{*i}_{\bpi}$ and $W^{*i}_{\bpi}$ were in fact the state value and action value functions, respectively, for an MDP with state space $\{ \varphi^i  (\bs) : \bs \in \bX \} \subseteq \yy$. An analogous remark holds for $V^{*j}_{\bpi}$ and $W^{*j}_{\bpi}$ and player $j$. The MDP in question was called an \emph{approximate belief MDP} by \cite{kara2022convergence}, and can be characterized in terms of the implied MDP construction of Appendix~\ref{appendix:implied-MDPs}.

\

\noindent We will argue that the approximate belief MDP on $\{ \varphi^i ( \bs ) : \bs \in \bX \}$ faced by player $i$ is equivalent to the approximate belief MDP on $\{ \varphi^j ( \bs ) : \bs \in \bX \}$ faced by player $j$. Then, since player $i$ and $j$'s policies are symmetric, they correspond to the same stationary policy for this approximate belief MDP, and are therefore both either $\epsilon$-optimal for that MDP or not; the result will follow.

To see that the approximate belief MDP facing player $i$ is equivalent to that facing player $j$, we observe that the construction of \cite{kara2022convergence} depends only on the stage cost function, the observation channel, and the unique invariant distribution on the underlying state space. The stage cost and observation channel are symmetric and shared by players $i$ and $j$, so it suffices to show that the unique invariant distribution on $\bX$, say $\nu_{\bpi}$, is symmetric in the following sense:
\begin{equation} \label{eq:symmetric-invariant-measure}
\nu_{\bpi}  ( \swap_{ij} ( \bs ) ) = \nu_{\bpi} ( \bs ), \quad \forall \bs \in \bX,
\end{equation}
where, for any $\bs \in \bX$, $\swap_{ij} (\bs) \in \bX $ is the global state satisfying (1)~$\swap_{ij} (\bs)^p = s^p$ for each $p \in \NN \setminus \{ i, j \}$, and (2)~we have $\swap_{ij} (\bs)^j = s^i$ and $\swap_{ij}(\bs)^i = s^j$.

To see that \eqref{eq:symmetric-invariant-measure} holds, note that by Assumption~\ref{ass:state-visitation}, $\{ \bx_t \}_{t \geq 0}$ is an irreducible, aperiodic Markov chain on $\bX$ under the soft policy $\bpi$. Thus, we have for any $\tilde{\nu} \in \Delta ( \bX)$. 
\[
\P^{\bpi}_{\tilde{\nu}} ( \bx_t \in \cdot  ) \to \nu_{\bpi} ( \cdot ),
\] 
as $t \to \infty$, where $\nu_{\bpi}$ is the (unique) invariant measure on $\bX$ induced by $\bpi$. In particular, putting the initial measure $\tilde{\nu} = \uniform ( \bX)$, we  have that for each $t \geq 0$ and each $\bs \in \bX$, 
\[
\P^{\bpi}_{\tilde{\nu}} ( \bx_t = \bs ) = \P^{\bpi}_{\tilde{\nu}} ( \bx_t = \swap_{ij} ( \bs  ) ).
\]

It follows that 
\[
\nu_{\bpi} ( \bs ) := \lim_{t \to \infty} \P^{\bpi}_{\tilde{\nu}} ( \bx_t = \bs ) = \lim_{t \to \infty} \P^{\bpi}_{\tilde{\nu}} ( \bx_t = \swap_{ij} ( \bs ) ) =: \nu_{\bpi} ( \swap_{ij} (\bs ) ).
\]
From this, it follows that the approximate belief MDP on $\yy$ faced by player $i$ is the same as the approximate belief MDP on $\yy$ faced by player $j$.
\end{proof}

\subsubsection*{Proof of Lemma~\ref{lemma:mean-field-oracle}}

We have that $\{ \bpi_k \}_{k \geq 0}$ is a time homogeneous Markov chain on $\widehat{\bPi} $. For any subjective $\epsilon$-equilibrium $\bpi^{*} \in \widehat{\bPi}  \cap \Subj_{\epsilon} (\VV^{*}, \WW^{*} )$, we have that the singleton $\{ \bpi^{*} \}$ is an absorbing set for this Markov chain. By part 2 of Lemma~\ref{lemma:mean-field-quantized-paths}, the game $\bG$ has the $(\VV^{*}, \WW^{*})$-subjective $\epsilon$-satisficing paths property in $\widehat{\bPi} $. For any $\bpi \in \widehat{\bPi} $, let $L_{\bpi} < \infty$ denote the length of a shortest $(\VV^{*} , \WW^{*})$-subjective $\epsilon$-satisficing path within $\widehat{\bPi} $ that starts at $\bpi$ and terminates at a policy in $\widehat{\bPi}  \cap \Subj_{\epsilon} (\VV^{*}, \WW^{*} )$. Call such a policy $\bpi^*$ and note it depends on $\bpi$. We also define $p_{\bpi} > 0 $ as the probability the Markov chain follows this path when starting at $\bpi$, i.e.
\[
p_{\bpi} := \P ( \bpi_{ L_{\bpi}} = \bpi^* | \bpi_0 = \bpi) > 0, \quad \forall \bpi \in \widehat{\bPi} .
\]

\noindent Define $L := \max\{ L_{\bpi} : \bpi \in \widehat{\bPi}  \}$ and $\hat{p} := \min \{ p_{\bpi} : \bpi \in \widehat{\bPi}   \} > 0$. For any $m \geq 0$, we have  
\[
\P \left( \bigcap_{j = 1}^{m} \{ \bpi_{jL} \notin \widehat{\bPi}  \cap \Subj_{\epsilon} (\VV^{*}, \WW^{*} )   \} \right) \leq ( 1 - \hat{p} )^m .
\]
Taking $m \to \infty$ gives the result.

\section{Approximation Results on the Sequences of Learning Iterates}

\textbf{Remark:} The contents of this and the next section closely resembles that of \cite[Appendix A]{yongacoglu2023satisficing}. The proof technique used here parallels the proof technique of \cite[Theorem 5.1]{yongacoglu2023satisficing}.

In the coming sections, we prove that Algorithm~\ref{algo:main} leads to the convergence of joint policies as described in Theorems~\ref{theorem:mean-field-state-observability}, \ref{theorem:global-state-observability}, and \ref{theorem:compressed-observability}. Since the evolution of the policy process $\{ \bpi_k \}_{k \geq 0}$ depends on the evolution of the learning iterates  $\{ \widehat{J}^i_t , \widehat{Q}^i_t \}_{t \geq 0, i \in \NN }$, we begin by studying the convergence behaviour of these iterates. We argue that if parameters are suitably selected, then these learning iterates sampled at the end of each exploration phase will closely approximate the subjective functions for that exploration phase, and consequently the policy process of Algorithm~\ref{algo:main} approximates the policy process of the Markov chain resulting from Algorithm~\ref{algo:oracle}.

We note that when each agent $i \in \NN$ uses Algorithm~\ref{algo:main}, it is actually following a particular randomized, non-stationary policy. When all agents use Algorithm~\ref{algo:main}, we use $\P$  (with no policy index in the superscript and optional initial distribution in the subscript, e.g. $\P_{\nu}$ for some $\nu \in \Delta (\bX )$) to denote the resulting probability measure on trajectories of states and actions. For all other policies $\tilde{\bpi} \in \bPi$, we continue to use $\P^{\tilde{\bpi}}_{\nu}$, as before.

The policy process $\{ \bpi_k \}_{k \geq 0}$ depends on the sequences $\{ \widehat{J}^i_t, \widehat{Q}^i_t\}_{ t \geq 0, i \in \NN}$ only 
through these sequences sampled at the end of exploration phases; that is, the iterate sequences are relevant to the updating of policies only along the subsequence of times $\{ t_k \}_{k \geq 0}$. Recall that we used $\{ \bar{Q}^i_t \}_{t\geq 0}$ and $\{ \bar{J}^i_t \}_{t \geq 0}$ to denote the naively learned stochastic iterates obtained when player $i\in\NN$ employed Algorithm~\ref{algo:Q-learning} and all players followed a soft stationary policy. We now analyze the sequences $\{ \widehat{Q}^i_t \}_{t \geq 0}$ and $\{ \widehat{J}^i_t \}_{t \geq 0}$ by comparison to the sequences $\{ \bar{Q}^i_t\}_{t \geq 0}$ and $\{ \bar{J}^i_t \}_{t \geq 0}$.

The sequences $\{ \widehat{Q}^i_t \}_{t \geq 0}$ and $\{ \bar{Q}^i_t \}_{t \geq 0}$ are related through the Q-factor update. There are, however, two major differences. First, Algorithm~\ref{algo:main} instructs player $i$ to reset its counters at the end of the $k^{th}$ exploration phase (i.e. after the update at time $t_{k+1}$, before the update at time $t_{k+1} + 1$), meaning the step sizes differ for the two iterate sequences $\{ \bar{Q}^i_t \}_{ t \geq 0} $ and $\{ \widehat{Q}^i_t \}_{ t \geq 0}$ even when the state-action-cost trajectories observed by player $i$ are identical. Second, Algorithm~\ref{algo:main} instructs player $i$ to reset its Q-factors at the end of the $k^{th}$ exploration phase, while Algorithm~\ref{algo:Q-learning} does not involve any resetting. 

Consequently, one sees that the process $\{ \widehat{Q}^i_t \}_{t \geq 0}$ depends on the initial condition $\widehat{Q}^i_0 = 0 \in \rr^{\yy \times \aa}$, the state-action trajectory, and the resetting times $\{ t_k \}_{k \geq 0}$. In contrast, the process $\{ \bar{Q}^i_t \}_{t \geq 0}$ depends only on the initial value $\bar{Q}^i_0 = 0 \in \rr^{\yy \times \aa}$ and the state-action trajectory. Analogous remarks hold relating $\{ \widehat{J}^i_t \}_{ t \geq 0}$ and $\{ \bar{J}^i_t \}_{ t \geq 0}$. 

Recall: exploration phase $k$ begins with the stage game at $t_k$ and ends before the stage game at $t_{k+1} = t_k + T_k$. During exploration phase $k$, the sequences $\{ \widehat{Q}^i_t \}_{t = t_k }^{ t_k + T_k }$ and $\{ \widehat{J}^i_t \}_{t = t_k}^{t_k + T_k }$ depend only on the state-action trajectory $\bx_{t_k}, \ba_{t_k}, \cdots, \ba_{t_k + T_k - 1 }, \bx_{t_k + T_k}$. This leads to the following useful observation: 
\begin{align*}
	&\P \left( \{ \bx_{t_k + T_k} = \bs_{T_k} \} \bigcap_{j = 0}^{T_k - 1} \{ \bx_{t_k + j } = \bs_{j} , \ba_{t_k + j } = \tilde{\ba}_j  \}  \middle| \bx_{t_k} = \bs , \bpi_k = \bpi \right)  \\ 
= 	&\P^{\bpi}_{\bs} \left( \{ \bx_{T_k} = \bs_{T_k} \} \bigcap_{j = 0}^{T_k-1} \{ \bx_j = \bs_j , \ba_j = \tilde{\ba}_j \}  \right) ,
\end{align*}
holds for any $( \bs_0, \bs_1, \cdots, \bs_{T_k} ) \in \bX^{ T_k + 1 }$ and $( \tilde{\ba}_0, \cdots, \tilde{\ba}_{T_k - 1 } ) \in \bA^{ T_k }$, where $\bA = \times_{i \in \NN} \aa$. 

In words, once players following Algorithm~\ref{algo:main} select a policy $\bpi$ for the $k^{th}$ exploration phase, then the conditional probabilities of the trajectories restricted to time indices in that exploration phase can be described by $\P^{\bpi}$, with the indices of the events suitably shifted to start at time 0. This leads to a series of useful lemmas, which we include below for completeness.

\begin{lemma} \label{lemma:same-measures}
For any $i \in \NN, \bpi \in \widehat{\bPi}, k \geq 0$ and global state $\bs \in \bX$, we have 
\emph{
\[
\P \left( \widehat{Q}^i_{t_{k+1}} \in \cdot \middle| \bpi_k = \bpi, \bx_{t_k} = \bs \right) = \P^{\bpi}_{\bs} \left( \bar{Q}^i_{T_k } \in \cdot  \right)
\]
}
and
\emph{
\[
\P \left( \widehat{J}^i_{t_{k+1}} \in \cdot \middle| \bpi_k = \bpi, \bx_{t_k} = \bs \right) = \P^{\bpi}_{\bs} \left( \bar{J}^i_{T_k } \in \cdot  \right).
\]
}
\end{lemma}

Combining Lemma~\ref{lemma:same-measures} with Theorem~\ref{theorem:naive-learning}, we get the following result.

\begin{lemma} \label{lemma:learning-error}
For any joint policy $\bpi \in \widehat{\bPi}$, global state $\bs \in \bX$, and player $i \in \NN$, we have
\begin{enumerate}	
	\item $ \P^{\bpi}_{\bs} \left( \lim_{t \to \infty} \bar{Q}^i_t = W^{*i}_{\bpi} \right) = 1$.
	
	\item $ \P^{\bpi}_{\bs} \left( \lim_{t \to \infty} \bar{J}^i_t = V^{*i}_{\bpi} \right) = 1$.
	
	\item For any $\xi > 0$, there exists $T = T ( i , \bpi, \xi ) \in \nn$ such that 
		\[
			\P^{\bpi}_{\bs} \left( \sup_{t \geq T } \left\| \bar{Q}^i_t - W^{*}_{\bpi} \right\|_{\infty} < \xi \right) \geq 1 - \xi , \text{ and } \P^{\bpi}_{\bs} \left( \sup_{t \geq T } \left\| \bar{J}^i_t - V^{*}_{\bpi} \right\|_{\infty} < \xi \right) \geq 1 - \xi 
		\]
\end{enumerate}
\end{lemma}

Finally, we combine Lemmas~\ref{lemma:same-measures} and \ref{lemma:learning-error} to obtain the following useful result on conditional probabilities.

\begin{lemma}  \label{lemma:learning-iterate-accuracy}
Let $k, \ell \in \zz_{\geq 0}$ and suppose $k \leq \ell$. Let $\mathcal{F}_k$ denote the $\sigma$-algebra generated by the random variables $ \bpi_k$ and \emph{$\bx_{t_k}$}. For any $\xi > 0$, there exists $T = T(\xi) \in \nn$ such that if $T_{\ell} \geq T$, then $\P$-almost surely, we have
\[
\P \left(	\bigcap_{i \in \NN} \left\{  \left\| \widehat{Q}^i_{t_{ \ell+1}} - W^{*i}_{\bpi_\ell} \right\|_{\infty} < \xi \right\} \cap \left\{  \left\| \widehat{J}^i_{t_{\ell+1}} - V^{*i}_{\bpi_\ell} \right\|_{\infty} < \xi \right\}	\middle| \mathcal{F}_k \right) \geq 1 - \xi .
\]
\end{lemma}

\section{Proof of Theorem~\ref{theorem:mean-field-state-observability}}     \label{appendix:proof-mean-field-state}

We begin by introducing the quantity $\bar{d}_{\mf}$,  which will serve as the upper bound for the tolerance parameters $d^i, i \in \NN$. The quantity $\bar{d}_{\mf}$,  depends on both $\epsilon > 0$ and the subset of policies $\widehat{\bPi} \subset \bPi_{S}$ as follows: $\bar{d}_{\mf} := \min  \mathcal{O}_{\mf} $, where $\mathcal{O}_{\mf} := S_{\mf} \setminus \{ 0 \}$, and  $S_{\mf}$ is given by 
\[
S_{\mf} := \left\{   \left| 	\epsilon - \left(   V^{*i}_{\bpi } ( y ) - \min_{a^i \in \aa } W^{*i}_{\bpi  } ( y, a^i ) \right) \right| : i \in \NN, \bpi \in \widehat{\bPi} , y \in \yy \right\}.
\]

That is, $S_{\mf}$ is the collection of (subjective) $\epsilon$-optimality gaps of the various joint policies in $\widehat{\bPi}$, and $\bar{d}_{\mf}$ is the minimum non-zero separation between $\epsilon$ and the (subjective) suboptimality gap of some player $i$'s performance.

In Assumption~\ref{ass:mean-field-delta}, we assumed that each player's $d^i$ parameter, which represents tolerance for suboptimality in their policy evaluation step and is included to account for noise in the learning iterates, is positive and small: $d^i \in ( 0 , \bar{d}_{\mf} )$ for each $i \in \NN$. 

We define  $\Xi := \frac 1 2 \min_{i \in \NN} \{ d^i , \bar{d}_{\mf} - d^i \}$. The quantity $\Xi$ will serve as a desirable upper bound on learning error: if players jointly follow a joint policy $\bpi \in \widehat{\bPi}$ and engage in Q-learning and value function estimation, then convergence to within $\Xi$ of the limiting values ensures that each player correctly assesses whether it is $(\VV^{*}, \WW^{*})$-subjectively $\epsilon$-best-responding by using the comparison of Line 17 of Algorithm~\ref{algo:main}. We formalize this below.

For $k \geq 0$, we define ${\rm Event} ( \Xi , k )$ as 
\[
{\rm Event} ( \Xi, k ) := \left\{ \max \left\{ \| \widehat{J}^i_{t_{k+1}} - V^{*i}_{\bpi_k} \|_{\infty} , \| \widehat{Q}^i_{t_{k+1}} - W^{*i}_{\bpi_k} \|_{\infty} : i \in \NN \right\} < \Xi \right\} .
\]

Given ${\rm Event} (\Xi, k )$, we have that for any player $i \in \NN$, 
\[
\pi^i_k \in \SubjBR^i_{\epsilon} ( \bpi^{-i}_k , \VV^{*}, \WW^{*}) \iff \widehat{J}^i_{t_{k+1}} ( \varphi^i (\bx) ) \leq \min_{a^i \in \aa } \widehat{Q}^i_{t_{k+1}} ( \varphi^i (\bx), a^i ) + \epsilon + d^i \quad \forall \bx \in \bX. 
\]  
For convenience, we also define the following intersection of events. For any $k, \ell \in \zz_{\geq 0}$, let 
\[
E_{k: k+\ell} := \cap_{j = 0}^{\ell} {\rm Event} ( \Xi, k + j ). 
\] 
That is, $E_{k:k+\ell}$ is the event where all agents obtain $\Xi$-accurate learning estimates in each of the exploration phases $k, k+1, \cdots, k + \ell$. For $k \in \zz_{\geq 0}$, we let $G_k$ denote the event that the policy $\bpi_k$ is an $\epsilon$-equilibrium: $G_k := \{ \bpi_k \in \bPi \cap \Subj_{\epsilon} (\VV^{*}, \WW^{*}) \}$.

From the preceding discussion on $\bar{d}_{\mf}, \{ d^i \}_{i \in \NN}$ and the choice of $\Xi$, we have that for any $\ell \geq 0$, 
\begin{equation} \label{ineq:1}
\P \left( G_{k+\ell} \middle| G_k \cap E_{k:k+\ell} \right) =1 .
\end{equation}

Recall the quantity $L := \max\{ L_{\bpi_0} : \bpi_0 \in \widehat{\bPi} \}$, where for each $\bpi_0 \in \widehat{\bPi}$, the number $L_{\bpi_0} < \infty$ is defined as the shortest $(\VV^{*}, \WW^{*})$-subjective $\epsilon$-satisficing path within $\widehat{\bPi}$ starting at $\bpi_0$ and ending in $\widehat{\bPi} \cap \Subj_{\epsilon} (\VV^{*}, \WW^{*})$. From our assumptions on $\widehat{\bPi}$, such a path exists for every $\bpi_0 \in \widehat{\bPi}$. If $L_{\bpi_0} < L$ for a particular initial policy $\bpi_0$, we may extend this path to have length $L$ by repeating the final term. Thus, for every $\bpi \in \widehat{\bPi}$, we obtain the inequality 
\begin{equation} \label{ineq:pmin}
\P \left( G_{k+ L} \middle| \{ \bpi_k = \bpi \} \cap E_{k:k+L}  \right) \geq p_{\min} > 0, 
\end{equation}

where $p_{\min} := \prod_{i \in \NN} \left( \frac{e^i}{ | \widehat{\Pi}^i | } \right)^{L} > 0$. The bound $p_{\min}$ is obtained through the following loose lower bounding argument: beginning at $\bpi_k = \bpi$, the joint policy process $\bpi_{k} , \dots, \bpi_{k+L}$ follows the $(\VV^{*}, \WW^{*})$-subjective $\epsilon$-satificing path of length $L$ described above with probability no less than the event that---at each step---the ``correct'' unsatisfied player updates to the ``correct'' policy uniformly at random, which occurs with probability no less than the probability given by the ratio in the product.

Fix $u^{*} \in (0,1)$ such that
\[
\frac{ u^{*} p_{\min} }{  1 - u^{*} + u^{*} p_{\min} } > 1 - \frac{\xi}{2}.
\]

Combining Lemma~\ref{lemma:learning-iterate-accuracy} with a union bound, we have that there exists $\tilde{T} \in \nn$ such that if $T_l > \tilde{T}$ for all $l \geq 0$, then $\P \left( E_{k:k+L}  | \bpi_k = \bpi \right) \geq u^{*}$ for all $k \geq 0$ and any $\bpi \in \widehat{\bPi}$. Thus, we have $\P \left( E_{k:k+L} \middle| G_k \right) \geq u^{*} $ and $\P \left( E_{k:k+L} \middle| G^c_k \right) \geq u^{*} $ for each $k \geq 0$.

We now lower bound $\P ( G_{k+L} )$ by conditioning on $G_k$ and $G_k^c$ as follows.
\begin{align*}
\P \left( G_{k+L} \right) = \P ( G_{k+L} | G_k ) \P (G_k ) + \P ( G_{k+L} | G_k^c) (1 - \P ( G_k )).
\end{align*}

We then lower bound each of the terms above by conditioning on $E_{k:k+L}$ and invoking inequalities \eqref{ineq:1} and \eqref{ineq:pmin}:
\[
\P ( G_{k+L} ) \geq 1 \cdot \P ( E_{k: k+L} | G_k ) \cdot \P (G_k ) + p_{\min} \P ( E_{k:k+L} | G_k^c ) (1 - \P (G_k) ).
\]

Assume now that $T_l > \tilde{T}$ for all $l \geq 0$. We have
\[
\P ( G_{k+L} ) \geq u^{*} \P ( G_k ) + u^{*} p_{\min} ( 1-  \P (G_k )), \quad \forall k \geq 0.
\]

For each $k \in \{ 0, 1, \dots, L-1\}$, define $y^{(k)}_0 := \P ( G_k )$, and for $m \geq 0$ define $y^{(k)}_{m+1} := u^{*} y^{(k)}_m + u^{*} p_{\min} \left(1 - y^{(k)}_m \right).$ By induction, one can show that 
\begin{equation} \label{ineq:y}
\P ( G_{k+ mL}  ) \geq y^{(k)}_m, \quad \forall m \geq 0.
\end{equation}

Observe that $y^{(k)}_{m+1}$ can be written as 
\[
y^{(k)}_{m+1} = \left( u^{*} - u^{*} p_{\min} \right)^{m+1} y^{(k)}_0 + u^{*} p_{\min} \sum_{j=0}^m \left( u^{*} - u^{*} p_{\min} \right)^j . 
\]

Since $0 < u^{*} - u^{*} p_{\min} < 1$, we have that $\lim_{m \to \infty} y^{(k)}_m = \frac{  u^{*} p_{\min} }{ 1 - u^{*} + u^{*} p_{\min} } > 1 - \frac{\xi}{2}$. Then, by \eqref{ineq:y}, we have that $\P ( G_{k+mL} ) \geq 1 - \xi/2$ holds for all sufficiently large $m$,  which completes the proof.

\section{An Approximation Result Relating to Mean-Field Equilibrium } \label{appendix:large-N}

In this section, we present results relating various solution concepts in mean-field games. The aim of this section is to suggest how one may use approximation results to establish the existence of {subjective $\epsilon$-equilibrium} in partially observed $n$-player mean-field games with local observability. We begin by introducing the concept of \textit{stationary mean-field equilibrium} (SMFE) and giving sufficient conditions for the existence of SMFE. We then argue that, under some additional conditions on the transition probabilities, a SMFE can be used to obtain an (objective) $\epsilon$-equilibrium for a particular partially observed $n$-player mean-field game with sufficiently large $n$. We conclude this section with high-level discussion on how these results can be used to establish existence of \textit{subjective} $\epsilon$-equilibrium.

\subsection{Mean-Field System}

To facilitate passing from a model with finitely many players to a model with infinitely many players, we now introduce the terminology of \emph{mean-field systems}, upon which the games will be defined or re-defined. The mean-field system underlying the mean-field game is a partial list of problem data, $G$, given by
\[
G =  ( \xx, \aa , c , P_{\loc}, \gamma).
\]
The symbols in $G$ retain their earlier meanings: $\xx$ is a finite set of states, $\aa$ is a finite set of actions, $c : \xx \times \Delta(\xx) \times  \aa \to \rr$ is a cost function, $P_{\loc} \in \PP ( \xx | \xx \times \Delta( \xx ) \times \aa )$ is a local state transition kernel, and $\gamma \in (0,1)$ is a discount factor.

\subsection{Mean-Field MDPs}
We now introduce a family of single-agent MDPs, denoted $\bM$, whose elements are indexed by mean-field terms in $\Delta ( \xx )$. For each $\nu \in \Delta ( \xx )$, we let
\[
\MM_{\nu} =  (  \xx , \aa , c_{\nu}, P_{\nu} , \gamma ) 
\]
be the MDP with state space $\xx$, action set $\aa$, discount factor $\gamma$, and where the stage cost function $c_{\nu}$ and transition kernel $P_{\nu}$ are given by
\[
P_{\nu} ( \cdot | x, a ) := P_{\loc} ( \cdot | x, \nu , a ) , \quad c_{\nu} ( x, a ) := c ( x, \nu, a ) , \quad \forall (x,a) \in \xx \times \aa. 
\]

To be clear, $\nu \in \Delta(\xx)$ need not be the initial distribution for the MDP $\MM_{\nu}$: it is used merely to define the cost function and state transition probabilities by fixing the mean-field term at $\nu$. In this section, any MDP of this form is called a \emph{mean-field MDP}.

\vspace{5pt}

We let $\bM := \{ \MM_{\nu} : \nu \in \Delta ( \xx ) \}$. {For each MDP $\MM_{\nu} \in \bM$, the history sets are defined as $H_0 := \xx$ and $H_{t+1} := H_t \times \aa \times \xx$ for each $t \geq 0$. An admissible policy is defined as a sequence of transition kernels $\pi = ( \pi_t )_{t \geq 0}$ such that $\pi_t \in \PP ( \aa | H_t )$ for each $t \geq 0$.} 
The set of all policies for any MDP $\MM_{\nu} \in \bM$ is denoted by $\Pi_{\bM}$, and we note that this set of admissible policies is common to every MDP in $\bM$. As usual, $\Pi_{\bM,S}$ denotes the set of stationary policies. Here, we identify stationary policies with transition kernels in $\PP ( \aa | \xx )$.

In accordance with the notation of Section~\ref{ss:MDPs}, the objective function for the MDP $\MM_{\nu}$ controlled by policy $\pi \in \Pi_{\bM}$ is denoted
\[
\JJ_{\MM_{\nu}} ( \pi, \nu_0 ) = E^{\pi}_{\nu_0} \left[ \sum_{t = 0}^{\infty} \gamma^t c_{\nu} ( x_t, a_t ) \right], 
\]
where the expectation $E^{\pi}_{\nu_0}$ denotes that $x_0 \sim \nu_0$, and that for each $t \geq 0$ we have that $x_{t+1} \sim P_{\nu} ( \cdot | x_t, a_t )$ and $a_t$ is selected according to $\pi$. When $\nu_0 = \delta_{s}$ for some $s \in \xx$, we simply write $\JJ_{\MM_{\nu}} ( \pi, s )$. {Later, it will be necessary to add further indices to distinguish limiting objects from a sequence indexed by $n$. For this reason, we will also denote the preceding expectation $E^{\pi}_{\nu_0}$ as $E^{\pi, \infty}_{\nu_0}$.}

\

\noindent We observe that the optimization problem faced by an agent controlling the single-agent MDP $\MM_{\nu}$ is equivalent to the optimization problem faced by an agent in a partially observed $n$-player mean-field game wherein the mean-field sequence $\{ \mu_t \}_{t \geq 0}$ is constant and given by $\mu_t = \nu$ for each $t \geq 0$.

\

\noindent\textbf{Optimality Multi-Function:} Since $\MM_{\nu}$ is an MDP for each $\nu \in \Delta ( \xx )$, the set of stationary optimal policies for $\MM_{\nu}$ is non-empty. We define ${\rm opt} : \Delta ( \xx ) \to 2^{\Pi_{\bM, S}}$ as
\[
{\rm opt} ( \nu ) := \left\{ 	\pi^{*} \in \Pi_{\bM, S} : \JJ_{\MM_{\nu}} ( \pi^{*} , x ) = \min_{\pi \in \Pi_{\bM}}  \JJ_{\MM_{\nu}} ( \pi, x )  \: \forall x \in \xx 	\right\}, \quad \forall \nu \in \Delta ( \xx ) . 
\]
Note that the set ${\rm opt} ( \nu )$ is compact and convex for each $\nu \in \Delta ( \xx )$.

\

\noindent\textbf{Population Update Function:} Lastly, we define a function $\Phi : \Delta ( \xx ) \times \Pi_{\bM, S} \to \Delta(  \xx )$ as follows: 
\[
\Phi \left( \nu, \pi \right) ( s' ) := \sum_{ s \in  \xx } \sum_{a \in \aa } P_{\loc} ( s' | s, a ) \pi ( a | s ) \nu ( s ) , \quad \forall s' \in \xx . 
\]
For each $s' \in \xx $, $\Phi ( \nu , \pi) ( s' ) $ may be interpreted as the probability of $ x_1 = s'$, where $x_0 \sim \nu$, $a_0 \sim \pi ( \cdot | x_0 )$, and $x_1 \sim P_{\loc} ( \cdot | x_0 , \nu, u_0)$.

\subsection{Stationary Mean-Field Equilibrium: Definition and Existence} 

We now formally introduce the notion of stationary mean-field equilibrium (c.f. \cite[Definition 2.1]{subramanian2019reinforcement}) and give sufficient conditions for its existence.

\begin{definition}
A pair $( \pi^{*} , \nu^{*} ) \in \Pi_{\bM, S} \times \Delta ( \xx )$ is called a \emph{stationary mean-field equilibrium} (SMFE) for the mean-field system $G = ( \xx  , \aa, c,  P_{\loc} , \gamma )$ if (i) $\Phi ( \nu^{*}, \pi^{*} ) = \nu^{*}$, and (ii) $\pi^{*} \in {\rm opt} ( \nu^{*} )$.
\end{definition}

\begin{assumption} \label{ass:unique-invariant-distribution}
For each stationary policy $\pi \in \Pi_{\bM, S}$, there exists a unique invariant probability measure $\nu_{\pi} \in \Delta ( \xx )$ satisfying $\Phi ( \nu_{\pi}, \pi ) = \nu_{\pi}$. Furthermore, the mapping $\pi \mapsto \nu_{\pi}$ is continuous on $\Pi_{\bM, S}$. 
\end{assumption} 

\begin{assumption} \label{ass:cost-continuous-in-measure}
For each $(s,a) \in \xx \times \aa $, the mapping $\nu \mapsto c ( s, \nu, a ) $ is continuous on $\Delta  ( \xx )$.  
\end{assumption} 

\begin{lemma}  \label{lemma:SMFE}
For the mean-field system $G = ( \xx, \aa , c,  P_{\loc}, \gamma )$ defined above, if Assumptions~\ref{ass:unique-invariant-distribution} and \ref{ass:cost-continuous-in-measure} hold, then there exists a stationary mean-field equilibrium $(\pi^{*}, \nu_{\pi^{*}}) \in \Pi_{\bM, S} \times \Delta ( \xx )$.
\end{lemma}

\begin{proof}
We define a point-to-set mapping $\BB: \Pi_{\bM, S} \to 2^{\Pi_{\bM, S}}$ as follows: for each $\pi \in \Pi_{\bM, S}$, $ \BB( \pi ) := {\rm opt} ( \nu_{\pi})$, where $\nu_{\pi}$ is the unique solution to $\Phi( \nu , \pi ) = \nu$. 

As previously noted, the correspondence $\nu \mapsto \opt( \nu )$ is convex-valued and compact-valued for each measure $\nu \in \Delta ( \xx )$, and therefore so is $\opt ( \nu_{\pi} )$ for each $\pi \in \Pi_{\bM, S}$. The continuity conditions imposed by Assumptions~\ref{ass:unique-invariant-distribution} and \ref{ass:cost-continuous-in-measure} ensure that $\BB$ is upper hemicontinuous. We then invoke Kakutani's fixed point theorem and obtain our result. 
\end{proof}

\

For the remainder of this section, we suppose Assumptions~\ref{ass:unique-invariant-distribution} and \ref{ass:cost-continuous-in-measure} hold, and we fix $(\pi^{*}, \nu_{\pi^{*}}) \in \Pi_{\bM, S} \times \Delta ( \xx )$  to be a stationary mean-field equilibrium.

\subsection{A Family of Partially-Observed $n$-Player Mean-Field Games} 

To state the approximation result of this section, in which SMFE is used to obtain an objective $\epsilon$-equilibrium in an $n$-player game when $n$ is sufficiently large, we now introduce a family $\newG$ of partially observed $n$-player mean-field games with local state observability. This family of games will be indexed both by the number of players, $n \in \nn$, as well as mean-field terms relevant to the initial distribution of the local states, $\nu \in \Delta( \xx )$.

For the analysis, it will also be necessary to introduce a second family of partially observed $n$-player mean-field games with local state observability. This second family of $n$-player games is denoted by $\bar{\newG}$.

For each $n \geq 1$ and $\nu \in \Delta( \xx )$, we define a partially observed $n$-player mean-field game 
\[
\GG_n ( \nu  ) = \left( \NN_{(n)}, \xx_{(n)}, \yy_{(n)}, \aa_{(n)},   \{ \varphi^i_{(n)} \}_{i \in \NN_{(n)} }, c_{(n)}, P^{(n)}_{\loc}, \gamma_{(n)}, \bnu^{(n)}  \right) .
\]
Here, $\NN_{(n)} = \{ 1, \dots, n \}$ is a set of $n$ players, and we use $(n)$ to note that the objects in the definition of $\GG_n ( \nu )$ depend on the number of players. We put $\xx_{(n)} = \xx$, $\yy_{(n)} = \xx$, $\aa_{(n)} = \aa$, $c_{(n)} = c$, $P^{(n)}_{\loc} = P_{\loc}$, $\gamma_{(n)} = \gamma$, and note that $\xx$, $\aa$, $c$, $P_{\loc}$, and $\gamma$ retain their meanings from the mean-field system $G$. The set of global states is therefore $\xx^n$, and the observation functions $\varphi^i_{(n)}: \xx^n \to \yy_{(n)}$ are given by $\varphi^i_{(n)} ( x^1 , \dots , x^n )  =  x^i$ for each $i \in \NN_{(n)}$ and $(x^1, \dots, x^n ) \in \xx^n$.

For the approximation results to follow, it will not be necessary to discuss $n$-player games with arbitrary initial distributions of global states. Our claims and analysis will be restricted to the special case where the local states of all agents are independently and identically distributed. We therefore let $\bnu^{(n)} \in \Delta( \xx^n )$ denote the product distribution with marginals given by $\nu$.

For any $n \in \nn$ and player $i \in \NN_n$, we let $\Pi^i_{(n)}$ denote player $i$'s set of policies for the game $\GG_n ( \nu )$. We note that $\Pi^i_{(n)}$ is in natural bijection with $\Pi_{\bM}$,  the set of admissible policies for any mean-field MDPs in the family $\bM$. The set of stationary policies for player $i \in \NN_n$ is denoted by $\Pi^i_{(n), S}$ and is in bijection with the set $\Pi_{\bM, S}$ of stationary policies for the mean-field MDPs in $\bM$. We then denote the joint policy set for $\GG_n ( \nu )$ by $\bPi^{(n)}$, and we let $\bPi^{(n)}_{S}$ denote the set of joint stationary policies for $\GG_n ( \nu )$. 

\

\noindent{}The family of $n$-player mean-field games $\newG$ is defined as $\newG = \left\{ \GG_n ( \nu ) : n \in \nn, \nu \in \Delta( \xx ) \right\}$. 

\

To distinguish state and action variables associated with different games in the family $\newG$, we index local states, local actions, and mean-field states with $(n)$: for instance, the local state of player $i \in \NN_n$ at time $t$ in the game $\GG_n ( \nu )$ is denoted $x^{i, (n)}_t$. Actions and mean-field states are similarly denoted $a^{i , (n)}_t$ and $\mu^{(n)}_t$. Dependence on $\nu$ will be reflected in the expectation operator and objective function, and is not included in the random variables.

The objective function for player $i \in \NN_n$ in the game $\GG_n ( \nu )$ is denoted by
\[
J^i_{(n)} \left( \pi^i_{(n)} , \bpi^{-i}_{(n)} , \bnu^{(n)}  \right)   = E^{\bpi_{(n)}}_{\bnu^{(n)}} \left[ \sum_{ t= 0}^{\infty} \gamma^t c \left( x^{i, (n)}_t , a^{i, (n)}_t , \mu^{(n)}_t \right)  \right], 
\]
for every $n \in \nn$, $i \in \NN_n$, $\bpi_{(n)} \in \bPi^{(n)}$, and $\nu \in \Delta(\xx)$.

\

\noindent Finally, we introduce special notation to capture the performance of a policy $\pi \in \Pi_{\bM}$ used in the game $\GG_n ( \nu )$ while all other agents in $\NN_n$ follow a fixed stationary policy. Let $(\pi^{*}, \nu_{\pi^{*}}) \in \Pi_{\bM, S} \times \Delta ( \xx )$ be the stationary mean-field equilibrium fixed after Lemma~\ref{lemma:SMFE}.  For each $n \geq 1$, we fix player $1 \in \NN_n$ and let $\bpi^{*-1}_{(n)}$ denote the $\left(n-1\right)$-player joint policy according to which all agents $j \not= 1$ use policy $\pi^{*}$. Then, we define
\[
\newJ^{(n)}_{\pi} ( \nu_{\pi^{*}} ) := J^{1}_{(n)} \left(  \pi, \bpi^{*-1}_{(n)}  ,  \bnu^{(n)}_{\pi^{*}} \right) , \quad \forall n \geq 1, \pi \in \Pi_{\bM}. 
\]

\subsection{A Second Family of Partially-Observed $n$-Player Mean-Field Games} 

For approximation purposes, we must introduce a second family of partially observed $n$-player mean-field games with local observability. The second family is denoted by $\bar{\newG}$, and game in this family will serve to approximate games in the family $\newG$. 

First, recall that for each $\nu \in \Delta ( \xx )$, we defined the transition kernel $P_{\nu} \in \PP ( \xx | \xx \times \aa)$ by $P_{\nu} ( \cdot | s,a ) = P_{\loc} ( \cdot | s, \nu , a )$ for every $(s,a) \in \xx \times \aa$. Now, we extend $P_{\nu}$ to obtain a transition kernel $\bar{P}_{\nu} \in  \PP ( \xx |  \xx \times \Delta(\xx) \times \aa)$: for each $\nu \in \Delta(\xx)$, define
\[
\bar{P}_{\nu} ( \cdot | s, \tilde{\nu}, a ) = P_{\nu} ( \cdot | s, a ), \quad \forall \tilde{\nu} \in \Delta(\xx) .
\]

For each $n \geq 1$ and $\nu \in \Delta ( \xx )$, we define a partially observed $n$-player mean-field game $\bar{\GG}_n ( \nu )$ with local observability: 
\[
\bar{\GG}_n ( \nu ) := \left( \NN_{(n)}, 	\xx_{(n)}, 	\yy_{(n)}, 		\aa_{(n)}, 	\{ \varphi^i_{(n)} \}_{i \in \NN_{(n)} }, 	c_{(n)}, 	\bar{P}_{\nu}, 		\gamma_{(n)}, 	\bnu^{(n)}  \right) .
\]

In the list above, the objects $\NN_{(n)}, \xx_{(n)}, \yy_{(n)}, \aa_{(n)}, \{ \varphi^i_{(n)} \}_{i \in \NN_{(n)} }, c_{(n)}$, $\gamma_{(n)}$, and $\bnu^{(n)}$ retain their meaning from the definition of $\GG_n ( \nu )$. The major difference between the definition of $\GG_n ( \nu ) $ and $\bar{\GG}_n ( \nu )$ is the transition kernel: in $\bar{\GG}_n ( \nu )$, the transition kernel has $\nu$ fixed as its mean-field term, and local state transitions do not depend on the empirical distribution of the population. We return to this point below.

As in the case of $\GG_n ( \nu )$, the game $\bar{\GG}_n ( \nu )$ has local observability, and the global state space is $\xx^n$. The joint action set is $\aa^n$. As before, the set of policies for a given player $i \in \NN_n$ coincide with the set of policies $\Pi_{\bM}$ for the mean-field MDPs in $\bM$. Due to this natural bijection, we use the symbols $\Pi^i_{(n)}$, $\Pi^{i}_{(n),S}$, $\bPi^{(n)}$, and $\bPi^{(n)}_{S}$, retaining their meanings from before.

\ 

\noindent The family of games $\bar{\newG}$ is given by $\bar{\newG} := \left\{ \bar{\GG}_n ( \nu ) : n \in \nn, \nu \in \Delta( \xx ) \right\}$. 

\

We now describe the construction of probability measures on sequences of global states and joint actions that describe probabilities for games in $\bar{\newG} := \{ \bar{\GG}_n ( \nu ) \}_{n , \nu }$. To avoid mixing notation between analogous games $\GG_n ( \nu ) $ and $\bar{\GG}_n ( \nu )$,  we employ the symbol ``{ }$\bar{ }${ }'' to identify that a quantity refers to $\bar{\GG}_n ( \nu ) $. That is, a sequence in $\xx^n \times \aa^n$ of global states and joint actions obtained from a game in $\bar{\newG}$ is expressed as
\[
\left( \bar{\bx}^{(n)}_t , \bar{\ba}^{(n)}_t \right)_{t \geq 0}, \text{ where } \left( \bar{\bx}^{(n)}_t, \bar{\ba}^{(n)}_t \right) =  \left( \bar{x}^{i, (n)}_t, \bar{a}^{i, (n)}_t \right)_{i \in \NN_n }, \: \forall \, t \geq 0.
\]
We define the empirical distribution of local states at time $t$ in the $n$-player game $\bar{\GG}_n ( \nu )$ in the natural way and denote it $\{ \bar{\mu}^{(n)}_t  : t \geq 0 \}$.

For $t \geq 0$, we use 
\[
\bar{\bh}^{(n)}_t = \left( \bar{\bx}^{(n}_0, \bar{\ba}^{(n)}_0, \dots, \bar{\bx}^{(n)}_{t-1}, \bar{\ba}^{(n)}_{t-1}, \bar{\bx}^{(n)}_t  \right)
\]
to denote the system history at time $t$, and for each $i \in \NN_n$ we let $\bar{h}^{i, (n)}_t$ denote player $i$'s suitably defined locally observable history. As before, the initial distribution $\bnu^{(n)} \in \Delta ( \xx^n )$ is defined such that $\{ \bar{x}^{i, (n)}_0 \}_{i \in \NN_n}$ is an i.i.d. family of random variables with common distribution $\nu \in \Delta ( \xx )$.

For each $\bpi \in \bPi^{(n)}$, $\overline{\P}^{ \bpi_{(n)} }_{\nu}$ is the unique canonically defined probability measure on $\left( \xx^n \times \aa^n \right)^{\infty}$ such that:
\begin{itemize}
	\item $\overline{\P}^{ \bpi_{(n)} }_{\nu} \left( \bar{\bx}^{(n)}_0 \in \cdot \right) = \bnu^{(n)} ( \cdot )$; 
	
	\item For every $i \in \NN_n$ and $t \geq 0$, $\overline{\P}^{ \bpi_{(n)} }_{\nu} \left( \bar{a}^{i, (n)}_t \in \cdot \middle| \bar{h}^{i, (n)}_t \right) = \pi^i_t ( \cdot | \bar{h}^{i, (n)}_t )$, and the collection $\{ \bar{a}^{j , (n)}_t \}_{j \in \NN_n }$ is jointly independent given $\bar{\bh}^{(n)}_t$;
	
	\item For every $i \in \NN_n$ and $t \geq 0$, $\bar{x}^{i, (n)}_{t+1} \sim \bar{P}_{\nu} \left( \cdot \middle| \bar{x}^{i, (n)}_t , \bar{\mu}^{(n)}_t , \bar{a}^{i, (n)}_t \right)$, and the collection $\{ \bar{x}^j_{t+1} \}_{j \in \NN_n }$ is jointly independent given $( \bar{\bh}^{(n)}_t, \bar{\ba}^{(n)}_t)$.

\end{itemize}

\ 

For each $n \geq 1$, $\nu \in \Delta (  \xx )$, player $i \in \NN_n$, and joint policy $\bpi \in \bPi^{(n)}$, we define player $i$'s objective function in the game $\bar{\GG}_n ( \nu )$ under joint policy $\bpi$ as 
\[
\bar{J}^{i}_{(n)} \left( \bpi  , \bnu^{(n)} \right) :=  \bar{E}^{ \bpi_{(n)} }_{\nu} \left[ \sum_{t \geq 0} \gamma^t c \left( \bar{x}^{i, (n)}_t, \bar{\mu}^{(n)}_t, \bar{a}^{i, (n)}_t \right) \right] .
\]
where $\bar{E}^{ \bpi_{(n)} }_{\nu}$ is the expectation associated with $\overline{\P}^{ \bpi_{(n)} }_{\nu}$. 

\ \\

\noindent\textbf{Remark:} Due to our definition of $\bar{P}_{\nu}$, we have that $\bar{x}^{i, (n)}_{t+1} \sim  \bar{P}_{\nu} \left( \cdot \middle| \bar{x}^{i, (n)}_t , \bar{\mu}^{(n)}_t , \bar{a}^{i, (n)}_t \right)$ is equivalent to $\bar{x}^{i, (n)}_{t+1} \sim  P_{\loc} \left( \cdot \middle| \bar{x}^{i, (n)}_t ,  \nu , \bar{a}^{i, (n)}_t \right)$. This distinction is crucial to the analysis below. In the game $\bar{\GG}_n ( \nu )$, the local state transitions for each player $i \in \NN_n$ are totally uncoupled from the mean-field state. Although each player's local state evolution is uncoupled from the mean-field state in the game $\bar{\GG}_n ( \nu )$, player $i$'s objective nevertheless depends on the mean-field sequence $\{ \bar{\mu}^{(n)}_{t} \}_{t \geq 0}$, which appears in the stage costs. 

\

Finally, for each $n \geq 1$ and policy $\pi \in \Pi_{\bM}$, we fix agent $1 \in \NN_n$, we define 
\[
\bar{\newJ}^{(n)}_{\pi} \left( \nu_{\pi^{*}} \right) := \bar{J}^{1}_{(n)} \left( \pi, \bpi^{*-1}_{(n)} , \bnu^{(n)}_{\pi^{*}} \right)  , \quad \forall n \geq 1, \pi \in \Pi_{\bM}, 
\]
where $( \pi^{*}, \nu_{\pi^{*}} )$ is the same SMFE as was used in the definition of $\newJ^{(n)}_{\pi} ( \nu_{\pi^{*}} )$, and $\bpi^{*-1}_{(n)}$ is the same joint policy as appeared in that definition.

\subsection{From Stationary Mean-Field Equilibrium to Objective Equilibrium in an $n$-Player Game}

We now proceed to the main result of this section, in which we show that for fixed $\epsilon > 0$, if $n$ is sufficiently large, then the joint policy $\bpi^{*}_{(n)} := ( \pi^{*}, \cdots, \pi^{*} ) \in \bPi^{(n)}$ is an $\epsilon$-equilibrium with respect to $\bnu^{(n)}_{\pi^{*}}$ for the game $\GG_n ( \nu_{\pi^{*}} )$. This result requires an additional assumption on the transition kernel $P_{\loc}$, as well as two intermediate lemmas to facilitate the analysis.

In what follows, we suppose Assumptions~\ref{ass:unique-invariant-distribution} and \ref{ass:cost-continuous-in-measure} hold, and we let $( \pi^{*}, \nu_{\pi^{*}} )$ denote the stationary mean-field equilibrium used in the preceding definitions (e.g. the definition of $\newJ^{(n)}_{\pi} ( \nu_{\pi^{*}})$).

\vspace{5pt}

\begin{assumption} \label{ass:uniform-continuity-of-transition-kernel}
There exists a function $F : \xx \times \Delta ( \xx ) \times \aa \times [0,1] \to \xx$ and a probability measure $\LL$ on $[0,1]$ such that
\[
P_{\loc} ( s' | s, \nu, a ) = \int_{[0,1]} \textbf{1}_{\{ \xi \in [0,1] : F (  s, \nu, a, \xi ) = s' \} } ( \tau ) \, d \LL ( \tau  ), 
\]
for each $s' \in \xx $ and every $(s, \nu, a ) \in \xx \times \Delta( \xx ) \times \aa$. Furthermore, there exists $\rho > 0 $ such that if $\| \nu - \nu_{\pi^{*}} \|_{\infty} < \rho$, then $F ( s, \nu, a , \xi ) = F ( s, \nu_{\pi^{*}}, a, \xi )$ holds for any $(s,a,\xi ) \in \xx \times \aa \times [0,1]$. 
\end{assumption}

\begin{lemma}  \label{lemma:real-vs-fake-game}
Let $\epsilon > 0$ and suppose Assumptions~\ref{ass:unique-invariant-distribution}, \ref{ass:cost-continuous-in-measure}, and \ref{ass:uniform-continuity-of-transition-kernel} hold. There exists $N ( \epsilon ) \in \nn$ such that if $n \geq N ( \epsilon )$, then for any policy $\pi \in \Pi_{\bM}$, we have 
\[
| \newJ^{(n)}_{\pi} ( \nu_{\pi^{*}} ) - \bar{\newJ}^{(n)}_{\pi} ( \nu_{\pi^{*}} ) | < \epsilon / 4 .
\]
\end{lemma}

\

\begin{proof} We use a coupling argument: for each $\pi \in \Pi_{\bM}$ and $n \geq 1$, we construct a probability space $( \Omega_{n, \pi}, \FF_{n, \pi} , \prob^{n, \pi} )$ on which both $\GG_n ( \nu_{\pi^{*}} )$ and $\bar{\GG}_n ( \nu_{\pi^{*}} )$ are defined, where agent $ 1 \in \NN_n$ uses policy $\pi$ and the remaining agents all use the stationary policy $\pi^{*}$.  The component projections give rise to the appropriate marginal probability measures, i.e. $\P^{ \bpi_{(n)} }_{\bm{\nu}_{\pi^{*}}}$ and $\overline{\P}^{ \bpi_{(n)} }_{\bm{\nu}_{\pi^{*}}}$ with suitably defined policy $\bpi \in \bPi^{(n)}$, but are otherwise closely coupled.

\

Let $\Omega_{n, \pi} := \left(  \xx^n \times \xx^n  \times[0,1]^n \times \aa^n  \times \aa^n \times [0,1]^n \right)^{\infty}$, and let $\FF_{n, \pi}$ be the $\sigma$-algebra generated by finite dimensional cylinders in $\Omega_{n, \pi}$. Components of $\Omega_{n, \pi}$ will be denoted by sequences $( \bar{\bx}^{(n)}_t, \bx^{(n)}_t, \bm{\eta}^{(n)}_t, \bar{\ba}^{(n)}_t, \ba^{(n)}_t, \bm{\xi}^{(n)}_t )_{t \geq 0}$. The probability measure $\prob^{n, \pi}$ is constructed as follows:

\begin{itemize}
	\setlength \itemsep{0.5em}

	\item For each $j \in \NN_n$, $\prob^{n, \pi} \left( \bar{x}^{j, (n)}_0 \in \cdot \right) =  \nu_{\pi^{*}} ( \cdot )$, and the collection $\left\{ \bar{x}^{j, (n)}_{0} \right\}_{j \in \NN_n }$ is jointly independent.
	
	\item $\prob^{n, \pi} \left( \bx^{(n)}_0 = \bar{\bx}^{(n)}_0 \right) = 1$.

	\item Let $\KK$ be a probability measure on $[0,1]$, let $g : \left( \cup_{t \geq 0} H_t  \right) \times [0,1] \to \aa$, and let $g^{*} : \xx \times [0,1] \to \aa$ satisfy
		\[
		\pi_t ( a | h_t ) = \int_{[0,1]} \textbf{1}_{ \{ \xi \in [0,1] : g ( h_t , \xi ) = a \} } ( \tau ) \, d \KK ( \tau ), \quad \forall t \geq 0, h_t \in H_t, a \in \aa, 
		\]
	\item[] where $\pi = (\pi_t )_{t \geq 0}$, and
		\[
		\pi^{*} ( a | s ) = \int_{ [0,1] } \textbf{1}_{ \{ \xi \in [0,1] :  g^{*} ( s, \xi ) = a \} } ( \tau ) \, d \KK ( \tau ), \quad \forall s \in \xx , a \in \aa. 
		\]
	
	\item For any $t \geq 0$, we have $\prob^{n, \pi} \left( \eta^{j, (n)}_t \in \cdot \right) = \KK ( \cdot )$ for all $j \in \NN_n$, the collection $\{ \eta^{j , (n)}_t \}_{j \in \NN_n }$ is jointly independent, and $\bm{\eta}^{(n)}_t$ is independent of the history up to $\bx_t^{(n)}$.
	
	\item For $j \not= 1$ and $t \geq 0$, we have $\bar{a}^j_t = g^{*} \left( \bar{x}^j_t, \eta^{j, (n)}_t \right)$ and $a^j_t = g^{*} \left( x^j_t, \eta^{j, (n)}_t \right)$.
	
	\item For agent $1 \in \NN_n$ and $t \geq 0$, we have $\bar{a}^{1, (n)}_t = g \left(  \bar{h}^{1, (n)}_t , \eta^{1, (n)}_t \right)$ and $a^{1, (n)}_t = g \left( h^{1, (n)}_t , \eta^{1, (n)}_t \right)$. 
	
	\item For any $t \geq 0$ and agent $j \in \NN_n$, we have $\prob^{n, \pi} \left( \xi^{j , (n)}_t \in \cdot \right) = \LL ( \cdot)$, where $\LL$ is the probability measure introduced in Assumption~\ref{ass:uniform-continuity-of-transition-kernel}. Furthermore, the collection $\{ \xi^{j , (n)}_t \}_{j \in \NN_n }$ is jointly independent, and $\bm{\xi}^{(n)}_t$ is independent of the history up to $\ba^{(n)}_t$;
	
	\item For every $t \geq 0$ and agent $j \in \NN_n$, we have 
		\[
		\bar{x}^{j, (n)}_{t+1} = F ( \bar{x}^j_t, \nu_{\pi^{*}} , \bar{a}^{j , (n)}_t , \xi^{j ,(n)}_t ) \text{ and } x^{j, (n)}_{t+1} = F ( x^{j, (n)}_t, \mu^{(n)}_t , a^{j, (n)}_t, \xi^{ j, (n)}_t ) ,
		\]
		
	\item[] where $\mu^{(n)}_t$ is defined as the empirical measure of $\bx^{(n)}_t$. 
	
\end{itemize}

The measure $\prob^{n, \pi}$ is then the unique extension of the conditional probabilities above to $( \Omega_{ n, \pi}, \FF_{ n, \pi} ) $. We now observe some consequences of this construction. 

\ 

First, observe that the state transitions resulting in $\{ \bar{x}^{j, (n)}_{t+1} \}_{j \in \NN_n}$ are uncoupled. For any $\pi \in \Pi_{\bM}$, $n \geq 1$, and $j \not= 1 \in \NN_n$, since player $j$ uses policy $\pi^{*}$, since $\bar{x}^{j, (n)}_0 \sim \nu_{\pi^{*}}$, and since $\Phi ( \nu_{\pi^{*}}, \pi^{*} ) = \nu_{\pi^{*}}$, we have that (1) $\bar{x}^{j, (n)}_t \sim \nu_{\pi^{*}}$ for every $t \geq 0$; (2) the collection $\{ \bar{x}^{j , (n)}_t \}_{ j \in \NN_n }$ is independent; (3) the distribution of $\{ \bar{x}^{j , (n)}_t \}_{ j \in \NN_n }$ does not depend on the policy $\pi$ for player $1 \in \NN_n$. Together, this implies that for any $t \geq 0$ and $\epsilon' > 0$, we have
\[
\lim_{n \to \infty} \prob^{n, \pi} \left( \| \bar{\mu}^{(n)}_t - \nu_{\pi^{*}} \|_{\infty} < \epsilon' \right) = 1 , \quad \forall \pi \in \Pi_{\bM}.
\]

In particular, this holds for $\epsilon' = \rho$, where $\rho > 0$ is the radius introduced in Assumption~\ref{ass:uniform-continuity-of-transition-kernel}.

\ \\
We fix $T \in \nn$ such that $\frac{\gamma^T  \| c \|_{\infty} }{ 1 - \gamma } < \frac{\epsilon}{ 32 }$. By the preceding remarks, we have that for any $\epsilon_1 > 0 $, there exists $N_1 ( \epsilon_1 ) \in \nn$ such that if $n \geq N_1 ( \epsilon_1 )$, we have
\begin{equation} \label{eq:epsilon_1} 
\prob^{n, \pi} \left( \bigcap_{t = 0}^{T-1} \{ \| \bar{\mu}^{(n)}_t  - \nu_{\pi^{*}} \|_{\infty} < \rho \} \right) \geq 1 - \epsilon_1   , \quad \forall \pi \in \Pi_{\bM} .
\end{equation}

We now argue that the preceding inequalities enable the following claim: for any $\epsilon_2 > 0$, there exists $N_2 ( \epsilon_2 ) \in \nn$ such that if $n \geq N_2 ( \epsilon_2 )$, then
\begin{equation} \label{eq:epsilon_2}
\prob^{n , \pi } \left( \bigcap_{t = 0}^{T-1} \left\{ \bar{\bx}^{(n)}_t = \bx^{(n)}_t , \bar{\ba}^{(n)}_t = \ba^{(n)}_t \right\} \right) \geq 1 - \epsilon_2 , \quad \forall \pi \in \Pi_{\bM}. 
\end{equation}

To see this, we take $\tau > 0 $ such that $1 - \tau > \left( \max\{ \frac{1}{2} , 1 - \epsilon_2 \} \right)^{1/T}. $ From the discussion above, there exists $N_1 ( \tau / 2 ) \in \nn$ such that if $n \geq N_1(\tau / 2)$, then \eqref{eq:epsilon_1} holds with $\epsilon_1 = \tau  /2$. For each $k \in \{ 0, 1, \dots, T-1 \}$, let  ${\rm Event} (k , n ) := \{ \bar{\bx}^{(n)}_k   = \bx^{(n)}_k , \bar{\ba}^{(n)}_k = \ba^{(n)}_k \}$. Observe the following three facts, which hold for any $t \in \{ 0 ,1, \dots, T - 1 \}$:
\begin{enumerate}
	
	\item By construction,
		\[
		\prob^{n, \pi} \left( {\rm Event} (t+1, n) \middle| \cap_{k = 0}^t {\rm Event} (k,  n ) , \left\| \bar{\mu}^{(n)}_t - \nu_{\pi^{*}} \right\|_{\infty} < \rho \right) = 1.
		\]
	
	\item If $\prob^{ n , \pi } \left( \cap_{k = 0}^{t} {\rm Event} ( k , n ) \right) \geq ( 1-  \tau )^t$, then by our choice of $\tau$ and $t \leq T$, this implies $\prob^{ n , \pi } \left( \cap_{k = 0}^{t} {\rm Event} ( k , n ) \right) \geq \frac{1}{2}$. Then, by our choice of $n \geq N_1 ( \tau/2)$, we have $\prob^{ n , \pi} \left( \left\| \bar{\mu}^{(n)}_t - \nu_{\pi^{*}} \right\|_{\infty} < \rho \right) \geq 1 - \tau/2$. Putting the two together,\footnote{Here, we have used the fact that if events $A$ and $B$ satisfy $\Pr ( B ) \geq b > 0$, $\P(A) \geq 1 - ab $ with $a > 0$, then $\P ( A | B ) \geq 1 - a$.}, we obtain
		\[
		\prob^{n , \pi } \left( \left\| \bar{\mu}^{(n)}_t - \nu_{\pi^{*}} \right\|_{\infty} < \rho \middle| \bigcap_{k = 0}^t {\rm Event} ( k, n ) \right) \geq 1 - \tau .
		\]
		
	\item If $\prob^{ n , \pi } \left( \cap_{k = 0}^{t} {\rm Event} ( k , n ) \right) \geq ( 1-  \tau )^t$, then by the previous two items, we have
		\[
		\prob^{ n, \pi } \left(  {\rm Event} ( t + 1 , n ) \middle| \bigcap_{k = 0}^t {\rm Event} ( k , n ) \right) \geq ( 1-  \tau)^{t + 1 }.
		\]
		
	\item By construction, we have that the base case for item 2 holds with $t = 0$. 
 	Thus, we repeatedly apply item 3 to see that, indeed, the inequality of \eqref{eq:epsilon_2} holds for $n \geq N_2 ( \epsilon_2 ) := N_1 ( \tau / 2)$.
	
\end{enumerate}

Now, we argue that the existence result summarized in \eqref{eq:epsilon_2} implies the result. For each $n \geq 1$ and $\pi \in \Pi_{\bM}$, let $\ee^{n , \pi}$ be the expectation associated with $\prob^{n, \pi}$. Observe that, for any $n \geq 1$ and $\pi \in \Pi_{\bM}$, we have
\[
\newJ^{(n)}_\pi ( \nu_{\pi^{*}} ) = \ee^{n , \pi } \left[ \sum_{t = 0 }^{\infty} \gamma^t  c  \left(  x^{ 1, (n)}_t, \mu^{( n)}_t, a^{ 1, (n)}_t  \right) \right] ,  
\]
and 
\[
\bar{\newJ}^{(n)}_\pi ( \nu_{\pi^{*}} ) = \ee^{n , \pi } \left[ \sum_{t = 0 }^{\infty} \gamma^t  c  \left(  \bar{x}^{ 1, (n)}_t, \bar{\mu}^{( n)}_t, \bar{a}^{ 1, (n)}_t \right) \right] .
\]

Then, for each $n \geq 1, \pi \in \Pi_{\bM}$, we let $\newJ^{ (n), T }_{\pi} ( \nu_{\pi^{*}} )$ and $\bar{\newJ}^{(n), T }_{\pi} ( \nu_{\pi^{*}} )$ denote the associated series truncated after the summand with index $T-1$. That is, 
\[
\newJ^{ (n), T }_{\pi} ( \nu_{\pi^{*}} ) := \ee^{n , \pi } \left[ \sum_{t = 0 }^{T-1} \gamma^t c  (  x^{ 1, (n)}_t, \mu^{( n)}_t, a^{ 1, (n)}_t ) \right] ,
\]
and $\bar{\newJ}^{(n), T }_{\pi} ( \nu_{\pi^{*}} )$ is defined analogously. 

\

By our choice of $T$ to satisfy $\frac{ \gamma^T \| c \|_{\infty} }{ 1 - \gamma } < \frac{\epsilon}{32}$, we have the following for any $n \geq 1, \pi \in \Pi_{\bM}$:
\[
\left| 	\newJ^{(n)}_{\pi } (\nu_{\pi^{*}})  - 	\newJ^{(n) , T}_{\pi } (\nu_{\pi^{*}})  \right| < \frac{ \epsilon } { 16}   \text{ and } \left| 	\bar{\newJ}^{(n)}_{\pi } (\nu_{\pi^{*}})  - 	\bar{\newJ}^{(n) , T}_{\pi } (\nu_{\pi^{*}})  \right| < \frac{ \epsilon } { 16}  .
\]

Applying the triangle inequality twice, we then obtain
\[
\left| 	\newJ^{(n)}_{\pi } (\nu_{\pi^{*}})  - 	\bar{\newJ}^{(n) }_{\pi } (\nu_{\pi^{*}})  \right| \leq \frac{ \epsilon}{8} + \left| 	\newJ^{(n),T}_{\pi } (\nu_{\pi^{*}})  - 	\bar{\newJ}^{(n) , T}_{\pi } (\nu_{\pi^{*}})  \right| . 
\]

By the deductions above culminating in \eqref{eq:epsilon_2}, we have that for any $\epsilon_3 > 0$, there exists $N_3 ( \epsilon_3 )$ such that if $n \geq N_3 ( \epsilon_3)$, we have 
\[
\left| 	\newJ^{(n),T}_{\pi } (\nu_{\pi^{*}})  - 	\bar{\newJ}^{(n) , T}_{\pi } (\nu_{\pi^{*}})  \right| < \epsilon_3.
\]

In particular, this is true for $\epsilon_3 = \frac{\epsilon}{8}$. This proves the result, with $N ( \epsilon ) = N_3 ( \epsilon/8)$. 
\end{proof}

For the next approximation result, recall that for $\nu \in \Delta (\xx)$, $\JJ_{\MM_{\nu}} ( \pi , \nu^{\prime} )$ is the objective performance of policy $\pi \in \Pi_{\bM}$ when controlling the MDP $\MM_{\nu}$ with initial state distributed according to $\nu^{\prime}$. In the lemma below, we fix both $\nu = \nu^{\prime} = \nu_{\pi^{*}}$ to be the probability distribution appearing in the stationary mean-field equilibrium $( \pi^{*}, \nu_{\pi^{*}} )$.

\begin{lemma} \label{lemma:MDP-vs-fake-game} 
Let $\epsilon > 0$ and suppose Assumptions~\ref{ass:unique-invariant-distribution}, \ref{ass:cost-continuous-in-measure}, and \ref{ass:uniform-continuity-of-transition-kernel} hold. There exists $N ( \epsilon ) \in \nn$ such that if $n \geq N ( \epsilon )$, then for any policy $\pi \in \Pi_{\bM}$, we have 
\[
| \JJ_{\MM_{\nu_{\pi^{*}}}} ( \pi ,  \nu_{\pi^{*}} ) - \bar{\newJ}^{(n)}_{\pi} ( \nu_{\pi^{*}} ) | < \epsilon / 4 .
\]
\end{lemma}

\begin{proof}
For each $n \geq 1$ and $\pi \in \Pi_{\bM}$, recall the probability spaces $( \Omega_{n ,\pi}, \FF_{n, \pi}, \prob^{n, \pi} )$ used in the coupling argument of the proof of Lemma~\ref{lemma:real-vs-fake-game} and recall the definition of $T \in \nn$. For the MDP $\MM_{\nu_{\pi^{*}}}$, we let $\P^{\pi, \infty}_{\nu_{\pi^{*}}}$ denote the probability measure on sequences of local states and actions when the MDP is controlled by $\pi$. 

We note that, by construction of $( \Omega_{n ,\pi}, \FF_{n, \pi}, \prob^{n, \pi} )$, the marginal distributions on (local) state-action trajectories for player $1 \in \NN_n$ are equal to those in the MDP $\MM_{\nu_{\pi^{*}}}$ controlled by policy $\pi$. That is,
\[
\prob^{n, \pi} \left( \{ \bar{x}^{ 1, ( n)}_t, \bar{a}^{1, (n)}_t \}_{t = 0}^{T-1} \in \cdot \right) = \P^{\pi, \infty}_{\nu_{\pi^{*}}} \left( \{ x_t , a_t\}_{t =0}^{T-1} \in \cdot \right). 
\] 

Thus, the expected costs $\JJ_{\MM_{\nu_{\pi^{*}}}} ( \pi,  \nu_{\pi^{*}} )$ and $\bar{\newJ}^{(n)}_{\pi} ( \nu_{\pi^{*}} )$ differ in the mean-field term, which is posited to be constant at $\nu_{\pi^{*}}$ in the former, while in the latter the mean-field sequence $\{ \bar{\mu}^{(n)}_t \}_{t \geq 0}$ is random. 

For each $n \geq 1$ and $\pi \in \Pi_{\bM}$, we recall the object $\bar{\newJ}^{(n), T }_{\pi} ( \nu_{\pi^{*}} )$ from the proof of Lemma~\ref{lemma:real-vs-fake-game}. We then define an analogous quantity involving the MDP $\MM_{\nu_{\pi^{*}}}$ controlled by policy $\pi$:
\[
\JJ^T_{\MM_{\nu_{\pi^{*}}}} ( \pi,  \nu_{\pi^{*}} ):= E^{\pi, \infty}_{\nu_{\pi^{*}}} \left[ \sum_{t = 0}^{T-1}  \gamma^t c ( x_t , \nu_{\pi^{*}} , a_t ) \right]. 
\]

\noindent By our choice of $T$, $\left| \bar{\newJ}^{(n)}_{\pi} (\nu_{\pi^{*}}) - \bar{\newJ}^{(n),T}_{\pi} (\nu_{\pi^{*}} )	\right| < \frac{ \epsilon}{16}$ and $\left| \JJ_{\MM_{\nu_{\pi^{*}}}} ( \pi,  \nu_{\pi^{*}} ) - \JJ^T_{\MM_{\nu_{\pi^{*}}}} ( \pi,  \nu_{\pi^{*}} )    \right| < \frac{ \epsilon}{16}$, thus 
\[
\left| \bar{\newJ}^{(n)}_{\pi} (\nu_{\pi^{*}})   -  \JJ_{\MM_{\nu_{\pi^{*}}}} ( \pi,  \nu_{\pi^{*}} )	\right| \leq \frac{ \epsilon}{8} + \left|   \bar{\newJ}^{(n),T}_{\pi} (\nu_{\pi^{*}} )  -  \JJ^T_{\MM_{\nu_{\pi^{*}}}} ( \pi,  \nu_{\pi^{*}} )    \right|.
\]

\ \\
As we argued in the proof of Lemma~\ref{lemma:real-vs-fake-game}, for any $\xi_1, \xi_2 > 0$, there exists $N = N ( \xi_1, \xi_2)  \in \nn$ such  that if $n \geq N$, we have 
\[
\prob^{n, \pi} \left( \bigcap_{t =0}^{T-1} \left\{ \left\| \bar{\mu}^{(n)}_t - \nu_{\pi^{*}} \right\|_{\infty} < \xi_1 \right\} \right) \geq 1 - \xi_2 , \quad \forall \pi \in \Pi_{\bM}. 
\]

By Assumption~\ref{ass:cost-continuous-in-measure}, the mapping $\nu \mapsto c ( s, \nu, a )$ is continuous on $\Delta ( \xx )$ for all $(s,a) \in \xx \times \aa $. Thus, there exists $\xi > 0 $ such that $\| \nu - \nu_{\pi^{*}} \|_{\infty} < \xi$ implies $| c(s,\nu, a ) - c ( s, \nu_{\pi^{*}} , a ) | \leq \frac{\epsilon}{8T}$ for any $(s,a) \in \xx \times \aa$. The result then follows another mechanical computation. 
\end{proof}

\begin{theorem} \label{theorem:large-N-equilibrium}
Let $(\pi^{*}, \nu_{\pi^{*}})$ be a stationary mean-field equilibrium for $( \xx , \aa, c, P_{\loc}, \gamma)$, and suppose Assumptions~\ref{ass:unique-invariant-distribution}, \ref{ass:cost-continuous-in-measure}, and \ref{ass:uniform-continuity-of-transition-kernel} hold. Let $\epsilon > 0$ be given. There exists $N ( \epsilon ) \in \nn$ such that if $n \geq N ( \epsilon )$, then, in the game $\GG_n ( \nu_{\pi^{*}} )$, the policy $\bpi^{*}_{(n)} \in \bPi^{(n)}_{S}$, in which every agent uses $\pi^{*}$, is an $\epsilon$-equilibrium with respect to $\bm{\nu}_{\pi^{*}}$. 
\end{theorem}

\begin{proof}
By Lemmas~\ref{lemma:real-vs-fake-game} and \ref{lemma:MDP-vs-fake-game}, there exists $\bar{N} ( \epsilon) \in \nn$ such that if $n \geq \bar{N} ( \epsilon)$, then both
\begin{equation} \label{eq:closeness}
\left|  \newJ^{(n)}_{\pi} (\nu_{\pi^{*}}) - \bar{\newJ}^{(n)}_{\pi} ( \nu_{\pi^{*}} )		\right| < \frac{\epsilon}{4} \; \text{ and }  \left| 	 \bar{\newJ}^{(n)}_{\pi} ( \nu_{\pi^{*}} ) -  \JJ_{\MM_{\nu_{\pi^{*}}}} ( \pi,  \nu_{\pi^{*}} ) 	\right| < \frac{\epsilon}{4} \quad   \forall \pi \in \Pi_{\bM}.
\end{equation}

Since $(\pi^{*}, \nu_{\pi^{*}})$ is a SMFE for the system, we know $\JJ_{\MM_{\nu_{\pi^{*}}}} ( \pi^{*},  \nu_{\pi^{*}} )   \leq \JJ_{\MM_{\nu_{\pi^{*}}}} ( \pi,  \nu_{\pi^{*}} )  $ for any $\pi \in \Pi_{\bM}$. Then, by \eqref{eq:closeness}, we have
\[
\JJ_{\MM_{\nu_{\pi^{*}}}} ( \pi,  \nu_{\pi^{*}} )      \leq    \newJ^{(n)}_{\pi} (\nu_{\pi^{*}} ) + \frac{\epsilon}{2}, \quad \forall \pi \in  \Pi_{\bM}.
\]

Also by \eqref{eq:closeness}, we have $\newJ^{(n)}_{\pi^{*}} ( \nu_{\pi^{*}} ) \leq \JJ_{\MM_{\nu_{\pi^{*}}}} ( \pi^{*},  \nu_{\pi^{*}} )  + \epsilon/2.$  Combining the previous observations, we have that 
\[
\newJ^{(n)}_{\pi^{*}} ( \nu_{\pi^{*}} ) \leq  \newJ ^{(n)}_{\pi} (  \nu_{\pi^{*}} )    + \epsilon, \quad \forall \pi \in \Pi_{\bM}.
\]

This shows that $\pi^{*}$ is an $\epsilon$-best-response to $\bpi^{*-1}_{(n)}$ with respect to the initial measure $\bnu^{(n)}_{\pi^{*}}$ in the game $\GG_n ( \nu_{\pi^{*}})$.   
\end{proof}

\subsection*{Remarks}

As we have observed in the preceding analysis, under some regularity conditions on the partially observed $n$-player mean-field game with local state observability, if the number of agents is sufficiently large and the agents happen to be using the policy $\pi^{*}$ given by a stationary mean-field equilibrium, \textit{and} if the initial global state is distributed according to the associated invariant measure $\nu_{\pi^{*}}$, then the environment faced by a learning agent is approximated by the MDP $\MM_{\nu_{\pi^{*}}}$. 

Although the summary above is highly qualified, we observe that if every agent uses policy $\pi^{*}$ during a given exploration phase during the running of Algorithm~\ref{algo:main} and (for one reason or another) does not switch policies at the end of that exploration phase, then the initial state of the subsequent exploration phase will be distributed according to a measure that is close to $\nu_{\pi^{*}}$, due to the assumed ergodicity of the system. Thus, it is conceivable that the naive learning iterates of an agent in the $n$-player game will be approximated by the Q-function and state value function for the MDP $\MM_{\nu_{\pi^{*}}}$, leading to a possible avenue for establishing {subjective $\epsilon$-equilibrium} in the truly decentralized problem.

\section{Satisficing Under Other Observation Channels}  \label{appendix:satisficing-under-other-observation-channels}

\subsubsection*{Subjective Satisficing Paths under Global Observability}

\begin{definition}
Let $\bG$ be a partially observed $n$-player mean-field game for which Assumption~\ref{ass:global-state-observability} is satisfied, and let $i \in \NN$. A stationary policy $\pi^i \in \Pi^i_{S}$ is said to be \emph{of the mean-field type} if there exists $f^i \in \PP ( \aa | \xx \times \Emp_N )$ such that $\pi^i ( \cdot | \bs ) = f^i \left( \cdot \middle| s^i , \upmu ( \bs )  \right)$ for every global state $\bs \in \bX$. 
\end{definition}

We identify each stationary policy of the mean-field type with its associated transition kernel in $\PP ( \aa | \xx \times \Emp_N )$.

\begin{definition}
Let $\bG$ be a partially observed $n$-player mean-field game for which Assumption~\ref{ass:global-state-observability} is satisfied. For $i, j \in \NN$, let $\pi^i \in \Pi^i_{S}$ and $\pi^j \in \Pi^j_{S}$ both be of the mean-field type. We say that $\pi^i$ and $\pi^j$ are \emph{mean-field symmetric} if they are identified with the same transition kernel in $\PP ( \aa | \xx \times \Emp_N)$.
\end{definition}

\begin{lemma}   \label{lemma:global-subjective-BR-iff}
Let $\bG$ be a partially observed $n$-player mean-field game for which Assumptions~\ref{ass:global-state-observability} and \ref{ass:state-visitation} are satisfied, and let $\epsilon \geq 0$. Let $\bpi \in \bPi_{S}$ be a {soft} stationary joint policy for which $\pi^p$ is of the mean-field type for each player $p \in \NN$. Suppose that, for some $i, j \in \NN$, we have that $\pi^i$ and $\pi^j$ are mean-field symmetric, then we have 
\[
\pi^i \in \SubjBR^i_{\epsilon} ( \bpi^{-i} , \VV^{*} ,\WW^{*} ) \iff  \pi^j \in \SubjBR^j_{\epsilon} ( \bpi^{-j} , \VV^{*} ,\WW^{*} ).
\]
\end{lemma}

\begin{proof}   \label{proof:global-subjective-BR-iff}
Let $i, j \in \NN$. For each $\bs \in \bX$, we define $\swap_{ij} (\bs) \in \bX $ to be the global state such that (1) $\swap_{ij} (\bs)^p = s^p$ for each $p \in \NN \setminus \{ i, j \}$, and (2) we have $\swap_{ij} (\bs)^j = s^i$ and $\swap_{ij}(\bs)^i = s^j$.

By Theorem~\ref{theorem:naive-learning}, we have that $V^{*i}_{\bpi} (\bs) = J^i ( \bpi , \bs)$ for each $\bs \in \bX$, $W^{*i}_{\bpi} = Q^{*i}_{\bpi^{-i}}$,  $V^{*j}_{\bpi} (\bs) = J^j ( \bpi , \bs)$ for each $\bs \in \bX$, and $W^{*j}_{\bpi} = Q^{*j}_{\bpi^{-j}}$. Thus, we have that $\pi^i \in   \SubjBR^i_{\epsilon} ( \bpi^{-i} , \VV^{*} ,\WW^{*} )$ if and only if $\pi^i \in \BR^i_{\epsilon} ( \bpi^{-i})$, and analogously for $j$.

Toward obtaining a contradiction, assume that $\pi^i \in  \SubjBR^i_{\epsilon} ( \bpi^{-i} , \VV^{*} ,\WW^{*} )$ while $\pi^j \notin  \SubjBR^j_{\epsilon} ( \bpi^{-j} , \VV^{*} ,\WW^{*} )$. Equivalently, assume that $\pi^i \in \BR^i_{\epsilon} (\pi^{-i} )$ while  $\pi^j \notin \BR^i_{\epsilon} (\pi^{-j} )$. That is, we have that
\[
J^i (\bpi ,\bs) \leq \inf_{ \tilde{\pi}^i \in \Pi^i_{S}} J^i ( \tilde{\pi}^i, \bpi^{-i}  ,  \bs ) + \epsilon, \quad \forall \bs \in \bX,
\]
while there exists $\bs^{*} \in \bX$ such that $J^j ( \bpi, \bs^{*})  > \inf_{\tilde{\pi}^j \in \Pi^j_{S}} J^j ( \tilde{\pi}^j , \bpi^{-j}  , \bs^{*} ) + \epsilon. $

Observe that if $\bpi \in \bPi_{S}$ is of the mean-field type and $\tilde{\pi}^i \in \Pi^i_{S}$ is any stationary policy for player $i$ (not necessarily of the mean-field type), then defining a policy $\tilde{\pi}^j \in \Pi^j_{S}$ state-wise by $\tilde{\pi}^j ( \cdot | \bs ) = \tilde{\pi}^i ( \cdot | \swap_{ij} (\bs ))$ for all $\bs$, we have that 
\[
J^i ( \tilde{\pi}^i, \bpi^{-i}   , \bs ) = J^j  ( \tilde{\pi}^j, \bpi^{-j}  ,  \swap_{ij} ( \bs ) ), \quad \forall \bs \in \bX. 
\]

\noindent It follows that, first, we have $J^i ( \bpi , \bs) = J^j ( \bpi ,  \swap_{ij} ( \bs ) )$ for each $\bs \in \bX$ and furthermore
\[
\inf_{ \tilde{\pi}^i \in \Pi^i_{S} } J^i ( \tilde{\pi}^i , \bpi^{-i} ,  \bs ) = \inf_{ \tilde{\pi}^j \in \Pi^j_{S} } J^j  ( \tilde{\pi}^j , \bpi^{-j} ,  \swap_{ij} (\bs ) ), \quad \forall \bs \in \bX.
\]
In particular, this holds for $\swap_{ij} ( \bs^{*} )$, which yields
\begin{align*}
J^j ( \bpi ,  \bs^{*} ) > \inf_{ \tilde{\pi}^j \in \Pi^j_{S} } J^j ( \tilde{\pi}^j , \bpi^{-j}  ,  \bs^{*} ) + \epsilon &= \inf_{\tilde{\pi}^i \in \Pi^i_{S} } J^i ( \tilde{\pi}^i , \bpi^{-i}  ,  \swap_{ij} (\bs^{*} ) ) + \epsilon  \\
&\geq J^i ( \bpi ,  \swap_{ij} (\bs^{*} )) = J^j ( \bpi ,  \bs^{*} ),
\end{align*}
a contradiction, which completes the proof. 
\end{proof}

\begin{lemma}   \label{lemma:global-quantized-paths}
Let $\bG$ be a partially observed $n$-player mean-field game satisfying Assumptions~\ref{ass:global-state-observability} and \ref{ass:state-visitation}. Let $\epsilon > 0$. There exists $\xi = \xi ( \epsilon ) > 0 $ such that if $\widehat{\bPi} \subset \bPi_{S}$ is any soft $\xi$-quantization of $\bPi_{S}$, then we have 
\begin{enumerate}
	\item[1)] $\eq[\epsilon] \cap \widehat{\bPi} \not= \varnothing$ and $\Subj_{\epsilon} ( \VV^{*} , \WW^{*} ) \cap \widehat{\bPi} \not= \varnothing$.
\end{enumerate}

Moreover, if $\tilde{\bPi} \subset \bPi_{S}$ is the set of joint stationary policies of the mean-field type, there exists $\xi = \xi ( \epsilon ) > 0 $ such that if $ {\bPi}^{\prime}$ is a soft, symmetric $\xi$-quantization of $\tilde{\bPi}$, then we have
\begin{enumerate}
	\item[2)] $\Subj_{\epsilon} ( \VV^{*} , \WW^{*} ) \cap \bPi^{\prime} \not= \varnothing$;
	\item[3)] $\bG$ has the $(\VV^{*}, \WW^{*} )$-subjective $\epsilon$-satisficing paths property within $\bPi^{\prime}$. 
\end{enumerate}
\end{lemma}

\begin{proposition}  \label{proposition:global-paths-property}
Let $\bG$ be a partially observed $n$-player mean-field game for which Assumptions~\ref{ass:global-state-observability} and \ref{ass:state-visitation} hold. Let $\epsilon > 0$, and let $(\VV^{*}, \WW^{*})$ be the subjective function family for $\bG$. Suppose $\widehat{\bPi} \subset \bPi_{S}$ is a subset of stationary joint policies satisfying the following properties: (i) Every $\bpi \in \widehat{\bPi} $ is of the mean-field type; (ii) $\widehat{\bPi}  \cap \Subj_{\epsilon} ( \VV^{*}, \WW^{*} ) \not= \varnothing$; (iii) the set $\widehat{\bPi} $ is symmetric, i.e. $\widehat{\Pi} ^i = \widehat{\Pi} ^j$ for each $i, j \in \NN$; (iv) every $\bpi \in \widehat{\bPi} $ is soft. 

Then, $\bG$ has the $(\VV^{*} , \WW^{*})$-subjective $\epsilon$-satisficing paths property within $\widehat{\bPi} $. 
\end{proposition}

\noindent The proof of Proposition~\ref{proposition:global-paths-property} is omitted, as it parallels that of \cite[Theorem 3.6]{yongacoglu2023satisficing}, which was in the context of symmetric, $n$-player stochastic games with global observability. One important difference, which explains the need for Lemma~\ref{lemma:global-subjective-BR-iff} and restriction to policies of the mean-field type, is that here the global state dynamics do not satisfy the conditions outlined in \cite{yongacoglu2023satisficing}. This occurs because of the added structure on the global state space present in mean-field games.

\subsubsection*{Subjective Satisficing Paths under Compressed Observability}

In light of our previous discussion on the possible non-existence of subjective $\epsilon$-equilibrium in games with compressed observability, we now give a qualified result analogous to Proposition~\ref{proposition:global-paths-property} for the case with compressed observability. In contrast to the earlier results, here we must assume the existence of subjective equilibrium.

\begin{definition}
Let $\bG$ be a partially observed $n$-player mean-field game and let $i \in \NN$. A policy $\pi^i \in \Pi^i_{S}$ is said to be \emph{of the local type} if there exists a transition kernel $g^i \in \PP ( \aa | \xx )$ such that 
\[
\pi^i ( \cdot | \varphi^i ( \bs ) ) = g^i ( \cdot  | s^i ), \quad \forall \bs \in \bX.
\]
\end{definition}

Each policy of the local type is identified with the corresponding transition kernel in $\PP ( \aa | \xx )$. For players $i, j \in \NN$, if the policies $\pi^i \in \Pi^i_{S}, \pi^j \in \Pi^j_{S}$ are both of the local type and are identified with the same transition kernel, we say $\pi^i$ and $\pi^j$ are \emph{locally symmetric}. 

\begin{lemma} \label{lemma:compressed-subjective-BR-iff}
Let $\bG$ be a partially observed $n$-player mean-field game for which Assumptions~\ref{ass:compressed-observability} and \ref{ass:state-visitation} hold. Let $\bpi \in \bPi_{S}$ be soft. For $i, j \in \NN$, suppose $\pi^i$ and $\pi^j$ are locally symmetric. Then, we have $\pi^i \in \SubjBR^{i}_{\epsilon} ( \bpi^{-i} , \VV^{*}, \WW^{*} )$ if and only if $\pi^j \in \SubjBR^{j}_{\epsilon} ( \bpi^{-j} , \VV^{*}, \WW^{*} ).$
\end{lemma}

\noindent The proof of Lemma~\ref{lemma:compressed-subjective-BR-iff} mirrors the proof of Lemma~\ref{lemma:mean-field-subjective-BR-iff} and is therefore omitted.

\begin{proposition}  \label{proposition:compressed-paths-property}
Let $\bG$ be a partially observed $n$-player mean-field game for which Assumptions~\ref{ass:compressed-observability} and \ref{ass:state-visitation} hold. Let $\epsilon > 0$. Assume that $\Subj_{\epsilon}  ( \VV^{*}, \WW^{*} ) \not= \varnothing$. Suppose $\widehat{\bPi} \subset \bPi_{S}$ is a subset of policies satisfying (i) $\widehat{\bPi} \cap \Subj_{\epsilon} ( \VV^{*}, \WW^{*} ) \not= \varnothing$; (ii) the set $\widehat{\bPi}$ is symmetric, i.e. $\widehat{\Pi}^i = \widehat{\Pi}^j$ for each $i, j \in \NN$; (iii) each policy $\bpi \in \widehat{\bPi}$ is soft. Then, $\bG$ has the $(\VV^{*} , \WW^{*})$-subjective $\epsilon$-satisficing paths property within $\widehat{\bPi} $. 
\end{proposition}

As with Proposition~\ref{proposition:global-paths-property}, Proposition~\ref{proposition:compressed-paths-property} can be proved using the argument of \cite[Theorem 3.6]{yongacoglu2023satisficing}, suitably modified to use Lemma~\ref{lemma:compressed-subjective-BR-iff} instead of \cite[Corollary 2.9]{yongacoglu2023satisficing}.

\subsubsection*{Convergence of the Oracle Algorithm under Global Observability}

\begin{lemma} \label{lemma:global-oracle}
Let $\bG$ be a partially observed $n$-player mean-field game satisfying Assumptions~\ref{ass:global-state-observability} and \ref{ass:state-visitation}, and let $\epsilon > 0$. Let $\widehat{\bPi} \subset \bPi_{S}$ be a quantization of the subset of policies of the mean-field type, $\{ \bpi \in \bPi_{S} : \pi^i \text{ is of the mean-field type } \forall i \in \NN \}$. Suppose $\widehat{\bPi}$ satisfies (1) $\widehat{\bPi}  \cap \Subj_{\epsilon} ( \VV^{*}, \WW^{*} ) \not= \varnothing$; (2) $\widehat{\Pi}^i = \widehat{\Pi}^j$ for all $i,j \in \NN$; and (3) every policy $\bpi \in \widehat{\bPi} $ is soft.  

Suppose that each agent $i \in \NN$ updates is policy sequence $\{ \pi^i_k \}_{ k \geq 0}$ according to Algorithm~\ref{algo:oracle} and that, for each $k \geq 0$, the policy updates for $\bpi_{k+1}$ are conditionally independent across agents given $\bpi_k$. Then, $\lim_{k \to \infty} \P ( \bpi_k \in \widehat{\bPi}  \cap \eq[\epsilon] ) = 1. $
\end{lemma}

The proof of Lemma \ref{lemma:global-oracle} is essentially the same as that of Lemma~\ref{lemma:mean-field-oracle}. One important difference between Lemmas \ref{lemma:global-oracle} and \ref{lemma:mean-field-oracle} is such: in the former, convergence to an objective equilibrium is guaranteed, while in the latter one only has convergence to a subjective equilibrium.

\section{Learning Results Under Other  Observation Channels}   \label{appendix:learning-results-under-other-observation-channels}

\subsection{Learning with Global Observability} \label{ss:global-state}
We now present convergence results for Algorithm~\ref{algo:main} under global observability, the richest of the information structures that we consider. Under Assumption~\ref{ass:global-state-observability}, strong convergence guarantees can be made. These are presented below in Theorem~\ref{theorem:global-state-observability}. In order to state this result, we now fix $\epsilon > 0$ and make the following assumptions on the various parameters of Algorithm~\ref{algo:main}.

\begin{assumption} \label{ass:global-quantized-set}
Fix $\epsilon > 0$. For each $i \in \NN$ define $\Pi^i_{S, \mf}$ as
\[
\Pi^i_{S, \mf} := \{ \pi^i \in \Pi^i_{S}: \pi^i \text{ is of the mean-field type.}\}.
\]
Assume that $\widehat{\bPi} \subset \times_{i \in \NN } \Pi^i_{S, \mf}$ is a fine quantization of $\times_{i \in \NN} \Pi^i_{S,\mf}$ satisfying: (1) $\widehat{\Pi}^i = \widehat{\Pi}^j$ for all $i,j \in \NN$; (2) $\widehat{\bPi} \cap \eq[\epsilon] \not= \varnothing$; (3) For any $\bpi \in \widehat{\bPi}$, $\bpi$ is soft.
\end{assumption}

Next we present a restriction on the parameters $\{ d^i \}_{i \in \NN}$. For each player $i \in \NN$, the tolerance parameter $d^i$ is taken to be positive, to account for noise in the learned estimates, but small, so that poorly performing policies are not mistaken for $\epsilon$-best-responses. The bound $\bar{d}_{G}$ below is defined analogous to the term $\bar{d}$ in \cite{yongacoglu2023satisficing} and depends on both $\epsilon$ and $\widehat{\bPi}$. %

\begin{assumption} \label{ass:rho-delta}
For each player $i \in \NN$, $d^i \in ( 0, \bar{d}_G)$, where $\bar{d}_G = \bar{d}_G ( \epsilon, \widehat{\bPi})$ is  defined as $\bar{d}_{G} := \min  \mathcal{O}_{G} $, where $\mathcal{O}_{G} := S_{G} \setminus \{ 0 \}$, and  $S_{G}$ is given by 
\[
S_{G} := \left\{   \left| 	\epsilon - \left(   V^{*i}_{\bpi } ( y ) - \min_{a^i \in \aa } W^{*i}_{\bpi  } ( y, a^i ) \right) \right| : i \in \NN, \bpi \in \widehat{\bPi} , y \in \yy \right\}.
\]

\end{assumption}

\begin{theorem} \label{theorem:global-state-observability} 
Let $\bG$ be a partially observed $n$-player mean-field game satisfying Assumptions~\ref{ass:global-state-observability} and \ref{ass:state-visitation}, and let $\epsilon > 0$. Suppose the policy set $\widehat{\bPi}$ and the tolerance parameters $\{ d^i \}_{i \in \NN}$ satisfy Assumptions~\ref{ass:global-quantized-set} and \ref{ass:rho-delta}, and suppose all players follow Algorithm~\ref{algo:main}. For any $\xi > 0$, there exists $\tilde{T} = \tilde{T} ( \xi , \epsilon, \widehat{\bPi} ,  \{ d^i \}_{i \in \NN} )$ such that if $T_k \geq \tilde{T}$ for all $k$, then 
\[
\P \left( \bpi_k \in \widehat{\bPi} \cap \eq[\epsilon] \right) \geq 1 - \xi, 
\] 
for all sufficiently large $k$. 
\end{theorem}

The details of the proof of Theorem~\ref{theorem:global-state-observability} resemble those of Theorem~\ref{theorem:mean-field-state-observability}, with the following exceptions: here, $\bar{d}_G$ replaces $\bar{d}_{\mf}$, and the term $\Xi$ is now defined as $\Xi := \frac 1 2 \min_{i \in \NN} \{ d^i , \bar{d}_G - d^i \}$. With these modifications, the mechanics of the proofs are identical.

\subsection{Learning with Compressed Observability State} \label{ss:compressed-observability}

We conclude with a discussion of convergence guarantees for Algorithm~\ref{algo:main} under compressed observability (Assumption~\ref{ass:compressed-observability}). As we have discussed previously, there is no guarantee that the set $\bPi_{S}$ contains $\epsilon$-equilibrium policies---either in the objective sense or in the subjective sense using the naively learned subjective function family $(\VV^{*}, \WW^{*})$. As a result, the convergence guarantees in this setting are highly qualified, and must be made with a potentially restrictive assumption that subjective $\epsilon$-equilibrium exist within the set of policies $\widehat{\bPi}$. 

\begin{assumption}  \label{ass:subjective-local-equilibria-exist} 
Fix $\epsilon > 0$. Assume that the set of $(\VV^{*}, \WW^{*})$-subjective $\epsilon$-equilibrium is non-empty. That is, $ \Subj_{\epsilon} ( \VV^{*} , \WW^{*}  ) \not= \varnothing . $
\end{assumption}

\begin{assumption} \label{ass:compressed-quantized-set}
Fix $\epsilon > 0$. Assume that $\widehat{\bPi}$ is a fine quantization of $\bPi_{S}$ satisfying: (1) $\widehat{\Pi}^i = \widehat{\Pi}^j$ for all $i,j \in \NN$; (2) $\widehat{\bPi} \cap \Subj_{\epsilon} ( \VV^{*}, \WW^{*}) \not= \varnothing$; (3) For any $\bpi \in \widehat{\bPi}$, $\bpi$ is soft.
\end{assumption}

\begin{assumption} \label{ass:compressed-delta}
For all $i \in \NN$, $d^i  \in (0, \bar{d}_{ \rm comp } )$, where $\bar{d}_{\rm comp} = \bar{d}_{ \rm comp} ( \epsilon, \widehat{\bPi})$ is defined as $\bar{d}_{\comp} := \min \mathcal{O}_{\comp}$, where $\mathcal{O}_{\comp} := S_{\comp} \setminus \{ 0 \}$ and 
\[
S_{\comp} := \left\{   \left| 	\epsilon - \left(   V^{*i}_{\bpi } ( y ) - \min_{a^i \in \aa } W^{*i}_{\bpi  } ( y, a^i ) \right) \right| : i \in \NN, \bpi \in \widehat{\bPi}  , y \in \yy \right\}.
\]
\end{assumption}

\begin{theorem} \label{theorem:compressed-observability}
Let $\epsilon > 0$ and let $\bG$ be a partially observed $n$-player mean-field game satisfying Assumptions~\ref{ass:compressed-observability}, \ref{ass:state-visitation}, and \ref{ass:subjective-local-equilibria-exist}. Suppose the policy set $\bPi$ and the tolerance parameters $\{ d^i \}_{i \in \NN}$ satisfy Assumptions~\ref{ass:compressed-quantized-set} and \ref{ass:compressed-delta}, and suppose all players follow Algorithm~\ref{algo:main}. For any $\xi > 0$, there exists $\tilde{T} = \tilde{T} ( \xi , \epsilon, \widehat{\bPi} ,  \{ d^i \}_{i \in \NN} )$ such that if $T_k \geq \tilde{T}$ for all $k$, then 
\[
\P \left( \bpi_k \in \widehat{\bPi} \cap \Subj_{\epsilon} ( \VV^{*}, \WW^{*} ) \right) \geq 1 - \xi, 
\] 
for all sufficiently large $k$. 
\end{theorem}

The proof of Theorem~\ref{theorem:compressed-observability} parallels the proof of Theorem~\ref{theorem:mean-field-state-observability}, with $\bar{d}_{\comp}$ replacing $\bar{d}_{\mf}$ and $\Xi$ redefined as  $\Xi := \frac 1 2 \min_{i \in \NN} \{ d^i , \bar{d}_{\comp} - d^i \}$. The rest of the proof goes through unchanged. We note that the requisite satisficing paths structure holds by Assumption~\ref{ass:compressed-quantized-set} and the usual line of argument following Proposition~\ref{proposition:compressed-paths-property}

\end{document}